\documentclass[
aip,jcp,amsmath,amssymb, reprint, twocolumn, floatfix
]{revtex4-2}
\usepackage{amsmath}
\usepackage{mathtools}
\usepackage{float}
\usepackage{amsfonts}
\usepackage{amssymb}
\usepackage{dcolumn} %% tables cols aligned at decimal point
\usepackage{array}
\newcolumntype{P}[1]{>{\centering\arraybackslash}p{#1}}
\newcolumntype{M}[1]{>{\centering\arraybackslash}m{#1}}
\newcolumntype{C}[1]{>{\centering\arraybackslwash}p{#1}}
\usepackage{float}
\usepackage{psfrag}
\usepackage{tabularx}
\usepackage{stackengine}
\usepackage{amssymb}
\usepackage{mathtools}  
\usepackage{xfrac} 
\usepackage[T1]{fontenc}
\usepackage{graphicx}
\usepackage{diagbox}
\usepackage[caption=false]{subfig}
\usepackage{tikz}
\usetikzlibrary{positioning}
\usetikzlibrary{arrows}
\usetikzlibrary{trees}
\usepackage{amssymb}
\usetikzlibrary{decorations.pathmorphing}
\usetikzlibrary{decorations.markings}
\usetikzlibrary{automata,positioning}
\usepackage{braket}
\usepackage{hyperref}
\usepackage{rotating}
\usepackage{adjustbox}
\usepackage{ragged2e}
\usepackage{simplewick}
\usepackage{simpler-wick}
\usepackage{booktabs}
\usepackage{multirow}
\usepackage[]{lineno}

\usepackage{float}  
\usepackage{algorithm}
\usepackage{algpseudocode}

\usepackage{csquotes}
\usepackage{physics,amsmath}
\usepackage{xcolor}
\usepackage{fancyhdr}
\UseRawInputEncoding
\usepackage[normalem]{ulem}
\usepackage{scalerel}
\usepackage{algorithm}
\usepackage{algpseudocode}

\begin{document}

\author{Sonaldeep Halder}
\affiliation{ Department of Chemistry,  \\ Indian Institute of Technology Bombay, \\ Powai, Mumbai 400076, India}

\author{Chayan Patra}
\affiliation{ Department of Chemistry,  \\ Indian Institute of Technology Bombay, \\ Powai, Mumbai 400076, India}

\author{Rahul Maitra}
\email{rmaitra@chem.iitb.ac.in}
\affiliation{ Department of Chemistry,  \\ Indian Institute of Technology Bombay, \\ Powai, Mumbai 400076, India}
\affiliation{Centre of Excellence in Quantum Information, Computing, Science \& Technology, \\ Indian Institute of Technology Bombay, \\ Powai, Mumbai 400076, India}

\title{Quantum Wavefunction Augmentation via Variational Autoencoders}
%\linenumbersCorrections

\begin{abstract}
Sample-based quantum diagonalization (SQD) has emerged as a promising route for quantum-centric supercomputing, relying on classical diagonalization of the molecular Hamiltonian within a hardware-sampled determinant subspace. However, its accuracy degrades in strongly correlated regimes where the relevant determinant space exceeds what finite-shot sampling can capture. In this work, we introduce Quantum Wavefunction Augmentation via Variational Autoencoders (Q-WAVE), a hybrid method that combines determinants sampled via SqDRIFT Krylov circuits and configuration interaction singles and doubles (CISD) determinants with generative machine learning. Using a custom $\beta$-annealed variational autoencoder (VAE) model, Q-WAVE iteratively expands this basis toward the variational ground state. The VAE learns the wavefunction's primary support structure from the combined hardware and CISD seed in a continuous latent space, generating new dominant determinants beyond any fixed excitation hierarchy. The resulting compact wavefunction exceeds what can be extracted from raw hardware samples alone. We demonstrate sub-millihartree accuracy compared to full configuration interaction for $\text{H}_2\text{O}$ and $\text{N}_2$ dissociation. Finally, we establish Q-WAVE's scalability on a 52-qubit ethylene system (achieving sub-millihartree accuracy versus CCSD(T)) and a highly correlated 60-qubit $\text{Cr}_2$ stress test that attains chemical accuracy upon a final perturbative correction.
\end{abstract}

\maketitle
\section{Introduction}
Computing the ground-state wavefunction for many-body systems remains a core challenge in fields ranging from high-energy physics to quantum chemistry. While fault-tolerant quantum computers promise exact solutions, current pre-fault-tolerant devices, despite exceeding the 100-qubit threshold, remain fundamentally constrained by noise. Early variational quantum eigensolver (VQE) frameworks \cite{peruzzo2014variational, McClean_2016, Kandala2017-qp, Romero_2019, Grimsley2019, Cerezo2021} showed initial promise in noisy settings but were fundamentally hindered by deep quantum circuits, impractically long optimization cycles, local traps\cite{anschuetz2022quantum} and barren plateaus \cite{McClean2018,larocca2025barren}, and require significant improvements for practical scalability\cite{Grimsley2023,zzawa2020jastrow,yordanov2020efficient,rivera2021avoiding,mondal2023development,sonaldeep2023,halder2024noise,patra2024projective,patra2024toward,10.1039/d3sc05807g,patra2025energy, ding2026simple, wang2025shadow,patel2026quantum,mondal2026advancing}. Consequently, recent efforts have shifted toward quantum-centric supercomputing methods, which utilize quantum processors for state preparation and sampling, followed by classical diagonalization. Methods such as Krylov quantum diagonalization (KQD) \cite{Yoshioka2025}, quantum selected configuration interaction (QSCI)\cite{kanno2026quantum},
sample-based quantum diagonalization (SQD) \cite{doi:10.1126/sciadv.adu9991} and its different variants\cite{shajan2025toward,danilov2025enhancing,mikkelsen2025quantum,merz2026crossing,shirakawa2026closed,shajan2026molecular,wang2026localized,yamamoto2026quantum,kamoshita2026qsci} have successfully pushed molecular simulations beyond the limits of pure exact classical diagonalization. A combination of the above-mentioned KQD and SQD methods culminated in sample-based KQD (SKQD) \cite{yu2025quantumcentricalgorithmsamplebasedkrylov} and its randomized compilation variant, SqDRIFT \cite{piccinelli2026quantumchemistryprovableconvergence}, which achieves provable convergence while maintaining circuit depths independent of Hamiltonian term counts. 

Despite these advances, the accuracy of sample-based methods remains strictly bounded by finite shot budgets. In strongly correlated regimes, such as bond dissociations and transition-metal clusters, the true ground state exhibits a long \enquote{tail} of determinants with non-negligible weights. Realistic shot budgets cannot sample this combinatorially vast space, and classical configuration-recovery steps of SQD cannot reconstruct determinants that were never originally sampled \cite{doi:10.1021/acs.jctc.5c00375}. Moreover, repeated measurements tend to re-select previously known configurations, making the discovery of new determinants prohibitively difficult. Classical selected CI methods, such as heat-bath CI (HCI) \cite{10.1021/acs.jctc.6b00407} and its stochastic variants (SHCI)\cite{10.1021/acs.jctc.6b01028}, approach this problem from the opposite direction, iteratively expanding the variational space based on a computationally efficient, deterministic connectivity criterion. However, because candidates at each step are strictly drawn from the first-order interacting space of the current basis, a determinant lying several excitations away can only be reached through a continuous chain of intermediates. Importantly, each intermediate in this chain must independently survive the selection threshold. This stepwise discovery route becomes highly unreliable in strongly multireference regimes, where critical dominant configurations are rarely connected to the reference by a smooth, heavily weighted excitation path. This leaves a critical gap -- the need for a scalable mechanism to systematically grow the variational determinant space toward the wavefunction's true support, bypassing the limitations of both finite hardware sampling and rigid classical excitation cutoffs.

In this work, we introduce a hybrid framework, called Quantum Wavefunction Augmentation via Variational Autoencoders (Q-WAVE), that combines hardware-sampled and configuration interaction singles and double (CISD) determinants with a generative machine learning (ML) architecture to iteratively expand the variational space. Use of ML models to represent and expand the quantum wave functions has previously been reported\cite{doi:10.1126/science.aag2302, 10.1039/d3sc05807g, 10.1021/acs.jpca.5c02346, doi:10.1021/acs.jctc.2c01216,patra2026physicsinformedgenerativemachinelearning, patra2026machinelearnedcompactsubspacegeneration, vargas2026machinelearningsamplebasedquantum,10.1021/acs.jctc.9b00828}. The classical CIgen approach \cite{doi:10.1021/acs.jctc.2c01216} iteratively trains a Restricted Boltzmann machine (RBM) on a CISD level wavefunction to generate important determinants. However, being seeded purely classically, the learned distribution remains confined to the manifold reachable from a single-reference excitation
hierarchy only. This limitation becomes decisive for strongly multireference systems. In a complementary direction, our earlier
PIGen-SQD framework \cite{patra2026physicsinformedgenerativemachinelearning} introduced physics-informed generative model into the quantum-centric pipeline, but more towards a \textit{recovery} setting -- reconstructing the dominant configurations
from noisy hardware measurements, with a reach ultimately bounded by the information content of the samples themselves. The present work unifies
and extends both directions. Hardware samples inject a multireference character that no classically seeded generative model can reach, while the generative loop expands the subspace far beyond anything the hardware ever measured. 

Our method uses SqDRIFT circuits to gather relevant samples from the quantum hardware, which offers the advantage of being tunable in circuit depth. We augment these hardware samples with deterministic CISD determinants to act as a base for the initial wavefunction. This combined set of determinants, filtered to conserve particle and $S_z$ (the spin projection along $z$; $S_z=0$ for all closed-shell systems studied here), is then fed into a custom $\beta$-annealed Variational Autoencoder (VAE)\cite{kingma2022autoencodingvariationalbayes, rezende2014stochasticbackpropagationapproximateinference, higgins2017betavae, bowman2016generatingsentencescontinuousspace}. The VAE encodes the training data (set of determinants) onto a continuous latent space. By regularizing the continuous latent space via Kullback-Leibler (KL) divergence, the VAE enforces an overlapping, densely packed representation of the determinant space. In this latent geometry, chemically related determinants occupy neighboring coordinates, transforming determinant discovery from a blind combinatorial search into a targeted geometric exploration. We can thus sample random coordinates from this space and decode them to generate highly realistic, novel determinants not originally present in the initial training set. Specifically, we use a tailored $\beta$-annealed VAE with clever prior sampling and posterior perturbation techniques for both broad and local exploration of new determinants. This exploration-and-exploitation mechanism enables VAE to efficiently reach completely new important determinants, as well as examine determinants closer to those it has already seen during training. These generated determinants are evaluated for importance through classical Hamiltonian diagonalization and again added to the VAE training set. This way, the wavefunction slowly grows towards the true support of the ground state through a generation-diagonalization loop.
Our method has been implemented to compute the ground-state energies of water, nitrogen, and ethylene molecules. As an extreme stress test, we also applied it to the strongly correlated metallic $\mathrm{Cr_2}$ dimer. For all systems, the method achieves chemical accuracy relative to FCI (where feasible) or other reference methods. For $\mathrm{Cr_2}$, a standard second-order perturbative energy correction was required (like other classical selected CI methods) to reach this accuracy threshold; for all other systems, the variational energy at the end of the ML loop was already within sub-millihartree accuracy. More broadly, these results suggest that the route to scalability in quantum-centric electronic structure methods lies not in ever-larger shot budgets but in learning the wavefunction's structure from a limited number of quantum samples.

\section{Theory}
This section outlines the theoretical and algorithmic foundations of the Q-WAVE framework. In Subsection A, we briefly describe the SqDRIFT circuit preparation, which represents the sole quantum hardware-dependent step in our pipeline. Subsection B details the core generative methodology, explaining how our custom VAE infrastructure identifies and generates dominant determinants. Subsection C outlines an iterative end-to-end protocol involving Q-WAVE together with the second-order perturbative correction that yields highly accurate molecular ground-state energetics. Finally, Subsection D describes the scalability of our method. This leads to the Results and Discussion section, where we apply our method to several molecular systems.

\subsection{The SqDRIFT method}
The initial samples from the quantum hardware are obtained using the SqDRIFT circuits together with an efficient $F2Q$-style circuit mapping strategy that reduces the two-qubit gate cost by placing orbitals that appear together in the same excitation term on nearby qubit indices. This strategy follows the locality-optimization principle described in Ref. \cite{piccinelli2026quantumchemistryprovableconvergence}, to which we refer the reader for technical details. SqDRIFT is a combination of SKQD and qDRIFT methods. Let $\hat{H}$ be an $n$-qubit Hamiltonian with the exact ground state $\ket{\phi_o}$, $\ket{\psi_{init}}$ be a reference wavefunction having a non-negligible overlap with $\ket{\phi_o}$, and $t$ be a reference time. The SKQD protocol constructs $d$ time-evolved states in a quantum hardware - 
\begin{equation}\label{skqd_parent}
    \ket{\psi_u}=(\prod_{j=1}^ue^{-i\hat{H}t})\ket{\psi_{init}}=e^{-i\hat{H}ut}\ket{\psi_{init}}
\end{equation}

for $u \in \{0,1,2,...,d-1\}$. Samples are taken from the quantum hardware for each prepared state, $\ket{\psi_u}$, then merged together, and the Hamiltonian is diagonalized in this subspace. However, the quantum circuits required to prepare Eq. \ref{skqd_parent} are deep for a highly non-local molecular Hamiltonian. Thus, a quantum stochastic drift (qDRIFT) method is used to compile these time-evolved circuits. Let the Hamiltonian be written as a sum of $\mathcal{N}$ terms:

  \begin{figure*}
  \includegraphics[width=\textwidth]{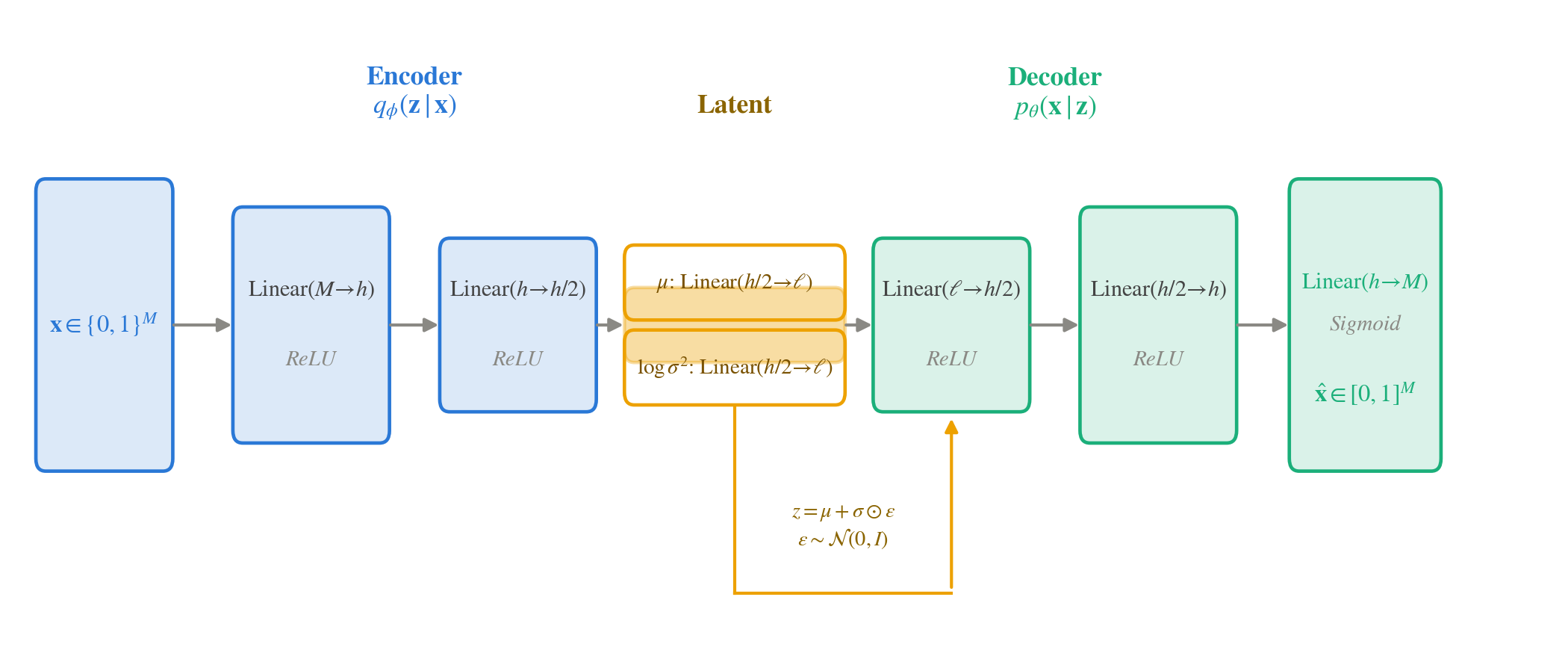}
  \caption{\label{fig:vae_arch}
  Architecture of the VAE used to propose new Slater determinants. Each
  Linear$(A\!\to\!B)$ block is a fully-connected neural network layer, $y=Wx+b$ for a
  trainable weight matrix $W$ and bias $b$, mapping an $A$-dimensional
  input to a $B$-dimensional output; Linear layers are followed by a
  ReLU (rectified linear unit) nonlinearity, $f(x)=\max(0,x)$. The encoder
  $q_\phi(\mathbf{z}\mid\mathbf{x})$ passes $\mathbf{x}\in\{0,1\}^M$
  ($M$ = qubit count of the target system, Table \ref{tab:setup}) through
  two Linear+ReLU layers to a hidden representation of width $h$, then
  splits into two parallel Linear heads producing the mean $\mu$ and
  log-variance $\log\sigma^2$ of the approximate posterior. A latent
  sample is drawn via the reparameterization trick,
  $z=\mu+\sigma\odot\epsilon$ with $\epsilon\sim\mathcal{N}(0,I)$, and
  passed to the decoder $p_\theta(\mathbf{x}\mid\mathbf{z})$, which mirrors
  the encoder in reverse and ends in a Linear layer followed by a Sigmoid
  activation, $f(x)=1/(1+e^{-x})\in(0,1)$, producing
  $\hat{\mathbf{x}}\in[0,1]^M$, the per-spin-orbital reconstruction
  probabilities of Eq. \eqref{eq:bernoulli_decoder}. The two networks are
  trained jointly to maximize the $\beta$-annealed ELBO of
  Eq. \eqref{eq:beta_elbo}. The hidden width, $h=\max(256,8M)$, follows a fixed, system-independent rule; the two Linear+ReLU encoder layers have output widths $h$ then $h/2$, and the latent dimension $\ell=180$ is identical for every system. The overall architecture is a standard VAE neural network (e.g., Ref. \cite{kingma2022autoencodingvariationalbayes}) fused with a custom $\beta$-annealing schedule and the clever dual prior-sampling/posterior-perturbation generation schemes that make Q-WAVE highly efficient for dominant determinant generation.}
  \end{figure*}

\begin{equation}\label{Hamiltonian_decomposition}
    \hat{H}=\sum_{i=1}^\mathcal{N}c_i\hat{h}_i
\end{equation}
where, without loss of generality, $c_i>0$ and the largest eigenvalue of $\hat{h}_i$ be equal to 1 in absolute value. For example, in the decomposition of $\hat{H}$ in the Pauli basis, $\hat{h}_i$ represents a Pauli string with a scalar weight $c_i$. Another way could be to decompose $\hat{H}$ in terms of creation-annihilation operators. In this work, $\hat{h}_i$ is taken to be a combination of creation-annihilation string and its hermitian conjugate. For example, a two-body term in $H$ could be illustrated as:
\begin{equation}
    \hat{h}_i = \hat{a}_p^{\dagger}\hat{a}_q^{\dagger}\hat{a}_r\hat{a}_g+\hat{a}_g^{\dagger}\hat{a}_r^{\dagger}\hat{a}_q\hat{a}_p
\end{equation}

where $\hat{a}^\dagger$ and $\hat{a}$ are the standard fermionic creation and annihilation operators acting on the respective spin-orbitals indexed by $p, q, r,$ and $g$. This construction ensures that each individual $\hat{h}_i$ remains Hermitian. The qDRIFT protocol aims to construct the time-evolution operator $e^{-i\hat{H}ut}$ through the use of a unitary operator $\hat{V}_k$. Instead of using the entire Hamiltonian to form the $e^{-i\hat{H}ut}$,  it randomly samples terms from Eq. (\ref{Hamiltonian_decomposition}) weighted by their interaction strengths, $c_i$. Let $\lambda=\sum_{i}c_i$. Then $\hat{V}_k$ is constructed as 

\begin{equation}\label{qDRIFT_parent}
    \hat{V}_k = \prod_{j=1}^Ne^{-i\hat{h}_{k_j}ut(\lambda/N)}
\end{equation}
where the sequence of indices $k=
(k_1,k_2...,k_N)$ is obtained by randomly sampling terms $\hat{h}_i$
from the distribution defined by $p_i = c_i/\lambda$. Thus, in the qDRIFT method, the deterministic state preparation of $\ket{\psi_u}$ (Eq. \ref{skqd_parent}) is replaced by an ensemble of $N_r$ stochastic states obtained through the use of $\hat{V}_k$. It has been shown in Ref. \cite{piccinelli2026quantumchemistryprovableconvergence} that one can trade circuit depth (lowering $N$ in Eq. \ref{qDRIFT_parent}) for a higher randomized circuit
realizations (increasing $N_r$), significantly reducing the gate depths of the resulting circuits, while preserving the
convergence guarantees of KQD. Measuring these $N_r$ states corresponding to each $\ket{\psi_u}$ (Eq. \ref{skqd_parent}, $u\in\{1, 2, \dots,d-1\}$)in the computational basis yields a distribution of bitstrings representing physically relevant Slater determinants. Note that in this work, for $u=0$, we get Hartree-Fock (HF) state, which is our chosen reference. Thus we do not need to replace $\ket{\psi_0}$ with a stochastic ensemble of $\hat{V}_k$. After particle and $S_z$ conservation filtering, these determinants form the foundational dataset for the generative machine learning model.

\subsection{The Q-WAVE Method: Wavefunction Augmentation via Variational Autoencoders}

The SqDRIFT protocol uses quantum hardware to generate an initial pool of Slater determinants in the computational basis. This set, augmented with classically generated CISD determinants, is used to form the initial dataset $\mathcal{D}^{(0)}$. While this subspace captures critical features of the wavefunction, the remaining determinants corresponding to the true ground state, as well as the combinatorially long tail of determinants characteristic of strongly correlated systems, cannot be exhaustively sampled due to finite shot budgets and hardware noise. To systematically grow the variational subspace toward the true support of the wavefunction, we introduce the Quantum Wavefunction Augmentation via Variational Autoencoders (Q-WAVE) method.

Within this framework, each Slater determinant is mapped to a binary vector $\mathbf{x} \in \{0, 1\}^M$, where $M$ is the number of spin-orbitals (mapped via the Jordan-Wigner transformation). The core generative model is a $\beta$-annealed Variational Autoencoder (VAE), which maps this discrete, high-dimensional physical space into a continuous stochastic latent space, $\mathbf{z} \in \mathbb{R}^{\ell}$, where $\ell$ is the latent dimension. The VAE consists of two neural networks parameterized by $\phi$ and $\theta$: an encoder (the approximate posterior), $q_\phi(\mathbf{z} \mid \mathbf{x})$, which converts the input determinant into a Gaussian distribution in the latent space, and a decoder, $p_\theta(\mathbf{x} \mid \mathbf{z})$, which reconstructs a determinant from a given latent coordinate. A diagrammatic representation of the VAE neural network used in this work is given in Fig. \ref{fig:vae_arch}

To ensure the VAE allocates its representational capacity to the most chemically significant regions of the Hilbert space, the training data is explicitly biased. The dataset $\mathcal{D}^{(s)}$ ($s=0$ signifies the initial dataset, $s\ge1$ denotes subsequent datasets during the iterative Q-WAVE method) is not passed to the VAE directly. First, the data is thresholded to discard any determinants with a squared coefficient $\vert{}c\vert{}^2 \le \epsilon_{\text{thresh}}$ ($\vert{}c\vert{}^2$ obtained via classical diagonalization of the Hamiltonian within the $\mathcal{D}^{(s)}$ subspace). The Hartree-Fock (HF) reference determinant itself is discarded as well. The surviving determinants are then resampled with replacement to form a fixed-size batch of $10^5$ determinants. Importantly, each determinant is drawn with a frequency proportional to a temperature-tempered weight, $\vert c_i \vert^{2/T}$. This tempering is essential because the threshold is set considerably low ($\epsilon_{\text{thresh}}=10^{-10}$). Resampling a batch of $10^5$ determinants using the untempered Born-rule weights ($\vert c_i \vert^2$) would statistically suppress determinants lying in the $10^{-10} < \vert c_i \vert^2 < 10^{-5}$ region, leading to an expected draw count of near zero. Setting the temperature parameter to $T=2$ flattens the probability distribution, ensuring these lower-weight configurations are represented with appreciable frequency. The tempered probabilities are renormalized to unity before this resampling takes place. This batch of $10^5$ determinants is used during training. Notably, the explicit exclusion of the HF determinant is a deliberate design choice: since its CI coefficient is typically much larger than those of the remaining dominant configurations, even after tempering it would dominate the resampled batch, which we anticipate would bias the VAE's learned distribution toward over-representing the HF state.

For training, the $\beta$-annealed VAE maximizes the Evidence Lower Bound (ELBO) for a given determinant instance $\mathbf{x}_i$ at training epoch $\tau$,defined as:
\begin{equation}\label{eq:beta_elbo}
\begin{split}
\mathcal{L}_{\beta}(\theta, \phi; \mathbf{x}_i)
= {}& \mathbb{E}_{q_\phi(\mathbf{z} \mid \mathbf{x}_i)}\!\left[ \log p_\theta(\mathbf{x}_i \mid \mathbf{z}) \right] \\
&- \beta(\tau)\, D_{\text{KL}}\!\left( q_\phi(\mathbf{z} \mid \mathbf{x}_i) \,\big\|\, p(\mathbf{z}) \right)
\end{split}
\end{equation}

The first term is the reconstruction loss, forcing the decoder to faithfully reproduce the dominant physical states. The second term is the Kullback-Leibler (KL) divergence, which acts as a regularizer. It penalizes the individual posteriors $q_\phi(\mathbf{z} \mid \mathbf{x}_i)$ for deviating from the standard normal prior distribution, $p(\mathbf{z}) = \mathcal{N}(\mathbf{0}, \mathbf{I})$.

In standard VAE implementations, $\beta = 1$. However, mapping discrete combinatorial data (such as Slater determinants represented as $\mathbf{x} \in \{0, 1\}^M$) to a continuous space frequently suffers from posterior collapse, wherein the KL penalty overwhelms the reconstruction loss. To circumvent this, we employ a sub-1 target value ($\beta_{\text{target}} < 1$) coupled with an annealing schedule across training epochs $\tau$ ($\tau=0,1, \dots, \tau_{max}-1$; $\tau_{max}$ representing the total number of training epochs):
\begin{equation}\label{eq:beta_anneal}
\beta(\tau) = \beta_{\text{target}} \cdot f(\tau), \qquad
f(\tau) = \min\!\left(1,\ \frac{2\tau}{\tau_{\max}}\right),
\end{equation}
a saturating linear ramp that reaches $\beta_{\text{target}}$ at the halfway point of training ($\tau=\tau_{\max}/2$) and remains fixed at $\beta_{\text{target}}$ thereafter. The exact numerical values of $\beta_{\text{target}}$ and  $\tau_{\max}$ are described in the subsequent sections. During the start of the training (when $\tau=0 \implies \beta(\tau)=0$), the autoencoder initially focuses on high-fidelity reconstruction, learning meaningful discrete features without topological constraints. As $f(\tau)$ increases, the KL penalty gently forces these representations into an overlapping, densely packed Gaussian mixture, ensuring structural continuity without destroying the mutual information between the discrete determinant $\mathbf{x}$ and its continuous latent coordinate $\mathbf{z}$.

Once trained, Q-WAVE generates new determinants to expand the variational subspace. Since the decoder parameterizes an independent Bernoulli distribution over each spin-orbital occupation (indexed by $m$), the reconstruction probability is:

\begin{equation}\label{eq:bernoulli_decoder}
p_\theta(\mathbf{x} \mid \mathbf{z}) = \prod_{m=1}^{M} p_{\theta,m}(\mathbf{z})^{x_m}\left[1-p_{\theta,m}(\mathbf{z})\right]^{1-x_m}
\end{equation}

where $p_{\theta,m}(\mathbf{z})$ is the decoded output probability that the $m$-th spin-orbital is occupied. Consequently, the decoder maps a latent coordinate $\mathbf{z}$ to an $M$-dimensional vector of orbital occupation probabilities. A candidate determinant is then stochastically generated by sampling from these individual probabilities: $x_m \sim \mathrm{Bernoulli}\!\left(p_{\theta,m}(\mathbf{z})\right)$.

New determinants are proposed through two complementary strategies that together realize an explore-exploit search of the latent manifold. In the exploratory mode, latent coordinates ($\mathbf{z}$) are drawn directly from the prior, $\mathbf{z} \sim \mathcal{N}(\mathbf{0}, \mathbf{I})$ and decoded. This probes regions of the latent space, and by extension, the determinant space, that are not directly represented in the current training data but are consistent with the aggregate overlapping structure learned by the encoder. In the exploitative mode, the entire batch of $10^5$ training determinants is re-encoded, and the training-time reparameterization $\mathbf{z} = \boldsymbol{\mu}_\phi(\mathbf{x}_i) + \boldsymbol{\sigma}_\phi(\mathbf{x}_i)\odot\boldsymbol{\epsilon}$ is reused directly, but with its noise term inflated by a factor $\kappa > 1$:

\begin{equation}\label{eq:inflated_sampling}
\mathbf{z} = \boldsymbol{\mu}_\phi(\mathbf{x}_i) + \kappa\boldsymbol{\sigma}_\phi(\mathbf{x}_i)\odot\boldsymbol{\epsilon}, \qquad \kappa > 1.
\end{equation}

This biases generation toward the local neighborhood of determinants already known to carry substantial CI weight. Because the amplitude-weighted training scheme concentrates the aggregate posterior density around the chemically dominant configurations of $\mathcal{D}^{(s)}$, both generation strategies preferentially propose determinants within the physically important region of Hilbert space rather than exploring it uniformly.

Finally, the decoded bitstrings are not guaranteed \textit{a priori} to have the correct number of $\alpha$- and $\beta$-electrons. The independent Bernoulli decoder places no constraint on how many of the $M$ spin-orbitals end up occupied. Each candidate determinant is therefore explicitly corrected before being added to the variational pool. This correction is applied independently to the $\alpha$ and $\beta$ orbital blocks: excess electrons are removed from orbitals with the lowest predicted occupation probabilities, while missing electrons are added to unoccupied orbitals with the highest predicted probabilities. Because the two spin channels are corrected independently rather than only their sum, the total electron count $N_e = N_\alpha + N_\beta$ is automatically correct as well, and every generated determinant is guaranteed the correct $S_z$, with minimal disturbance to the decoder's most confident predictions.

\subsection{End-to-End Iterative Q-WAVE Method}
The single generation step described above is embedded in a loop that repeatedly cycles between forming a training set, generating new determinants, and re-diagonalizing the expanded basis, progressively driving the subspace toward the true support of the ground state. Recall that $s = 0, 1, 2, \dots$ indexes each iteration of this loop, with $\mathcal{D}^{(s)}$ the working determinant basis at iteration $s$; we further denote its CI vector $c^{(s)}$ and variational energy $E^{(s)}$. The initial basis $\mathcal{D}^{(0)}$ is constructed in three stages: the Hamiltonian is first diagonalized in the hardware-sampled determinants alone (filtered for correct particle number and $S_z$), and separately in the full CISD determinant set generated classically from the Hartree-Fock reference; each resulting CI vector is thresholded independently at $\vert{}c\vert{}^2 > \epsilon_{\text{thresh}}$ (again excluding the Hartree-Fock determinant from the thresholding itself), and the union of the two thresholded sets, together with the Hartree-Fock determinant, is diagonalized once more, giving the final $\mathcal{D}^{(0)}$ and $c^{(0)}$ used to seed the iterative loop.

From here, each iteration advances through four steps - forming a training set, generating candidate determinants, expanding and re-diagonalizing the basis, and deciding whether to keep the result - before the cycle repeats.

\paragraph*{Step 1: Forming the training set.} Rather than retrain Q-WAVE on the full basis $\mathcal{D}^{(s)}$, only its chemically significant part is used. The basis is first thresholded,
\begin{equation}
\mathcal{D}^{(s)}_{\text{active}} =  \{\mathbf{x}\in\mathcal{D}^{(s)} : |c^{(s)}_{\mathbf{x}}|^2>\epsilon_{\text{thresh}}\}\setminus\{\text{HF}\} 
\end{equation}
explicitly removing the Hartree-Fock (HF) determinant from the training set, with $\epsilon_{\text{thresh}} = 10^{-10}$. $\mathcal{D}^{(s)}_{\text{active}}$ is then resampled with replacement into a fixed-size batch of $10^5$ determinants, each drawn with frequency proportional to $\vert{}c^{(s)}_{\mathbf{x}}\vert{}^{2/T}$ ($T=2$), producing the training data actually fed to Q-WAVE.

\paragraph*{Step 2: Generation.} Q-WAVE is trained using the batch of $10^5$ determinants drawn from $\mathcal{D}^{(s)}_{\text{active}}$. It then proposes a set of new determinants $\mathcal{Q}^{(s+1)}$ via the prior-sampling and posterior-perturbation modes described in previous Subsection B. The number of prior sampled points ($n_{\text{prior-samples}}$) is set as $500000$ and the number of determinants obtained from posterior-modes is same as the training batch size of $100000$. These values are tunable and user defined. Here, we have used values that worked well for all molecular systems we tested this method on. Since $\mathcal{D}^{(s)}_{\text{active}}$ changes only incrementally between iterations, Q-WAVE is warm-started rather than retrained from scratch: only $s=0$ trains a randomly-initialized network to convergence ($\tau_{max}=1000$ epochs, learning rate $=1.5\times10^{-3}$), while every subsequent iteration fine-tunes the previous iteration's weights for a fraction of the schedule ($\tau_{max}=300$ epochs, learning rate $=7.5\times10^{-4}$), converging quickly since the bulk of the determinant manifold is already learned.

\paragraph*{Step 3: Expansion and re-diagonalization.} The newly generated determinants are added back to the active seed they were trained on, together with the Hartree-Fock determinant itself (withheld from training in Step 1, but always required as a variational reference). Writing this expanded set as
\begin{equation}
\mathcal{B}^{(s+1)} \equiv \mathcal{D}^{(s)}_{\text{active}} \cup \mathcal{Q}^{(s+1)} \cup \{\text{HF}\},
\end{equation}
every member of $\mathcal{B}^{(s+1)}$ is further augmented by its spin-mirror image $\mathcal{M}(\mathbf{x})$ (obtained by exchanging the $\alpha$- and $\beta$-occupation strings of $\mathbf{x}$, which for $N_\alpha=N_\beta$ systems contribute comparably to the wavefunction and are therefore included proactively at negligible extra cost), giving the expanded trial basis
\begin{equation}
\tilde{\mathcal{D}}^{(s+1)} = \mathcal{B}^{(s+1)} \cup \mathcal{M}\left(\mathcal{B}^{(s+1)}\right).
\end{equation}
The Hamiltonian is then re-diagonalized in this expanded trial basis, yielding a trial energy $\tilde{E}^{(s+1)}$ and CI vector.

\paragraph*{Step 4: Accept or reject.} Because every subspace diagonalization yields a variational upper bound to the exact ground-state energy, a strictly lower energy unambiguously identifies a superior basis. The trial basis is therefore accepted or rejected based solely on this criterion:

\begin{equation}\label{eq:accept_reject}
\begin{split}
\mathcal{D}^{(s+1)}, E^{(s+1)} = \
&\quad \begin{cases}
\tilde{\mathcal{D}}^{(s+1)}, \tilde{E}^{(s+1)} & \text{if } \tilde{E}^{(s+1)} < E^{(s)},\\
\mathcal{D}^{(s)}, E^{(s)} & \text{otherwise.}
\end{cases}
\end{split}
\end{equation}
Importantly, this test is non-trivial. Step 1 discards every determinant in $\mathcal{D}^{(s)}$ whose weight falls below $\epsilon_{\text{thresh}}$, and Step 3 builds the trial basis from this truncated $\mathcal{D}^{(s)}_{\text{active}}$ rather than the full $\mathcal{D}^{(s)}$. Consequently, successive bases are \emph{not} strictly nested ($\tilde{\mathcal{D}}^{(s+1)} \nsupseteq \mathcal{D}^{(s)}$). Because the variational principle only guarantees a non-increasing energy for nested subspaces, the trial energy $\tilde{E}^{(s+1)}$ will actually rise if the correlation energy recovered by the newly generated determinants $\mathcal{Q}^{(s+1)}$ fails to offset the energy lost by pruning the sub-threshold tail. If the trial basis were instead built by expanding the unpruned $\mathcal{D}^{(s)}$, nesting would force $\tilde{E}^{(s+1)} \le E^{(s)}$ at every iteration, rendering the test obsolete and forcing the basis to monotonically accumulate unimportant determinants. Thresholding by $\epsilon_{\text{thresh}}$ keeps the variational space compact; the accept/reject rule makes that pruning safe.

By demanding a strict energy decrease, $E^{(s)}$ is guaranteed to be monotonically non-increasing and bounded below by the exact ground-state energy of the active space. The basis carried forward is thus always the best one found so far, eliminating the need to separately track a ``best-so-far'' state. If a trial basis is rejected, the loop does not terminate. Instead, it reverts to the current best basis and returns to Step 1, relying on stochastic resampling to provide the generative model with a fresh draw and another opportunity to discover energy-lowering determinants. This cycle continues until either the stagnation criterion or the maximum iteration count $S_{\max}$ is reached.

\paragraph*{Convergence.} The cycle repeats until $S_{\max}$ iterations have elapsed, or until the energy fails to improve by more than $E_{\text{conv}}$ for $W$ consecutive iterations:
\begin{equation}
\left|E^{(s)} - E^{(s-1)}\right| < E_{\text{conv}} \quad \text{for } W \text{ consecutive iterations}.
\end{equation}
One bookkeeping detail deserves explicit mention. Throughout the loop, we maintain a cumulative pool of every determinant that has ever entered an accepted basis,

\begin{equation}\label{eq:novelty_pool}
\mathcal{P}^{(s)} = \mathcal{D}^{(0)}_{\text{active}} \cup \{\text{HF}\} \cup \bigcup_{s'=1}^{s} \mathcal{D}^{(s')},
\end{equation}

where the union runs over all the previous accepted bases till the current iteration. The proposed candidate determinants generated in Step 2 are filtered against $\mathcal{P}^{(s)}$, so that $\mathcal{Q}^{(s+1)}$ contains only determinants genuinely new to the entire search history. Additionally, this has two other practical consequences. First, a determinant that once entered the basis but subsequently fell below $\epsilon_{\text{thresh}}$ is never re-proposed through the generative channel. Second, because rejected trial bases (Step 4) are never absorbed into $\mathcal{P}^{(s)}$, their proposals remain eligible in later iterations. 

\paragraph*{Step 5: Semistochastic Second Order Perturbative Correction.} The variational loop concludes with a converged variational subspace $\mathcal{V} \equiv \mathcal{D}^{(s^\star)}$, which spans the wavefunction $\vert\Psi_V\rangle = \sum_{I\in\mathcal V} c_I\vert D_I\rangle$ and yields a corresponding variational energy $E_V \equiv E^{(s^\star)}$. While $\mathcal{V}$ captures the strongly correlated character of the system, it may fail to include weakly coupled determinants, specifically in systems with strongly multi-reference wavefunctions (as seen in the case of $\text{Cr}_2$ in the Results and Discussion section). Rather than discarding this residual correlation energy, we recover it using a second-order perturbative (PT2) correction, following the semistochastic approach of Sharma \textit{et al.} \cite{10.1021/acs.jctc.6b01028}. The exact correction ($E^o_{\text{PT2}}$) is

\begin{equation}\label{eq:pt2_exact}
E^o_{\text{PT2}} = \sum_{a\notin\mathcal V}\frac{N_a^2}{E_V-H_{aa}}, \qquad N_a = \sum_{I\in\mathcal V} c_I H_{aI},
\end{equation}

where the sum runs over all external determinants $a$ that are connected to $\mathcal V$ by nonzero off-diagonal Hamiltonian matrix elements, but are not themselves contained in $\mathcal V$. The term $H_{aI}\equiv\langle D_a\vert\hat H\vert D_I\rangle$ is the matrix element between determinants $a$ and $I$, and $H_{aa}\equiv\langle D_a\vert\hat H\vert D_a\rangle$ is the diagonal energy of $D_a$. However, due to the large number of connected determinants, evaluating Eq. \eqref{eq:pt2_exact} exhaustively is intractable. To make this calculation feasible, two complementary approximations are employed \cite{10.1021/acs.jctc.6b01028}.

First, the \emph{deterministic} correction discards individual pairwise contributions that fall below a screening threshold $\epsilon_2$:

\begin{equation}\label{eq:pt2_screened}
E^{D}_{\text{PT2}}[\epsilon_2] = \sum_{a\notin\mathcal V}\frac{\big(\sum_{I}^{(\epsilon_2)} c_I H_{aI}\big)^2}{E_V-H_{aa}},
\end{equation}

where $\sum_I^{(\epsilon_2)} c_I H_{aI}$ includes only terms for which $\vert{}c_I H_{aI}\vert{}>\epsilon_2$. While this recovers Eq. \eqref{eq:pt2_exact} exactly in the limit $\epsilon_2\to0$, storing every surviving determinant $a$ becomes memory-prohibitive for tight values of $\epsilon_2$.

Second, the \emph{stochastic} correction circumvents this memory bottleneck by sampling $N_d$ determinants with replacement from the \emph{entire} space $\mathcal V$. Each determinant is drawn independently with probability $p_I=\vert{}c_I\vert{}/\sum_{J\in\mathcal V}\vert{}c_J\vert{}$. This yields $N_d^{\text{diff}}\le N_d$ distinct determinants (re-indexed as $I=1,\dots,N_d^{\text{diff}}$ within the sample), each appearing with an integer multiplicity $w_I$ such that $\sum_I w_I = N_d$. The unbiased estimate of the stochastic second-order perturbation correction($E^{S}_{\text{PT2}}$) is given as:

\begin{equation}\label{eq:pt2_stochastic}
\begin{split}
E^{S}_{\text{PT2}}[\epsilon_2] = {}& \frac{1}{N_d(N_d-1)}\bigg\langle\sum_{a\notin\mathcal V}\frac{1}{E_V-H_{aa}} \\
&\quad\times\bigg[\Big(\sum_{I=1}^{N_d^{\text{diff}}} w_I y_I\Big)^{2} \\
&\quad+ \sum_{I=1}^{N_d^{\text{diff}}} w_I\big[(N_d-1)p_I-w_I\big]\,y_I^2\bigg]\bigg\rangle_{\text{batch}},
\end{split}
\end{equation}

where $y_I \equiv c_I H_{aI}/p_I$ and $\langle\cdot\rangle_{\text{batch}}$ denotes the average over $N_{\text{batch}}$ independent batches, with each batch representing a fresh draw of $N_d$ determinants. As in Eq. \eqref{eq:pt2_screened}, each pairwise contribution $c_IH_{aI}$ entering $y_I$ is additionally screened at $\epsilon_2$ (omitted from the notation for brevity, following Sharma \textit{et al.}).

The \emph{semistochastic} correction ultimately combines both approaches to further reduce the classical computation time. A deterministic pass at a loose threshold $\epsilon_2^d\gg\epsilon_2$ captures the bulk of the correlation energy exactly, while a stochastic term recovers the difference between the loose and tight thresholds:

\begin{equation}\label{eq:pt2_semistochastic}
E_{\text{PT2}} = \Big(E^{S}_{\text{PT2}}[\epsilon_2] - E^{S}_{\text{PT2}}[\epsilon_2^d]\Big) + E^{D}_{\text{PT2}}[\epsilon_2^d].
\end{equation}

Importantly, both stochastic terms in Eq. \eqref{eq:pt2_semistochastic} are evaluated using the \emph{same} sampled batches. Because both thresholds are applied to identical sets of drawn determinants in every batch, their stochastic errors are strongly correlated and largely cancel out in the difference. This error cancellation is the primary source of the semistochastic estimator's efficiency over a purely stochastic approach.

We use $\epsilon_2=10^{-8}$ Ha and $N_d=200$ throughout our study, matching the values used in Sharma \textit{et al.}; for the loose threshold, we use $\epsilon_2^d=5\times10^{-6}$ Ha uniformly. The total correction is accumulated over $N_{\text{batch}}^{\text{PT2}}=20$ independent batches ($b=1,\dots,N_{\text{batch}}^{\text{PT2}}$). Each batch yields an independent estimate $E_{\text{PT2}}^{(b)}$ via Eq. \eqref{eq:pt2_semistochastic}. The reported PT2 correction is the mean over these batches, $E_{\text{PT2}} = \frac{1}{N_{\text{batch}}^{\text{PT2}}}\sum_b E_{\text{PT2}}^{(b)}$, with its standard error estimated directly from the batch-to-batch variance:

\begin{equation}
\sigma_{\text{PT2}} = \sqrt{\frac{1}{N_{\text{batch}}^{\text{PT2}}(N_{\text{batch}}^{\text{PT2}}-1)}\sum_{b=1}^{N_{\text{batch}}}\big(E_{\text{PT2}}^{(b)}-E_{\text{PT2}}\big)^2}.
\end{equation}

Finally, the total energy reported by the end-to-end framework is:

\begin{equation}
E_{\text{total}} = E_V + E_{\text{PT2}} \pm \sigma_{\text{PT2}}.
\end{equation}

Algorithm 1 summarizes the full cycle described above.

\begin{algorithm}[H]
\caption{Iterative Q-WAVE basis augmentation}
\begin{algorithmic}[1]
\State \textbf{Initialize:} diagonalize SqDRIFT and CISD pools separately, threshold each at $\epsilon_{\text{thresh}}$; $\mathcal{D}^{(0)} \gets$ union $\cup\ \{\text{HF}\}$, diagonalize $\to E^{(0)}, c^{(0)}$
\State $\mathcal{D}^{(0)}_{\text{active}} \gets \{\mathbf{x}\in\mathcal{D}^{(0)} : \vert{}c^{(0)}_{\mathbf{x}}\vert{}^2 > \epsilon_{\text{thresh}}\}\setminus\{\text{HF}\}$
\State $\mathcal{P}^{(0)} \gets \mathcal{D}^{(0)}_{\text{active}} \cup \{\text{HF}\}$;\quad $\text{stagnation} \gets 0$
\For{$s = 0, 1, \dots, S_{\max}-1$}
\State \textbf{Step 1:} Threshold $\mathcal{D}^{(s)}$ by $\vert{}c^{(s)}\vert{}^2 \to \mathcal{D}^{(s)}_{\text{active}}$; resample by $\vert{}c^{(s)}\vert{}^{2/T} \to$ training batch
\State \textbf{Step 2:} Train VAE on batch (from scratch if $s=0$, else warm-started); generate $\mathcal{Q}^{(s+1)}$ ($S_z$-corrected; novelty-filtered against $\mathcal{P}^{(s)}$)
\State \textbf{Step 3:} $\mathcal{B}^{(s+1)} \gets \mathcal{D}^{(s)}_{\text{active}} \cup \mathcal{Q}^{(s+1)} \cup \{\text{HF}\}$; $\tilde{\mathcal{D}}^{(s+1)} \gets \mathcal{B}^{(s+1)} \cup \mathcal{M}(\mathcal{B}^{(s+1)})$; diagonalize $\to \tilde{E}^{(s+1)}, \tilde{c}^{(s+1)}$
\State \textbf{Step 4:} $\Delta \gets E^{(s)} - \tilde{E}^{(s+1)}$
\If{$\Delta > 0$}
\State \textbf{Accept:} $(\mathcal{D}^{(s+1)}, E^{(s+1)}, c^{(s+1)}) \gets (\tilde{\mathcal{D}}^{(s+1)}, \tilde{E}^{(s+1)}, \tilde{c}^{(s+1)})$; $\mathcal{P}^{(s+1)} \gets \mathcal{P}^{(s)} \cup \tilde{\mathcal{D}}^{(s+1)}$
\State $\text{stagnation} \gets \text{stagnation}+1$ if $\Delta < E_{\text{conv}}$ else $0$
\Else
\State \textbf{Reject:} $(\mathcal{D}^{(s+1)}, E^{(s+1)}, c^{(s+1)}) \gets (\mathcal{D}^{(s)}, E^{(s)}, c^{(s)})$, $\mathcal{P}^{(s+1)} \gets \mathcal{P}^{(s)}$; $\text{stagnation} \gets \text{stagnation}+1$
\EndIf
\State \Comment{$\mathcal{D}^{(s+1)}, c^{(s+1)}$ now feed Step 1 of the next iteration}
\If{$\text{stagnation} \ge W$}
\State \textbf{break}
\EndIf
\EndFor
\State \textbf{Step 5:} $(\mathcal{V}, E_V, c) \gets (\mathcal{D}^{(s^\star)}, E^{(s^\star)}, c^{(s^\star)})$, $s^\star$ the final index reached; semistochastic PT2 $\to E_{\text{PT2}} \pm \sigma_{\text{PT2}}$
\State \Return $E_{\text{total}} = E_V + E_{\text{PT2}} \pm \sigma_{\text{PT2}}$
\end{algorithmic}
\end{algorithm}
\subsection{Scalability of Q-WAVE}

The scalability of Q-WAVE is governed by three components: the number of VAE model's parameters which are optimized through training and later used during determinant generation, the size of the basis in which Hamiltonian is diagonalized, and the final PT2 correction cost. We characterize the scaling of each below.

\paragraph*{Number of VAE parameters scales polynomially.}
Both the training and determinant generation processes are dependent on the parameters of the VAE neural network. During training, the parameters are optimized through ELBO maximization (Eq. \ref{eq:beta_elbo}). The total number of generated determinants is kept fixed (($5\times10^5$ prior samples plus $10^5$ posterior-perturbed samples)) across all studied systems. However, each generated determinant uses the trained parameters (Eq. \ref{eq:bernoulli_decoder}, Eq. \ref{eq:inflated_sampling}). VAE neural network's parameter count is dictated by $h=\max(256,8M)$ and
$\ell=180$. Concretely, each Linear layer in Fig. \ref{fig:vae_arch}
connects every one of its $A$ input values to every one of its $B$
output values through independently adjustable weights and biases,
totaling $(A\times B) + B$ numbers (the additional $B$ is due to the
bias of the neural network layer). The VAE's total parameter count is
just the sum of these weights and biases over all seven layers of
Fig. \ref{fig:vae_arch}. For $M\ge32$ (the $h=8M$ regime), the total
parameter count comes out as $N_\text{params}(M) = 80M^2 + (12\ell+25)M + 2\ell$.
In the strict asymptotic limit $M\gg\ell$, the total parameter count scales as
$\mathcal{O}(M^2)$. However, at the qubit counts studied here
($M=24$-$60$, comparable to $\ell$), the linear term
($(12\ell+25)M$) remains a sizable fraction of the total, and the
observed growth is sub-quadratic. For example, doubling $M$ from 40
to 80 grows the parameter count of our VAE neural network from
215,760 to 687,160, a factor of $\approx3.2$, approaching the
asymptotic factor of 4 only once $M\gg\ell$. Thus, the total VAE parameter
count scales polynomially in qubit count and remains independent of
$\vert\mathcal{H}_\text{FCI}\vert$ (the dimension of the full particle and $S_z$ conserving determinant space). Training cost is further reduced
by warm-starting: only the first iteration trains a randomly
initialized network to convergence, and every subsequent iteration
fine-tunes for a fixed, small number of epochs.

\paragraph*{Diagonalization happens in a compact, basis-restricted
representation, never the full Hilbert space.}
The Hamiltonian is diagonalized only within the adaptively selected
variational basis $\mathcal{D}^{(s)}$. The largest basis ever
diagonalized in the pipeline, whether the initial, separate
diagonalizations of the particle- and $S_z$-filtered hardware pool and
the CISD pool used to construct $\mathcal{D}^{(0)}$ (Sec. II.C), or the
largest basis reached during the ML loop itself, remains, by
construction, only a small fraction of the full CI space
$\vert\mathcal{H}_\text{FCI}\vert$ (Table \ref{tab:results}). A sparse
representation of the Hamiltonian matrix restricted to $\mathcal{D}^{(s)}$ is built once per iteration via Slater-Condon rules and never touches determinants outside $\mathcal{D}^{(s)}$. The ground state is then obtained by the Davidson iteration, a sequence of sparse matrix vector products against this stored (never dense, never full-qubit-space) Hamiltonian matrix. Per-iteration
diagonalization cost therefore tracks $\vert\mathcal{D} ^{(s)}\vert$ which is considerably lower than $\vert\mathcal{H}_\text{FCI}\vert$ (see Results and Discussion Section).

\paragraph*{The PT2 correction's cost is set by reference-space connectivity.}
Step 5's deterministic term $E^D_\text{PT2}[\epsilon_2^d]$
(Eq. \eqref{eq:pt2_screened}) evaluates, for every determinant in the
converged space $\mathcal{V}$, its screened connections to external
determinants. Sharma \textit{et al.} \cite{10.1021/acs.jctc.6b01028} establish that the number of such nonzero connections scales as
$\mathcal{O}(n^2v^2\vert\mathcal{V}\vert)$ (where $n$ and $v$ are the occupied/virtual orbital counts, and $\vert\mathcal{V}\vert$ is the number of
determinants in the variational space). This is linear in $\vert\mathcal{V}\vert$
with a connectivity factor polynomial in system size. The stochastic term (Eq. \eqref{eq:pt2_stochastic}) grows sublinearly with $\vert\mathcal{V}\vert$ (as reported in \cite{10.1021/acs.jctc.6b01028}), since low-weight
reference determinants are sampled only rarely. A purely deterministic evaluation at the tight threshold $\epsilon_2$ is memory-prohibitive at the reference-space sizes reached here, since every external determinant surviving screening must be stored. A purely stochastic
evaluation at $\epsilon_2$ avoids this, but needs many samples to control its statistical error at such a tight threshold. The semistochastic combination (Eq. \eqref{eq:pt2_semistochastic}) instead
evaluates the stochastic term at both $\epsilon_2$ and $\epsilon_2^d$ using the \emph{same} sampled batch of $N_d$ determinants, so the two stochastic errors are strongly correlated and largely cancel in their difference. Because the perturbative cost scales with $\vert\mathcal{V}\vert$
rather than $\vert\mathcal{H}_\text{FCI}\vert$, and $\vert\mathcal{V}\vert$ itself remains a vanishing fraction of $\vert\mathcal{H}_\text{FCI}\vert$ as system size grows (see Results and Discussion section), the PT2 correction benefits from Q-WAVE's compact variational wavefunction rather than becoming a bottleneck.

\section{Results and Discussion}

\subsection{Computational setup}

We assess the end-to-end iterative Q-WAVE protocol on four molecular systems of increasing difficulty: H$_2$O ($r_{\text{OH}}=0.958$ \AA, $\angle\text{HOH}=104.5^\circ$ at equilibrium) and N$_2$ ($r_{\text{NN}}=1.1$ \AA\ at equilibrium) in the 6-31G basis (canonical Restricted Hartree Fock orbitals and frozen core) across their dissociation profiles, C$_2$H$_4$ (6-31G basis, canonical Restricted Hartree Fock orbitals and all-electron) at its experimental equilibrium geometry ($r_{\text{C=C}}=1.339$ \AA, $r_{\text{C--H}}=1.086$ \AA, $\angle\text{HCH}=117.6^\circ$, $\angle\text{HCC}=121.2^\circ$) as a 52-qubit scalability test, and the Cr$_2$ dimer at $r=1.5$~\AA{} in the Ahlrichs VDZ basis as a strongly correlated stress test.

\begin{table*}[t]
  \caption{\label{tab:setup}
  Systems and quantum-sampling parameters. All runs use qDRIFT sequences of
  $N=15$ gates, evolution time $t=0.15$, and Krylov dimension $d=5$.
  $\vert\mathcal{H}_{\text{FCI}}\vert$ is the dimension of the full particle- and
  $S_z$-conserving determinant space. Depth is the transpiled circuit depth on
  the production backend (ibm\_kingston for H$_2$O/N$_2$/Cr$_2$, ibm\_boston for C$_2$H$_4$; optimization\_level=3),
  reported as the median (min-max) over 20 randomly sampled qDRIFT
  realizations per nontrivial Krylov index $k=1,\dots,4$; each realization
  draws an independent random Pauli-term sequence, so depth is a distribution
  rather than a fixed number. The $k=0$ circuit, which contains only the
  constant-depth Hartree-Fock state-preparation gates and no qDRIFT evolution
  blocks, is excluded.}
  \begin{ruledtabular}
  \begin{tabular}{lccccc}
  System & Space & Qubits & $\vert\mathcal{H}_{\text{FCI}}\vert$ & Shots / $N_r$ & Depth (median, range) \\
  \hline
  H$_2$O (6-31G)     & (8e,\,12o)  & 24 & $2.45\times10^{5}$  & 1024 / 100 & 532 (116--2194) \\
  N$_2$ (6-31G)      & (10e,\,16o) & 32 & $1.91\times10^{7}$  & 5120 / 100 & 1142 (244--2672) \\
  C$_2$H$_4$ (6-31G) & (16e,\,26o) & 52 & $2.44\times10^{12}$ & 5120 / 200 & 1757 (142--5596) \\
  Cr$_2$ (VDZ)       & (24e,\,30o) & 60 & $7.48\times10^{15}$ & 5120 / 200 & 1445 (217--5631) \\
  \end{tabular}
  \end{ruledtabular}
\end{table*}

For Cr$_2$, we work in a (24e,\,30o) active space of CASSCF(12,12) natural
orbitals (60 qubits). Reference energies are exact FCI for H$_2$O and N$_2$,
all-electron CCSD(T) for C$_2$H$_4$, and the extrapolated DMRG result of
Ref. \cite{10.1063/1.4905329} for Cr$_2$.
Table \ref{tab:setup} lists the per-system quantum-sampling parameters; the
qDRIFT sequence length ($N=15$), evolution time ($t=0.15$), and Krylov
dimension ($d=5$) are common to all systems, while the shot budget and
number of randomized realizations $N_r$ vary modestly, as noted. We gradually increased the shot size and $N_r$ with the increasing size of the molecular system to accommodate the increasing size of the Hilbert space and the Hamiltonian. 
All Q-WAVE hyperparameters are collected in Table \ref{tab:hyper}. They are identical across all systems studied; no per-molecule tuning was performed.
\begin{table*}[t]
\caption{\label{tab:hyper}
Q-WAVE hyperparameters. All values are identical for every system studied;
no per-molecule tuning was performed. The hidden width $h$ follows a fixed,
system-independent rule in the qubit count $M$ (Fig. \ref{fig:vae_arch}).
Sampling parameters specific to the SqDRIFT stage are given separately in
Table \ref{tab:setup}.}
\begin{ruledtabular}
\begin{tabular}{lc@{\hspace{2em}}lc}
Parameter & Value & Parameter & Value \\
\hline
\multicolumn{2}{l}{\textit{VAE architecture and training}}
  & \multicolumn{2}{l}{\textit{Iterative loop}} \\
Latent dimension, $\ell$                    & $180$
  & Selection threshold, $\epsilon_{\text{thresh}}$      & $10^{-10}$ \\
Hidden width, $h$                           & $\max(256,\,8M)$
  & Maximum iterations, $S_{\max}$                       & $100$ \\
Training batch size                         & $10^{5}$
  & Stagnation window, $W$                               & $5$ \\
Tempering temperature, $T$                  & $2$
  & Convergence tolerance, $E_{\text{conv}}$             & $10^{-5}$ Ha \\
\cline{3-4}
Annealing target, $\beta_{\text{target}}$   & $0.3$
  & \multicolumn{2}{l}{\textit{Semistochastic PT2}} \\
Epochs, cold start ($s=0$), $\tau_{\max}$   & $1000$
  & Tight screening threshold, $\epsilon_2$              & $10^{-8}$ Ha \\
Epochs, warm start ($s\ge1$), $\tau_{\max}$ & $300$
  & Sampled determinants per batch, $N_d$                & $200$ \\
Learning rate, warm start                   & $7.5\times10^{-4}$
  & Number of batches, $N_{\text{batch}}^{\text{PT2}}$   & $20$ \\
\cline{1-2}\cline{3-4}
\multicolumn{2}{l}{\textit{Determinant generation (per iteration)}}
  & \multicolumn{2}{l}{\textit{Implementation}} \\
Prior samples, $n_{\text{prior-samples}}$   & $5\times10^{5}$
  & Optimizer                                            & Adam \\
Posterior-perturbed samples                 & $10^{5}$
  & Framework                                            & PyTorch \\
Posterior-perturbation scale, $\kappa$      & $6$
  &                                                      & \\
\end{tabular}
\end{ruledtabular}
\end{table*}

Five determinant-selection strategies are compared throughout:
\begin{enumerate}
    \item \emph{Q-WAVE (SqDRIFT+CISD)} represents the full protocol. This strategy serves as the primary method and is seeded by a combination of both hardware samples and Configuration Interaction Singles and Doubles (CISD) determinants.

    \item \emph{Q-WAVE (CISD-only)} is the first of two seed ablations. This strategy is used to evaluate the impact of the hardware data by omitting the hardware samples entirely and relying strictly on the CISD support for the initial seed.

    \item \emph{Q-WAVE (HW-only)} is the complementary seed ablation. In contrast to the CISD-only approach, this version omits the CISD support completely and relies exclusively on the hardware samples to seed the protocol.

    \item \emph{Random proposals} serves as a matched baseline that replaces the generative model with uniform random draws from the particle- and $S_z$-conserving determinant space, explicitly excluding the cumulative pool $\mathcal{P}^{(s)}$ of every determinant that has ever entered an accepted basis (Eq. \eqref{eq:novelty_pool}). To ensure a rigorous comparison, this baseline is perfectly matched to the \emph{Q-WAVE (SqDRIFT+CISD)} run at every other level. It starts from the same seed basis $\mathcal{D}^{(0)}$ (the exact same hardware samples and CISD determinants) and, at each iteration $s{+}1$, draws exactly as many random determinants as the \emph{Q-WAVE (SqDRIFT+CISD)} run's generative model contributed as genuinely new accepted proposals. Mirror augmentation, the variational accept/reject rule (Eq. \eqref{eq:accept_reject}), and the final semistochastic PT2 correction (Sec. II.C, Step 5; with identical $\epsilon_2$, $\epsilon_2^d$, $N_d$, and $N_\text{batch}$) are all applied without modification. Therefore, the only difference from the full protocol is the source of new determinants -- a trained generative model versus random uniform sampling from the untouched determinant space.

    \item \emph{SQD} (Sample-based Quantum Diagonalization) is run for comparison using the baseline methodology of Ref. \cite{doi:10.1126/sciadv.adu9991}. This strategy applies the configuration-recovery algorithm of qiskit\_addon\_sqd \cite{Qiskit} directly to the identical raw hardware bitstrings used to seed Q-WAVE. SQD's configuration-recovery hyperparameters, specifically a batch count of $N_{\text{batch}}^{\text{SQD}}=10$, $300$ samples per batch, and a carryover probability threshold of $10^{-5}$, are held fixed across every system. Following standard SQD practice, the number of retained configuration strings is capped independently at 2000 per spin sector; because the diagonalized subspace is the direct product of the two sectors, this imposes a hard ceiling of $2000^2=4\times10^{6}$ on the total diagonalization dimension. SQD is run until it reaches either its own convergence test ($\vert\Delta E\vert<10^{-5}$ Ha for 5 consecutive iterations, applied identically for every system) or the $4\times10^{6}$ determinant ceiling, whichever comes first. Unlike the Q-WAVE protocols, SQD receives no PT2 correction.

\end{enumerate}

All molecular integrals, the Hartree-Fock reference, and the CASSCF natural-orbital active space (for Cr$_2$) are generated using PySCF \cite{10.1063/5.0006074}. The resulting fermionic Hamiltonian is mapped to qubits using the Jordan-Wigner transformation. Circuit construction and transpilation are handled by Qiskit \cite{Qiskit}, with execution performed on IBM Quantum hardware using their Heron r2/r3 quantum processors.

\subsection{Compact nature of variational Q-WAVE wavefunction}

\begin{figure*}
\includegraphics[width=\textwidth]{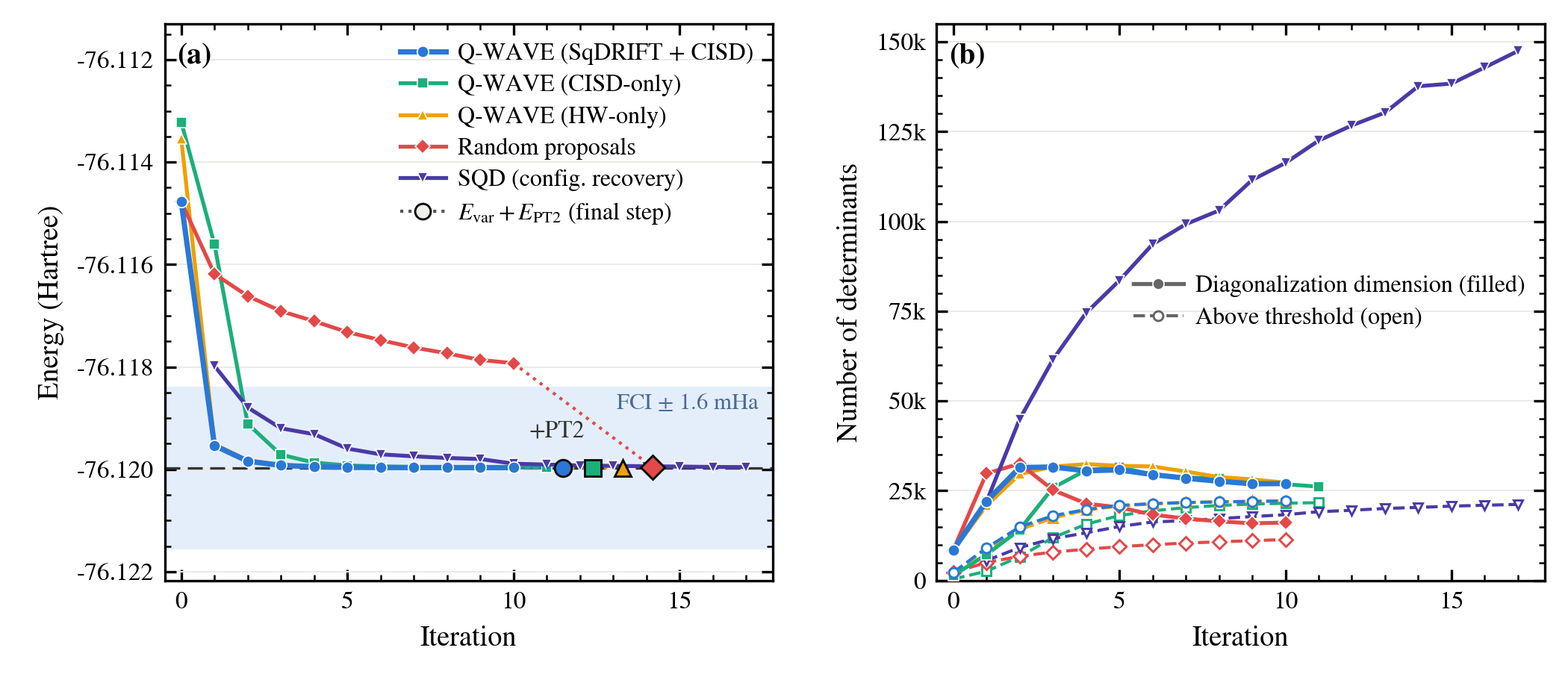}
\caption{\label{fig:water_eq}
(a) Ground-state energy of H$_2$O (6-31G, equilibrium) versus iteration for the five selection strategies. Iteration $0$ denotes the initial basis $\mathcal{D}^{(0)}$, prior to any generative expansion. The shaded band marks chemical accuracy
($\pm1.6$ mHa) around FCI (dashed line); the black-edged symbols beyond the
final iteration show each method's energy after the semistochastic PT2
correction, drawn as one additional step on its own trajectory. SQD does not
receive a PT2 step. (b) Subspace sizes: filled symbols with solid lines show
the diagonalization dimension; open symbols with dashed lines show the number
of determinants above the selection threshold
$\epsilon_{\text{thresh}}=10^{-10}$.}
\end{figure*}

Figure \ref{fig:water_eq} details a single Q-WAVE run for H$_2$O at
equilibrium and establishes how all subsequent figures are read.
All three Q-WAVE variants enter the chemical-accuracy window
($\pm1.6$ mHa around FCI, shaded) within one to two iterations and converge
to within $0.007$ mHa of FCI, terminating after 10-11 iterations with a maximum subspace that was ever diagonalized during the iterative procedure consisting of only $\sim31$k determinants
[Fig. \ref{fig:water_eq}(b)]. The subsequent perturbative step is
correspondingly tiny ($E_{\text{PT2}}=-8.4\,\mu$Ha). For an extremely good variational Q-WAVE basis, PT2 acts as a consistency check rather than a correction. Two comparisons isolate \emph{why} the method works.
First, the matched random baseline with the same seed, same budget, same
protocol, only difference being the generative model replaced with a random determinant generator, stalls at $+2.0$ mHa (variational energy). This proves that the Q-WAVE actually guides the growth of the wavefunction towards the dominant determinants.
Second, SQD requires $147$k determinants (60\% of this system's entire FCI space) to finally converge. Thus, Q-WAVE produces a compact wavefunction while maintaining high energy accuracy.

\subsection{Potential energy surfaces}

\begin{figure*}
\includegraphics[width=\textwidth]{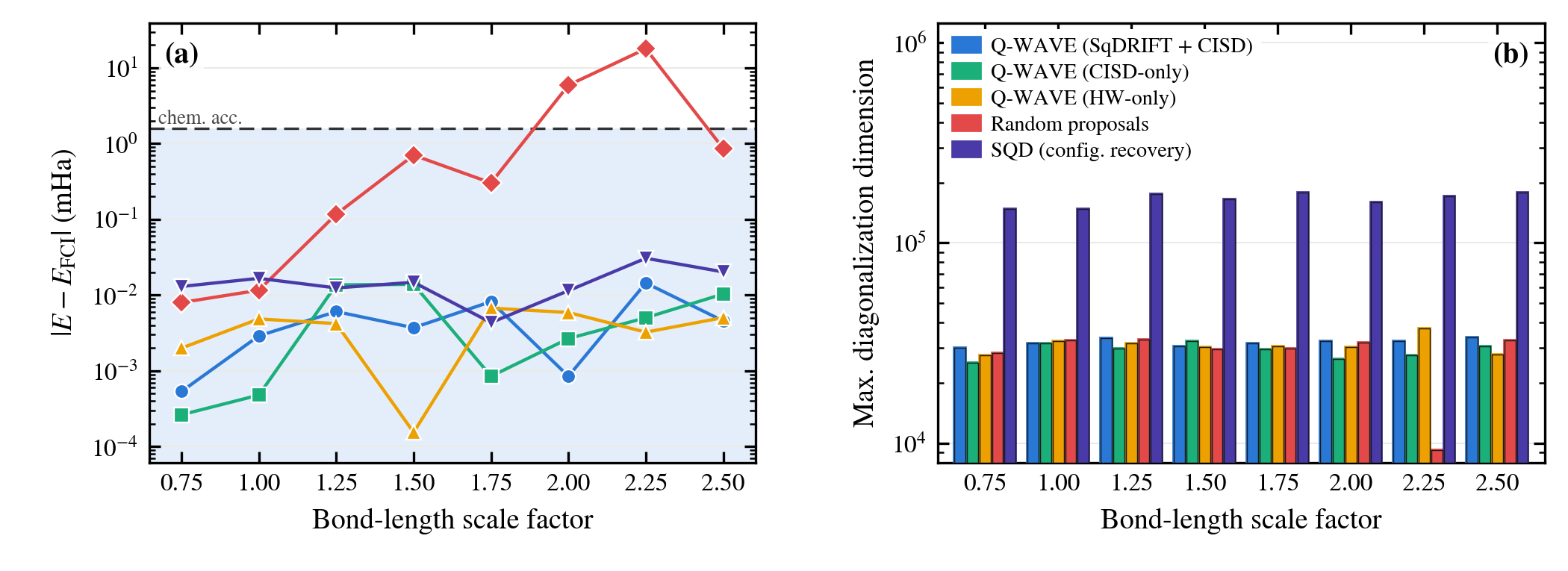}
\caption{\label{fig:pes_water}
Potential energy surface of H$_2$O (6-31G, symmetric O--H stretch), compared across all five selection strategies (color/marker key in panel (b), shared with panel (a)). Bond-length scale factor of 1.00 represents the equilibrium geometry and the other scales are factors by which both O--H bonds are stretched. (a) Absolute energy error $\vert E-E_{\text{FCI}}\vert$ (log scale) at each geometry, using each method's final reported energy ($E_{\text{var}}+E_{PT2}$). SQD has no PT2 correction. The chemical-accuracy threshold ($1.6$ mHa) is marked by the dashed line, with the region below it shaded. (b) Maximal diagonalization dimension reached by each method at each geometry, shown as grouped bars on a log scale.}
\end{figure*}

\begin{figure*}
\includegraphics[width=\textwidth]{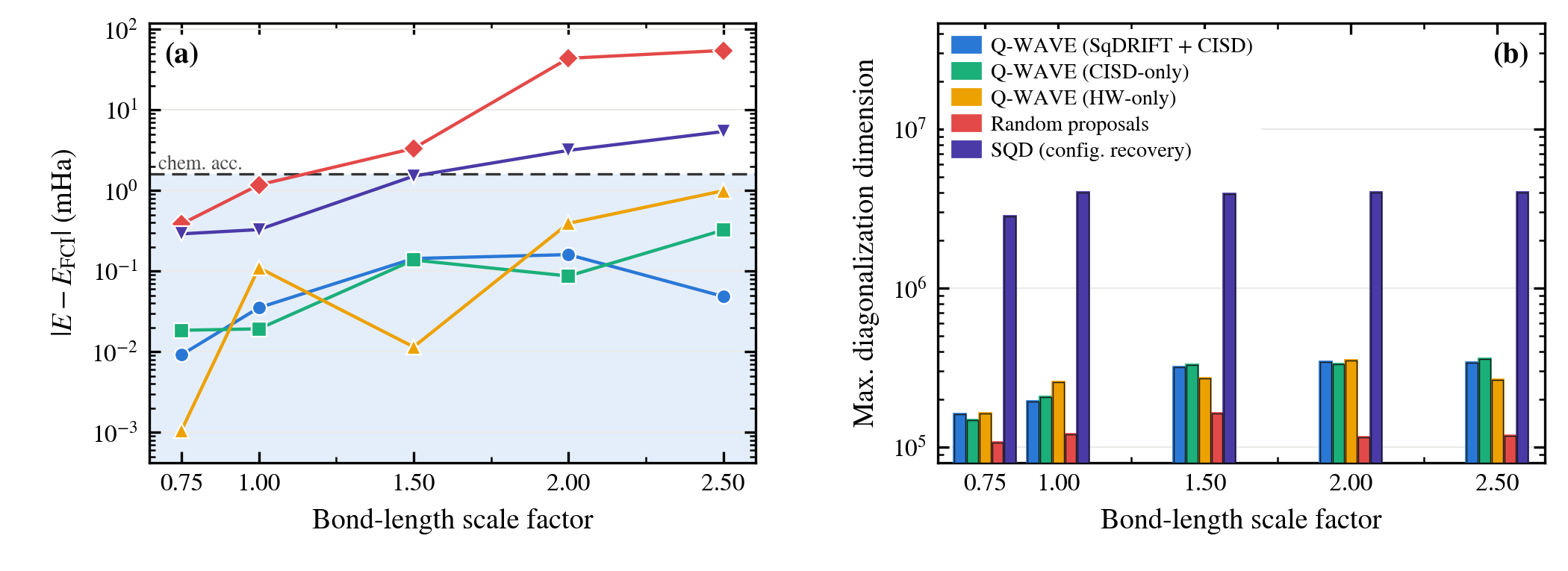}
\caption{\label{fig:pes_n2}
Potential energy surface for N$_2$ (6-31G), compared across all five selection strategies (color/marker key in panel (b), shared with panel (a)). Bond-length scale factor of 1.00 represents the equilibrium geometry and the other scales are factors by which N--N bond is stretched. (a) Absolute energy error $\vert E-E_{\text{FCI}}\vert$(log scale) at each geometry, using each method's final reported energy ($E_{\text{var}}+E_{PT2}$). SQD has no PT2 correction. The chemical-accuracy threshold ($1.6$ mHa) is marked by the dashed line, with the region below it shaded; the three Q-WAVE variants remain within or close to this band at every geometry, while random proposals and SQD grow substantially worse as the bond is stretched. (b) Maximal diagonalization dimension reached by each method at each geometry, shown as grouped bars on a log scale; SQD requires roughly an order of magnitude more determinants than any Q-WAVE variant at every geometry.}
\end{figure*}

Chemical accuracy at a single geometry is necessary but not sufficient; a
useful method must be uniformly accurate along a potential surface.
Figures \ref{fig:pes_water} and \ref{fig:pes_n2} show the dissociation
profiles of H$_2$O (symmetric O--H stretch) and N$_2$ over bond-length scale
factors of 0.75--2.5 using the developed Q-WAVE(SqDRIFT+CISD) method. For water [Fig. \ref{fig:pes_water}(a)], both the
variational and PT2-corrected errors are below $0.015$ mHa at every geometry.
For N$_2$ [Fig. \ref{fig:pes_n2}(a)], the variational error grows smoothly
from $0.12$ to $0.30$ mHa as the triple bond is stretched and the PT2 correction returns every geometry to within
$0.17$ mHa of FCI. Panels (b) illustrate the accompanying cost: the maximal subspace is
essentially geometry-independent for water (30--34k determinants, 12--14\%
of its FCI space), while for N$_2$ it grows from 161k at compressed geometry
to 338k toward dissociation. For N$_2$, this is around $1.8\%$ of the $1.9\times10^{7}$-determinant FCI
space. Thus throughout the potential energy surfaces of H$_2$O and N$_2$, we get highly accurate ground state energies with very compact wavefunctions.

\subsection{Beyond exact diagonalization and a stress test}

\begin{figure*}
\includegraphics[width=\textwidth]{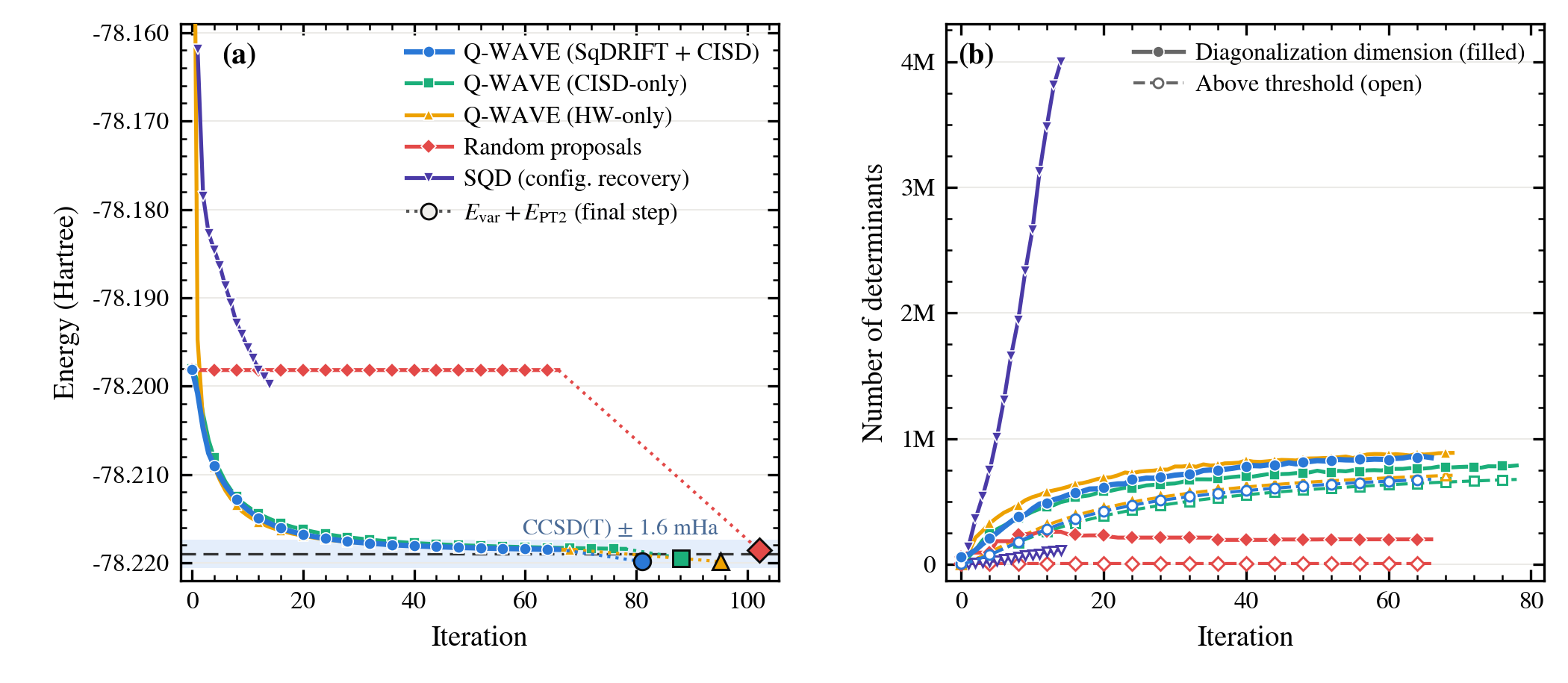}
\caption{\label{fig:c2h4}
(a) Ground-state energy of all-electron C$_2$H$_4$ (6-31G; $r_{\text{C=C}}=1.339$ \AA, $r_{\text{C--H}}=1.086$ \AA, $\angle\text{HCH}=117.6^\circ$, $\angle\text{HCC}=121.2^\circ$) versus iteration for the five selection strategies, referenced to CCSD(T). The shaded band marks chemical accuracy ($\pm1.6$ mHa) around CCSD(T) (dashed line); black-edged symbols beyond the final iteration show each method's energy after the semistochastic PT2 correction, drawn as one additional step on its own trajectory. SQD does not receive a PT2 step: its configuration-recovery iterations saturate the $4\times10^{6}$-determinant cap while stalling $17$ mHa above CCSD(T), whereas all three Q-WAVE variants converge into the CCSD(T) band with a roughly five-fold smaller basis. (b) Subspace sizes: filled symbols with solid lines show the diagonalization dimension; open symbols with dashed lines show the number of determinants above the selection threshold $\epsilon_{\text{thresh}}=10^{-10}$.}
\end{figure*}

Ethylene molecule (C$_2$H$_4$, 52 qubit) has a FCI determinant
space of $2.4\times10^{12}$, which is beyond exact
diagonalization [Fig. \ref{fig:c2h4}]. SQD saturates the  $4\times10^{6}$ determinant cap and stalls $17$ mHa above
CCSD(T), while all three Q-WAVE variants converge smoothly to under
$\sim+0.6$ mHa variationally with $\sim850k$ determinants. The PT2-corrected energies land below CCSD(T) for every variant, though not uniformly: Q-WAVE(SqDRIFT+CISD) and Q-WAVE(HW-only) land $\sim0.85$ mHa below ($-0.85$ and $-0.86$ mHa respectively), while Q-WAVE(CISD-only) lands closer to CCSD(T) at $\sim0.45$ mHa below. Since $E_{\text{var}}+E_{\text{PT2}}$ is not variational and CCSD(T) is itself approximate, we regard sub-mHa mutual agreement between these two independent methods on this single-reference system as cross-validation of both. Interestingly, the random baseline has a flat variational profile. This follows directly from how its per-iteration draw budget scales against the determinant space. Each iteration draws only as many random determinants as the novel determinants proposed by VAE in the corresponding Q-WAVE(SqDRIFT+CISD) run. This draw is very small compared to the $2.4\times10^{12}$ FCI determinant space. Random draws rarely produce determinants that lower the energy below the current minimum. Under the standard variational accept/reject rule (Step 4 in Section IIC), these proposals are almost universally rejected, causing the algorithm to revert to the previous best variational space and leaving the reported trajectory essentially flat.

\begin{figure*}
\includegraphics[width=\textwidth]{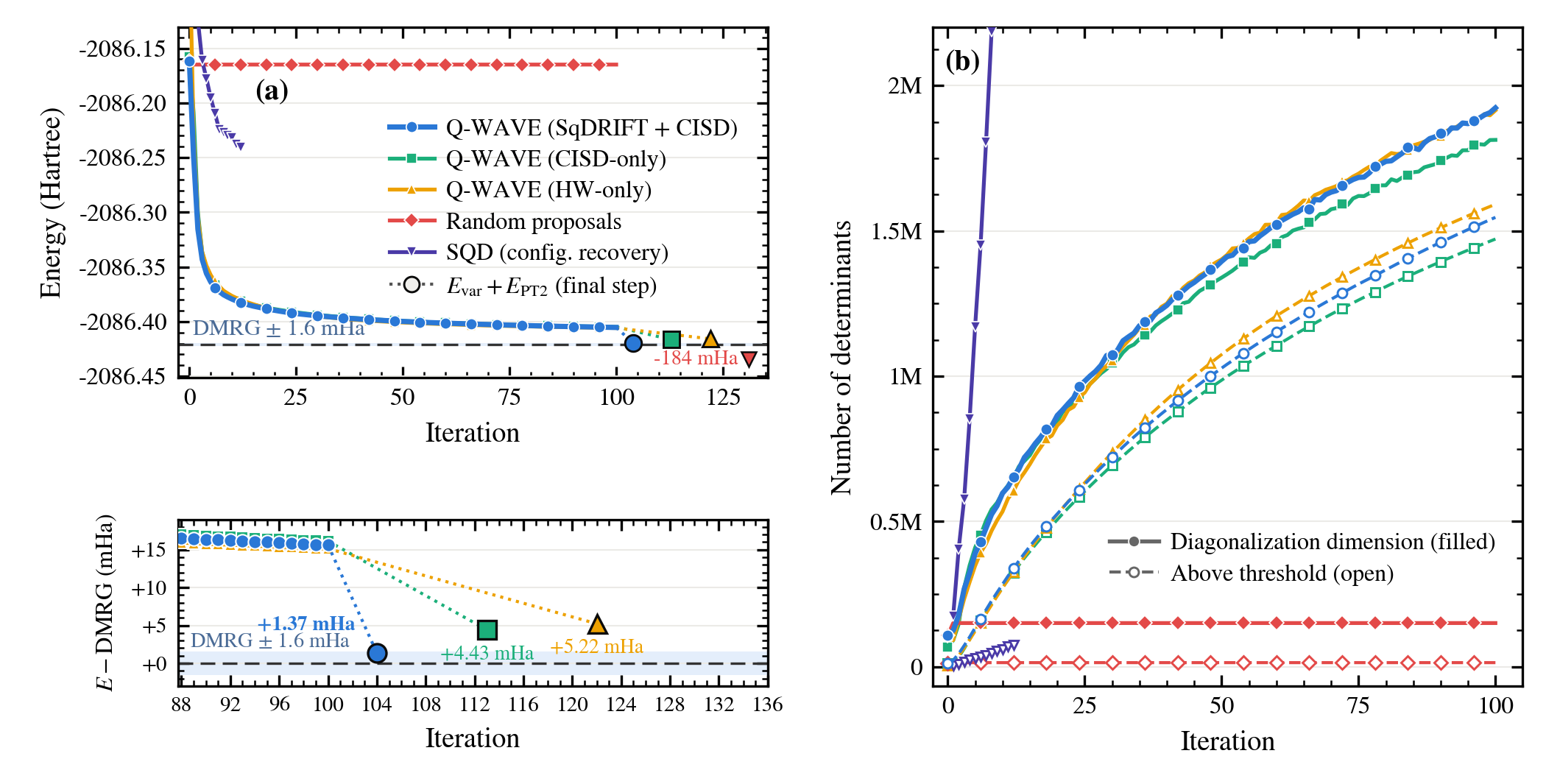}
\caption{\label{fig:cr2}
(a) Ground-state energy of Cr$_2$ ($r=1.5$ \AA, Ahlrichs VDZ, (24e,30o) CASSCF(12,12) natural orbitals) versus iteration for the five selection strategies, referenced to DMRG ($-2086.42095$ Ha). The shaded band marks chemical accuracy ($\pm1.6$ mHa) around DMRG (dashed line); black-edged symbols beyond the final iteration show each method's energy after the semistochastic PT2 correction, drawn as one additional step on its own trajectory. SQD does not receive a PT2 step. The three Q-WAVE variants are indistinguishable on this scale and overlap; the lower strip replays their last 12 iterations and final PT2-corrected points on a linear, zoomed scale (deviation from DMRG in mHa, with the same chemical-accuracy band shaded): only the full SqDRIFT+CISD variant lands inside the band after PT2 ($+1.37\pm1.74$ mHa), while CISD-only ($+4.43\pm1.61$ mHa) and HW-only ($+5.22\pm2.05$ mHa) fall just outside it. The random baseline's PT2 result lies far off scale on the main panel (marker pinned at the lower edge, labeled with its true value). (b) Subspace sizes: filled symbols with solid lines show the diagonalization dimension; open symbols with dashed lines show the number of determinants above the selection threshold $\epsilon_{\text{thresh}}=10^{-10}$.}
\end{figure*}

Cr$_2$ at $1.5$ \AA{} is a canonical worst case for electronic-structure
methods: in the (24e,\,30o) natural-orbital active space, the determinant
space contains $7.5\times10^{15}$ configurations. Figure \ref{fig:cr2} summarizes the outcome, measured
against the DMRG reference energy of $-2086.42095$ Ha. SQD, which performed respectably for the weakly correlated systems, degrades sharply: its configuration-recovery
iterations saturate the same $4\times10^{6}$-determinant cap used elsewhere
in this study after only 12 iterations, without approaching convergence -- an error of $+180$ mHa, two orders of magnitude outside chemical accuracy. The random baseline variationally saturates $256$ mHa above the reference and, notably, the end PT2 correction evaluated on that reference is not merely inaccurate but unphysical, overshooting far below the DMRG energy
($E_{\text{var}}+E_{\text{PT2}} = -2086.60477$ Ha, i.e.\ $-183.8$ mHa on the \emph{wrong side} of the reference). Because the PT2
denominators and the first-order interacting space are both inherited from the variational reference, PT2 amplifies rather than repairs a qualitatively wrong wavefunction. 

The three Q-WAVE variants all converge smoothly to variational errors of
$15$--$16$ mHa with $\sim1.8$--$1.9$ million determinants -- a mere
$2.6\times10^{-10}$ of the full determinant space. The decisive comparison,
however, is at the chemical-accuracy threshold: the CISD-only and
hardware-only seeds finish at central values $+4.43\pm1.61$ and $+5.22\pm2.05$ mHa respectively, just outside the window, while the full SqDRIFT+CISD protocol is the only variant to land \emph{inside}, at
\begin{multline*}
E_{\text{var}}+E_{\text{PT2}} = -2086.41958 \pm 0.00174 \text{Ha} \\
(+1.37\pm1.74 \text{mHa vs.\ DMRG}).
\end{multline*}
The full protocol gives the lowest error, and is the only variant whose central value falls inside the window, though the three are not separated beyond the PT2 statistical uncertainty for the case of Cr$_2$. Nevertheless, this highlights a key structural advantage of the method: the user-defined seed. While a user may choose for a purely classical (CISD) or purely quantum (hardware) initialization, combining them captures the best of both domains. Neither ingredient suffices alone, as hardware samples provide unique determinant combinations inaccessible to the CISD-seeded model. Although this synergistic advantage is partially masked by statistical noise in Cr$_2$, integrating both sources is designed to yield superior performance in larger, more challenging systems. Moreover, as quality of hardware improves, we will obtain more meaningful hardware samples, making the initial seed better.

\subsection{Classical resource cost: generative model versus diagonalization}

A natural objection to any generative model-assisted selected CI method is that the model itself might become the classical bottleneck it is meant to avoid. We address this directly for Cr$_2$ (the hardest system studied here) by recording, for the full 100-iteration Q-WAVE (SqDRIFT+CISD) run, two isolated measures: the
VAE's peak Graphical Processing Unit (GPU) memory and the diagonalizer's memory. The Cr$_2$ has the same geometry as studied in Figure \ref{fig:cr2}.

\begin{figure*}[t]
\centering
\includegraphics[width=\textwidth]{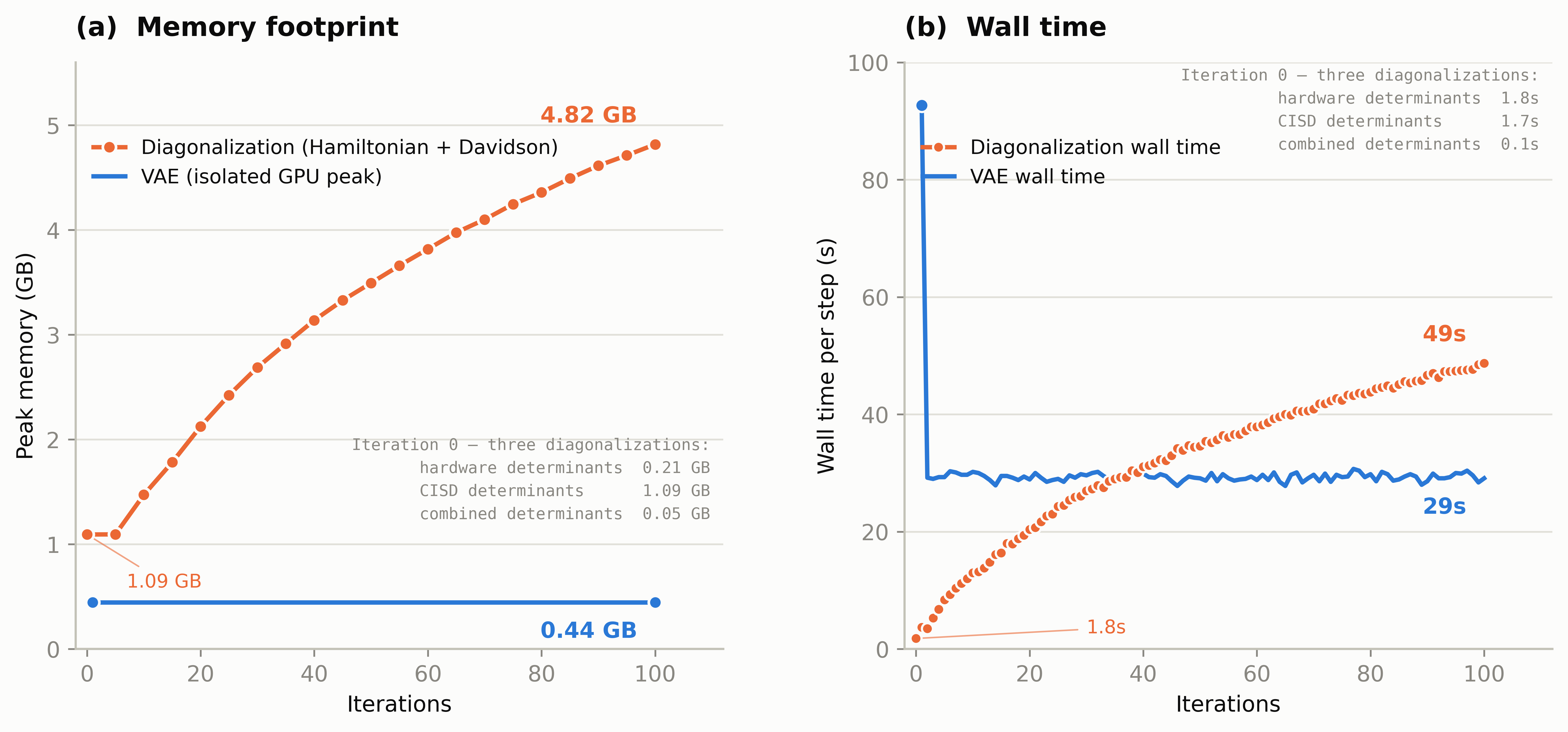}
\caption{\label{fig:vae_diag_resource}
VAE versus diagonalization classical resource cost, Cr$_2$ (Q-WAVE SqDRIFT+CISD protocol, no PT2 at the end). Iteration 0 comprises three separate diagonalizations run before the ML loop begins (hardware determinant diagonalization, CISD determinant diagonalization, and their combined/filtered basis); the value plotted at iteration 0 in each panel is the maximum of the three, with all three individual values given in the figure. (a) Peak memory per iteration: diagonalization footprint grows from $1.09$ GB at iteration 0 to $4.82$ GB at iteration 100 as the basis expands; the VAE's isolated GPU peak (blue) is constant at $0.44$ GB across iterations 1-100 (it has no iteration-0 call).
(b) Wall time per step: diagonalization grows from $1.8$s to $49$ s; the VAE settles to $\sim$$29$ s after a one-time $93$s cold-start training pass at iteration 1.}
\end{figure*}

Fig. \ref{fig:vae_diag_resource} records this study and tells a clear story. The GPU memory footprint of VAE (VAE lives in GPU) is low and remains constant throughout the iterative process. This is because the number of VAE parameters remain fixed for a given qubit number (see Section IID). The wall time for VAE also remains constant, except for the first iteration, where the model must undergo $1000$ training epochs. On the other hand, the diagonalization memory footprint and wall time grows as new dominant determinants inflate the subspace. We take this as direct evidence against the concern that motivates this section: not only is the VAE cost small, it remains almost fixed with the increase in the subspace size. On the other hand, the diagonalization cost is exactly where classical resource pressure concentrates. The end PT2 is a one step computation. As showcased in Ref. \cite{doi:10.1021/acs.jctc.6b00407}, it is highly parallelizable and very efficient implementation is described in the same reference. Hence, the study of PT2 cost is left out of this discussion of whether VAE cost becomes a bottleneck or not. 

\subsection{Comparative study with other methods}
All our studied molecules already contain a comparison with standard SQD (configuration recovery). Here we perform additional comparisons: (a) with HI-VQE \cite{pellowjarman2025hivqehandoveriterativevariational}, (b) comparison with another standard generative model - Restricted Boltzmann machine (RBM), and (c) the classical HCI\cite{10.1021/acs.jctc.6b00407}.

\subsubsection{Comparison with HI-VQE}
\begin{figure*}[t]
\centering
\includegraphics[width=\textwidth]{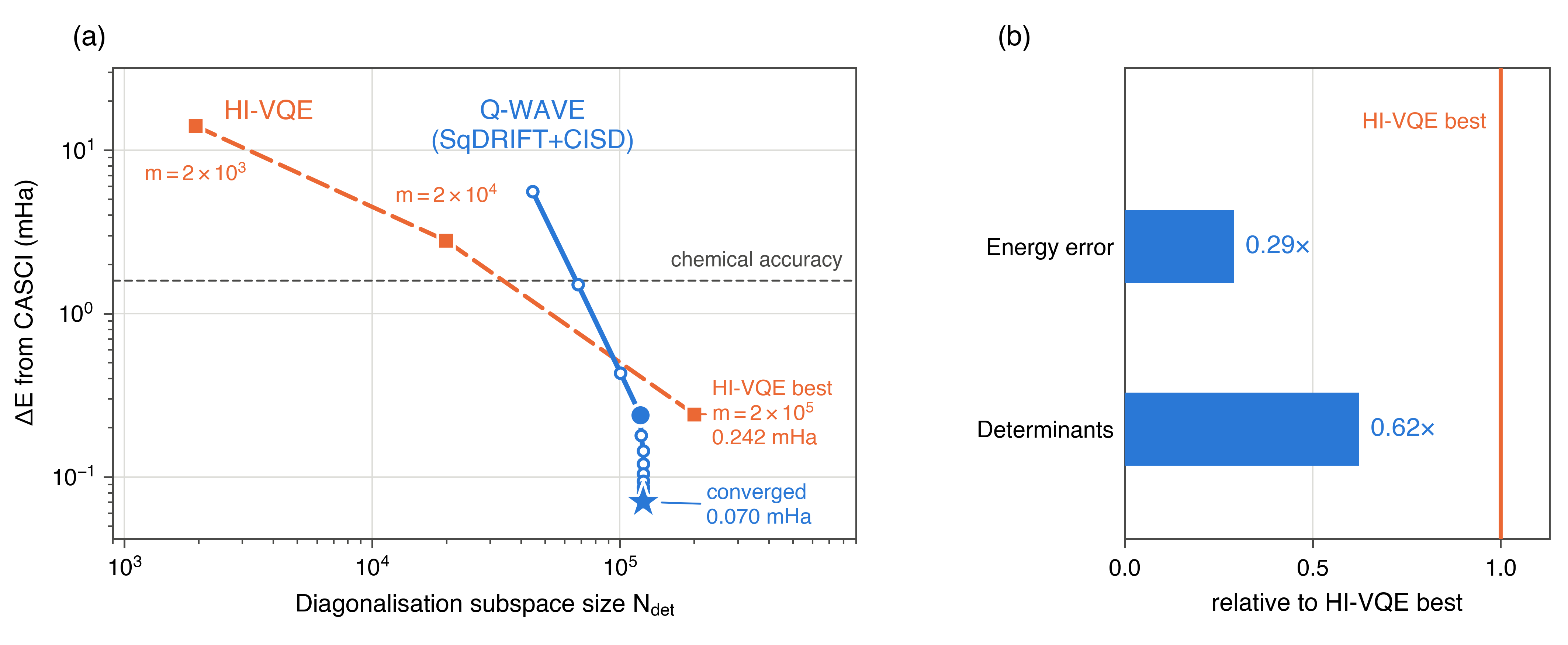}
\caption{Comparison with HI-VQE for NH$_3$ in the 6-31G basis
($15$ orbitals, $10$ electrons, $30$ qubits). (a) Error with respect to CASCI
against diagonalization subspace size. Both axes are logarithmic. Squares are the three HI-VQE calculations using their tuning parameter $m$. Circles trace the entire iterative Q-WAVE, with the filled circle marking the iteration
at which we match the best HI-VQE energy and the star, the converged Q-WAVE variational result. We do not invoke the end PT2 correction for Q-WAVE to keep it analogous to HI-VQE. (b) Converged Q-WAVE variational result relative to the best HI-VQE calculation. The number of determinants for Q-WAVE is taken to be the maximum diagonalization space encountered during its iterative run. Values below
unity indicate simultaneous improvement in both accuracy and subspace size.}
\label{fig:nh3_hivqe}
\end{figure*}

HI-VQE \cite{pellowjarman2025hivqehandoveriterativevariational} uses a variationally optimized circuit to propose electronic configurations that are then diagonalized classically. As HI-VQE manuscript contains a tabular representation of the Ammonia ($\text{NH}_3$) molecule's results, we run iterative Q-WAVE (SqDRIFT + CISD) for this molecule (Fig. \ref{fig:nh3_hivqe}). In particular, we take $\text{NH}_3$ at the coordinates $r(\mathrm{N\text{--}H}) = 1.0190$ \AA{}
and $\angle(\mathrm{H\text{--}N\text{--}H}) = 105.998^{\circ}$. We independently run Hartree-Fock and Complete Active Space CI (CASCI) which match with their reported values. This confirms our geometry is exactly same as theirs. The full CI space consists of $9{,}018{,}009$ determinants.

Figure \ref{fig:nh3_hivqe} reveals a clear story. Q-WAVE converges to a better accuracy using fewer determinants. The advantage appears well before convergence. By iteration 3,  Q-WAVE has already
matched HI-VQE's best energy with a maximum diagonalization dimension of $121{,}224$ determinants (HI-VQE takes $199{,}809$ to reach its best). Q-WAVE's remaining iterations take its energy error to $0.070$ mHa (from CASCI) compared to the $0.24$ mHa for HI-VQE.

\subsubsection{Comparison with Restricted Boltzmann Machine}
\label{sec:vae_vs_rbm}

The restricted Boltzmann machine (RBM) is a standard generative model that has been used extensively to describe molecular wavefunctions\cite{carleo2017solving, doi:10.1021/acs.jctc.2c01216, choo2020fermionic}. Moreover, it has been used as a determinant generator \cite{10.1039/d3sc05807g, 10.1021/acs.jpca.5c02346, patra2026physicsinformedgenerativemachinelearning}. For example, PIGen-SQD \cite{patra2026physicsinformedgenerativemachinelearning}, developed by our group, describes a novel method of utilizing an RBM together with clever perturbative determinant formation to produce accurate and compact molecular wavefunctions. In this method, a major fraction of the dominant determinants are folded into the variational space at the start by using efficient perturbative measures. Subsequently, the RBM acts as a standard generative engine, expanding this subspace to produce highly compact wavefunctions and accurate ground state energies. While the RBM is a well known generative model for wavefunction representation, the primary innovative step of Q-WAVE is the custom-built VAE architecture, which combines a $\beta$-annealing schedule with a two-step sampling process: sampling from the prior and from an inflated latent-space representation of the training determinants.

A natural question is whether the VAE's role as the generative determinant proposer is essential, or whether a simpler, standard generative model would perform comparably. We address this by replacing the VAE with an RBM while holding every other part of the iterative Q-WAVE protocol fixed: the same SqDRIFT+CISD seed, the same $600{,}000$ determinant generation budget per iteration, and the same accept/rollback convergence logic. PT2 is omitted from this comparison so that the generator is the sole variable. We evaluate the performance on N$_2$ stretched to $2.5\,R_\mathrm{eq}$ and on Cr$_2$, the same systems used in the previous sections. N$_2$ acts as a small control system, verifying that the RBM, set up at PIGen-SQD's published settings, functions as a sound standard generator within the iterative Q-WAVE pipeline. Cr$_2$ is a highly correlated system with a large $\vert\mathcal{H}_{\text{FCI}}\vert$; generating dominant determinants within such a large and strongly correlated space provides a demanding environment in which to distinguish the capabilities of the two generators. The RBM implementation (including its hyperparameters) is a direct, operation-for-operation copy of our published PIGen-SQD implementation \cite{patra2026physicsinformedgenerativemachinelearning}; we simply swap the VAE generator in the iterative Q-WAVE method (Section IIC) with this RBM. Table \ref{tab:vae_vs_rbm} reports the variational energy at termination, the absolute error against the respective reference (FCI for N$_2$, DMRG for Cr$_2$), and the maximum diagonalization dimension reached by each generator.

\begin{table*}[t]
\centering
\caption{VAE versus RBM generator, SqDRIFT+CISD variant, $E_\mathrm{var}$ only (no PT2). Errors are absolute deviations from FCI (N$_2$) or DMRG (Cr$_2$). Both Cr$_2$ runs terminated at the $S_{\max}=100$ iteration cap without satisfying the convergence criterion.}
\label{tab:vae_vs_rbm}
\begin{tabular*}{\textwidth}{@{\extracolsep{\fill}}llccc@{}}
\toprule
System & Generator & $E_\mathrm{var}$ at termination (Ha) & Error (mHa) & Max.\ diag.\ dim.\ \\
\midrule
\multirow{2}{*}{N$_2$ ($2.5\,R_\mathrm{eq}$)}
  & VAE & $-108.83948254$ & $0.299$ & $337{,}781$ \\
  & RBM & $-108.83953059$ & $0.251$ & $694{,}337$ \\
\midrule
\multirow{2}{*}{Cr$_2$}
  & VAE & $-2086.40533223$ & $15.618$ & $1{,}923{,}282$ \\
  & RBM & $-2086.27965233$ & $141.298$ & $1{,}332{,}171$ \\
\bottomrule
\end{tabular*}
\end{table*}

On N$_2$ the two generators are indistinguishable. Both reach well inside chemical accuracy -- $0.299$ mHa (VAE) and $0.251$ mHa (RBM). Moreover, during the iterative Q-WAVE run with the RBM for N$_2$, the calculation passes through a diagonalization dimension of $424{,}507$, the point in its trajectory closest to the VAE's final $337{,}781$ although still $26\%$ larger, and at that point its error relative to FCI is $0.266$ mHa. The RBM, at PIGen-SQD's published settings, therefore functions as a sound generator inside the Q-WAVE pipeline, and at this system size the choice between the two is immaterial.

On Cr$_2$, the larger (60-qubit) system, the picture changes. The RBM's error is $9.0\times$ larger than the VAE's, and it reaches this worse energy while diagonalizing a smaller maximum subspace despite an identical per-iteration generative budget of $600{,}000$. Both generators run to the same $S_{\max}=100$-iteration cap without satisfying the convergence criterion of energy improvement on Cr$_2$. So the RBM's smaller subspace cannot be attributed to premature stopping; it reflects fewer dominant determinants surviving the screening per iteration. We attribute this to the comparative diffuseness of the RBM's proposal distribution. In N$_2$'s modest Hilbert space, diffuse proposals still land on genuinely new dominant determinants, so the only cost is a larger maximum diagonalization subspace for comparable accuracy. In Cr$_2$'s far larger space, they land predominantly on determinants that are screened out, and the growth of the variational space stalls. Importantly, during the iterative Q-WAVE run with the VAE for Cr$_2$, the calculation passes through a comparable diagonalization dimension of $1{,}337{,}543$ (against the RBM's final $1{,}332{,}171$) at one of its iterative steps, and at that point its error is $21.9$ mHa, already far smaller than the RBM's final $141.3$ mHa at essentially the same subspace size. Returning to the question posed at the outset: the VAE is not needed everywhere, but it is needed exactly where Q-WAVE is meant to be deployed. At N$_2$'s system size, a standard RBM generator, at published settings and with no retuning, matches the VAE, and the choice of generative model is of no practical consequence. The separation appears only as the Hilbert space grows. On Cr$_2$, under an identical generation budget and an identical iteration cap, the RBM converges to an energy $9.0\times$ further from the DMRG reference, and remains worse by a factor of six even when the VAE is compared against it at a matched subspace dimension. Since the regime in which a quantum-assisted method is worth deploying at all is precisely the regime of large, strongly correlated Hilbert spaces, the generator's behavior there, not its behavior on moderate systems, is what decides the design. On that criterion, the custom VAE architecture is fully justified.

\subsubsection{Comparison with Heat-Bath Configuration Interaction}
\label{sec:vs_hci}

Heat-bath configuration interaction (HCI) \cite{10.1021/acs.jctc.6b00407} constructs its variational space deterministically, admitting determinants from the first-order interacting space whose estimated coefficient exceeds a threshold $\epsilon_1$. It is among the most determinant-efficient classical solvers available for strongly correlated systems, and Ref. \cite{10.1021/acs.jctc.6b01028} reports Cr$_2$ in the Ahlrichs VDZ basis at $r = 1.5$ \AA{} with a frozen Mg core (24 electrons in 30 orbitals), using natural orbitals from a (12e, 12o) CASSCF. This is the same Hamiltonian, active space, and orbital set employed here, so the two methods may be compared directly. This comparison is given in Table \ref{tab:qwave_vs_hci} Since Q-WAVE already adopts the semistochastic perturbative stage of SHCI \cite{10.1021/acs.jctc.6b01028}, the comparison isolates what actually differs between them: how the variational space is built. 

\begin{table}[t]
\centering
\caption{Cr$_2$ (24e, 30o, Ahlrichs VDZ, $r = 1.5$ \AA): Q-WAVE against HCI \cite{10.1021/acs.jctc.6b00407}, the latter runs at $\epsilon_1 = 1$ mHa, $\epsilon_2 = 10\,\mu$Ha. Deviations are signed, relative to the converged DMRG value \cite{10.1063/1.4905329}. The HCI variational energy is quoted in Ref. \cite{10.1021/acs.jctc.6b00407} to three decimal places, so the corresponding deviation carries an uncertainty of $\pm 0.5$ mHa.}
\label{tab:qwave_vs_hci}
\begin{ruledtabular}
\begin{tabular}{lrcc}
Method & $N_{\text{det}}$ & $E_{\text{var}}$ (Ha) & $E_{\text{var}} + E_{\text{PT2}}$ (Ha) \\
\hline
HCI ($\epsilon_1 = 1$ mHa) & $42{,}945$      & $-2086.368$   & $-2086.42130$ \\
Q-WAVE                     & $1{,}923{,}282$ & $-2086.40533$ & $-2086.41958$ \\
DMRG \cite{10.1063/1.4905329} & ---           & ---           & $-2086.42095$ \\
\end{tabular}
\end{ruledtabular}
\end{table}

The two methods sit at opposite ends of the same trade-off (Table \ref{tab:qwave_vs_hci}). On determinant count, HCI is decisively compact: its $42{,}945$-determinant wavefunction is smaller than the Q-WAVE seed subspace, and Q-WAVE requires $430{,}471$ determinants, a factor of ten more, before its variational energy falls below the HCI value. This is the expected consequence of the two selection mechanisms. HCI evaluates an importance criterion directly from the Hamiltonian matrix elements and admits only determinants connected to the current space, which is maximally efficient but confines growth to the first-order interacting space at every step. Q-WAVE proposes determinants from a learned latent geometry with no locality restriction and no explicit selection rule, and pays for that reach in compactness.

The ordering reverses on the quality of the variational wavefunction. Q-WAVE reaches $E_{\text{var}} = -2086.40533$ Ha, $37$ mHa below the HCI variational energy at $\epsilon_1 = 1$ mHa, and correspondingly requires a perturbative correction of only $-14.25$ mHa against HCI's $-53.3$ mHa. The total energies bracket the DMRG reference, HCI overshooting by $0.35$ mHa and Q-WAVE falling short by $1.4$ mHa, both within chemical accuracy. Because the second-order correction is a perturbative expansion about a space that is incomplete, it is the least controlled component of a selected-CI estimator; a variational energy standing closer to the reference before any correction is applied is a meaningful property in its own right. We note that $\epsilon_1 = 1$ mHa is one point on the HCI accuracy-cost curve, selected because it is the point for which Ref.\cite{doi:10.1021/acs.jctc.6b00407} reports the determinant count; smaller $\epsilon_1$ yields larger variational spaces and lower variational energies.

These two profiles, economical but local selection against expansive but nonlocal proposal, are complementary rather than competing, a point we return to in Section IV.

\subsection{Discussion}

\begin{table*}
\caption{\label{tab:results}
Summary of final energies and subspace sizes for some of the representative systems studied here. Errors are reported relative to
the reference in the second column ($\Delta E_\text{var}$). Parenthetical values in the $\Delta(E_{\text{var}}+E_{\text{PT2}})$ column give the statistical uncertainty ($\sigma_{\text{PT2}}$) in units of the last digit(s) of the preceding number [e.g. $-0.0029(46)$ mHa denotes $-0.0029\pm0.0046$ mHa].
$D_{\max}$ is the largest diagonalization dimension reached.
Q-WAVE denotes the full SqDRIFT+CISD protocol.}
\begin{ruledtabular}
\begin{tabular}{llcccccc}
System & Ref. & $\Delta E_{\text{var}}$ (mHa) & $E_{\text{PT2}}$ (mHa) &
$\Delta(E_{\text{var}}{+}E_{\text{PT2}})$ (mHa) & $D_{\max}^{\text{Q-WAVE}}$ &
$D_{\max}^{\text{SQD}}$ & $D_{\max}^{\text{Q-WAVE}}/\vert\mathcal{H}_{\text{FCI}}\vert$ \\
\hline
H$_2$O eq.          & FCI     & $+0.005$ & $-0.008$ & $-0.0029(46)$ & $31.6$k & $147$k  & $13\%$ \\
H$_2$O $2.5\times$  & FCI     & $+0.009$ & $-0.013$ & $-0.0046(82)$ & $33.6$k & $178$k  & $14\%$ \\
N$_2$ eq.           & FCI     & $+0.122$ & $-0.087$ & $+0.035(27)$ & $194$k  & $4.0$M  & $1.0\%$ \\
N$_2$ $2.5\times$   & FCI     & $+0.299$ & $-0.347$ & $-0.05(25)$  & $338$k  & $4.0$M  & $1.8\%$ \\
C$_2$H$_4$ eq.      & CCSD(T) & $+0.531$ & $-1.379$ & $-0.85(20)$  & $857$k  & $4.0$M  & $3.5\times10^{-7}$ \\
Cr$_2$ $1.5$ \AA    & DMRG    & $+15.62$ & $-14.25$ & $+1.4(17)$  & $1.92$M & $4.0$M & $2.6\times10^{-10}$ \\
\end{tabular}
\end{ruledtabular}
\end{table*}

Three observations organize the results (Table \ref{tab:results}).

\emph{The generative model, not basis growth, drives convergence.} For every system studied, the matched random baseline fails to enter
the chemical-accuracy window variationally, stalling at $2$ mHa (H$_2$O eq.)
to $256$ mHa (Cr$_2$). For weakly correlated systems, PT2 can still repair the random basis. As correlation strengthens, this rescue collapses and finally inverts into large overcorrection for Cr$_2$. PT2 is trustworthy when the underlying reference is sound, a regime the generative expansion reaches and uniform sampling does not.

\emph{Compactness compared to FCI space is the enabling property.} Q-WAVE's subspaces undercut
SQD's by factors of $\approx 2-21$ at equal or better accuracy, and the gap widens
with system size: at the Cr$_2$ scale, the working basis occupies ten orders
of magnitude less than the full determinant space. Because the classical
diagonalization dominates the wall time at these scales, compactness
translates directly into reach -- the largest Cr$_2$ diagonalizations
($\sim$1.9M determinants) remain routine for the Davidson solver, while the
quantum resource requirements stay fixed and modest (no more than 5120 shots per circuit, $N=15$ qDRIFT blocks).

\emph{Hardware samples earn their place.} While our method's modular initialization allows users to rely on a purely classical CISD seed, which proves sufficient for weakly correlated systems where the seed ablations are nearly degenerate, the value of combining CISD determinants with hardware sampled determinants becomes indicative in more challenging regimes. As we move to a larger, more difficult system (Cr$_2$), it is the only initialization that successfully drives the final energy (central value) to chemical accuracy. However, the central value energies using all three initialization (SqDRIFT+CISD, CISD only, and HW only) seeds follow closely, falling within a standard deviation of one another. For $\text{Cr}_2$, combining the seeds yields an overall improvement, even if neither individual seed clearly dominates the gain. Because seed selection in our framework is fully modular, accepting any determinant set to initialize the VAE, we anticipate this balance will shift over time. As hardware sampling quality advances, the SqDRIFT seed will become a more decisive contributor. Moreover, as molecular size and wavefunction complexity scale, a standalone CISD seed will ultimately prove inadequate.

\section{Conclusion and Future Directions}

We have introduced Q-WAVE, a hybrid quantum-classical framework in which a $\beta$-annealed variational autoencoder learns the support structure of a molecular ground-state wavefunction. Seeded by a synergistic combination of SqDRIFT-sampled hardware determinants and CISD space, the VAE generates new dominant determinants that lie strictly outside both the sampled distribution and any fixed excitation hierarchy. By regularizing the continuous latent space via the KL term in the ELBO, the model places chemically related determinants at neighboring latent coordinates. This converts determinant discovery from a blind combinatorial search to a targeted geometric one: prior sampling explores regions of the latent manifold not directly represented in the training data, while inflated posterior sampling exploits the neighborhoods of configurations that already carry large CI weight. Embedded in a generation-diagonalization loop, our method systematically grows a compact variational subspace toward the true support of the ground state.

Across four benchmark systems of increasing difficulty, our method achieves chemical accuracy with a very compact wavefunction representation. The hardware samples become important precisely where the classical excitation hierarchy ceases to be a useful organizing principle. This underscores the core advantage of our user-defined initialization: uniting the CISD and SqDRIFT seeds captures the best of both domains, injecting essential multireference determinants that enable the generative loop to explore a richer chemical space. Broadly, this suggests that the route to scalability in quantum-centric electronic structure lies not in ever-larger shot budgets but in learning the wavefunction's structure from a limited number of highly informative quantum samples. Measured against other quantum-centric methods, Q-WAVE is consistently more
accurate using a compact wavefunction. It achieves energies equal to or better than those of standard SQD with $2$-$21\times$ fewer determinants. Also, compared to the state-of-the-art HI-VQE, it achieves higher accuracy with fewer determinants. The generator (VAE), with seeds
drawn from hardware and electronic structure calculations, itself is what carries this. %Substituting by RBM produces a large energy error in highly correlated Cr$_2$ system. 
In summary, Q-WAVE therefore stands currently as the most determinant-efficient route among quantum-centric solvers.

Several methodological extensions follow naturally. First, the Bernoulli decoder used here places no constraint on particle number or spin projection, both of which are currently restored by an explicit post hoc correction applied separately to the $\alpha$ and $\beta$ blocks. Autoregressive or masked decoders that enforce $N_\alpha$ and $N_\beta$ by construction would eliminate this correction and likely improve the yield of accepted proposals per generated candidate.

Another direction is merging Q-WAVE directly with HCI. The two methods are complementary in a way that the present work only partially exploits: we currently adopt the semistochastic perturbative stage of SHCI, while VAE generation-diagonalize-threshold criterion for the variational stage. HCI proposes determinants from the first-order interacting space. Reaching a configuration several excitations away requires HCI to traverse a continuous chain of intermediates that must each survive the selection threshold, a route that routinely breaks down in strongly multireference regimes. The generative model possesses the exact opposite profile: its proposals carry no locality restriction and can jump far from the current support. A hybrid approach, in which the VAE supplies nonlocal candidates in combination with an HCI-like interacting space, would combine a globally aware proposal distribution. Combining the strengths of both HCI and Q-WAVE would result in a highly compact dominant determinant space that can be reached quickly.

Finally, because nothing in the generative loop is strictly limited to ground-state, extensions to excited states and extended transition-metal systems, all follow directly from the machinery established here.

\section{Acknowledgment}
We acknowledge the use of IBM Quantum Credits Program for this work. SH acknowledges the Council of Scientific and Industrial Research (CSIR) for their fellowship.
CP acknowledges the University Grants Commission (UGC) for the fellowship. RM acknowledges the financial support from the Industrial Research and
Consultancy Centre (IRCC), IIT Bombay.

\section*{AUTHOR DECLARATIONS}
\subsection*{Conflict of Interest:}
The authors have no conflict of interest to disclose.

 \section*{Data Availability}
The code implementing the Q-WAVE pipeline described in Sec.\ II, including the hardware-sampled data underlying the N$_2$ ($2.5\times$) result reported here, is openly available at \url{https://github.com/theomolsci/QWAVE}.
% \url{https://github.com/theomolsci/QWAVE}.

\section*{References:}
%aipnum4-2.bst 2019-01-14 (MD) hand-edited version of apsrev4-1.bst


\begin{thebibliography}{59}%
\makeatletter
\providecommand \@ifxundefined [1]{%
 \@ifx{#1\undefined}
}%
\providecommand \@ifnum [1]{%
 \ifnum #1\expandafter \@firstoftwo
 \else \expandafter \@secondoftwo
 \fi
}%
\providecommand \@ifx [1]{%
 \ifx #1\expandafter \@firstoftwo
 \else \expandafter \@secondoftwo
 \fi
}%
\providecommand \natexlab [1]{#1}%
\providecommand \enquote  [1]{``#1''}%
\providecommand \bibnamefont  [1]{#1}%
\providecommand \bibfnamefont [1]{#1}%
\providecommand \citenamefont [1]{#1}%
\providecommand \href@noop [0]{\@secondoftwo}%
\providecommand \href [0]{\begingroup \@sanitize@url \@href}%
\providecommand \@href[1]{\@@startlink{#1}\@@href}%
\providecommand \@@href[1]{\endgroup#1\@@endlink}%
\providecommand \@sanitize@url [0]{\catcode `\\12\catcode `\$12\catcode `\&12\catcode `\#12\catcode `\^12\catcode `\_12\catcode `\%12\relax}%
\providecommand \@@startlink[1]{}%
\providecommand \@@endlink[0]{}%
\providecommand \url  [0]{\begingroup\@sanitize@url \@url }%
\providecommand \@url [1]{\endgroup\@href {#1}{\urlprefix }}%
\providecommand \urlprefix  [0]{URL }%
\providecommand \Eprint [0]{\href }%
\providecommand \doibase [0]{https://doi.org/}%
\providecommand \selectlanguage [0]{\@gobble}%
\providecommand \bibinfo  [0]{\@secondoftwo}%
\providecommand \bibfield  [0]{\@secondoftwo}%
\providecommand \translation [1]{[#1]}%
\providecommand \BibitemOpen [0]{}%
\providecommand \bibitemStop [0]{}%
\providecommand \bibitemNoStop [0]{.\EOS\space}%
\providecommand \EOS [0]{\spacefactor3000\relax}%
\providecommand \BibitemShut  [1]{\csname bibitem#1\endcsname}%
\let\auto@bib@innerbib\@empty
%</preamble>
\bibitem [{\citenamefont {Peruzzo}\ \emph {et~al.}(2014)\citenamefont {Peruzzo}, \citenamefont {McClean}, \citenamefont {Shadbolt}, \citenamefont {Yung}, \citenamefont {Zhou}, \citenamefont {Love}, \citenamefont {Aspuru-Guzik},\ and\ \citenamefont {O’brien}}]{peruzzo2014variational}%
  \BibitemOpen
  \bibfield  {author} {\bibinfo {author} {\bibfnamefont {A.}~\bibnamefont {Peruzzo}}, \bibinfo {author} {\bibfnamefont {J.}~\bibnamefont {McClean}}, \bibinfo {author} {\bibfnamefont {P.}~\bibnamefont {Shadbolt}}, \bibinfo {author} {\bibfnamefont {M.-H.}\ \bibnamefont {Yung}}, \bibinfo {author} {\bibfnamefont {X.-Q.}\ \bibnamefont {Zhou}}, \bibinfo {author} {\bibfnamefont {P.~J.}\ \bibnamefont {Love}}, \bibinfo {author} {\bibfnamefont {A.}~\bibnamefont {Aspuru-Guzik}},\ and\ \bibinfo {author} {\bibfnamefont {J.~L.}\ \bibnamefont {O’brien}},\ }\bibfield  {title} {\enquote {\bibinfo {title} {A variational eigenvalue solver on a photonic quantum processor},}\ }\href@noop {} {\bibfield  {journal} {\bibinfo  {journal} {Nature communications}\ }\textbf {\bibinfo {volume} {5}},\ \bibinfo {pages} {4213} (\bibinfo {year} {2014})}\BibitemShut {NoStop}%
\bibitem [{\citenamefont {McClean}\ \emph {et~al.}(2016)\citenamefont {McClean}, \citenamefont {Romero}, \citenamefont {Babbush},\ and\ \citenamefont {Aspuru-Guzik}}]{McClean_2016}%
  \BibitemOpen
  \bibfield  {author} {\bibinfo {author} {\bibfnamefont {J.~R.}\ \bibnamefont {McClean}}, \bibinfo {author} {\bibfnamefont {J.}~\bibnamefont {Romero}}, \bibinfo {author} {\bibfnamefont {R.}~\bibnamefont {Babbush}},\ and\ \bibinfo {author} {\bibfnamefont {A.}~\bibnamefont {Aspuru-Guzik}},\ }\bibfield  {title} {\enquote {\bibinfo {title} {The theory of variational hybrid quantum-classical algorithms},}\ }\href {https://doi.org/10.1088/1367-2630/18/2/023023} {\bibfield  {journal} {\bibinfo  {journal} {New Journal of Physics}\ }\textbf {\bibinfo {volume} {18}},\ \bibinfo {pages} {023023} (\bibinfo {year} {2016})}\BibitemShut {NoStop}%
\bibitem [{\citenamefont {Kandala}\ \emph {et~al.}(2017)\citenamefont {Kandala}, \citenamefont {Mezzacapo}, \citenamefont {Temme}, \citenamefont {Takita}, \citenamefont {Brink}, \citenamefont {Chow},\ and\ \citenamefont {Gambetta}}]{Kandala2017-qp}%
  \BibitemOpen
  \bibfield  {author} {\bibinfo {author} {\bibfnamefont {A.}~\bibnamefont {Kandala}}, \bibinfo {author} {\bibfnamefont {A.}~\bibnamefont {Mezzacapo}}, \bibinfo {author} {\bibfnamefont {K.}~\bibnamefont {Temme}}, \bibinfo {author} {\bibfnamefont {M.}~\bibnamefont {Takita}}, \bibinfo {author} {\bibfnamefont {M.}~\bibnamefont {Brink}}, \bibinfo {author} {\bibfnamefont {J.~M.}\ \bibnamefont {Chow}},\ and\ \bibinfo {author} {\bibfnamefont {J.~M.}\ \bibnamefont {Gambetta}},\ }\bibfield  {title} {\enquote {\bibinfo {title} {Hardware-efficient variational quantum eigensolver for small molecules and quantum magnets},}\ }\href@noop {} {\bibfield  {journal} {\bibinfo  {journal} {Nature}\ }\textbf {\bibinfo {volume} {549}},\ \bibinfo {pages} {242--246} (\bibinfo {year} {2017})}\BibitemShut {NoStop}%
\bibitem [{\citenamefont {Romero}\ \emph {et~al.}(2018)\citenamefont {Romero}, \citenamefont {Babbush}, \citenamefont {McClean}, \citenamefont {Hempel}, \citenamefont {Love},\ and\ \citenamefont {Aspuru-Guzik}}]{Romero_2019}%
  \BibitemOpen
  \bibfield  {author} {\bibinfo {author} {\bibfnamefont {J.}~\bibnamefont {Romero}}, \bibinfo {author} {\bibfnamefont {R.}~\bibnamefont {Babbush}}, \bibinfo {author} {\bibfnamefont {J.~R.}\ \bibnamefont {McClean}}, \bibinfo {author} {\bibfnamefont {C.}~\bibnamefont {Hempel}}, \bibinfo {author} {\bibfnamefont {P.~J.}\ \bibnamefont {Love}},\ and\ \bibinfo {author} {\bibfnamefont {A.}~\bibnamefont {Aspuru-Guzik}},\ }\bibfield  {title} {\enquote {\bibinfo {title} {Strategies for quantum computing molecular energies using the unitary coupled cluster ansatz},}\ }\href {https://doi.org/10.1088/2058-9565/aad3e4} {\bibfield  {journal} {\bibinfo  {journal} {Quantum Science and Technology}\ }\textbf {\bibinfo {volume} {4}},\ \bibinfo {pages} {014008} (\bibinfo {year} {2018})}\BibitemShut {NoStop}%
\bibitem [{\citenamefont {Grimsley}\ \emph {et~al.}(2019)\citenamefont {Grimsley}, \citenamefont {Economou}, \citenamefont {Barnes},\ and\ \citenamefont {Mayhall}}]{Grimsley2019}%
  \BibitemOpen
  \bibfield  {author} {\bibinfo {author} {\bibfnamefont {H.~R.}\ \bibnamefont {Grimsley}}, \bibinfo {author} {\bibfnamefont {S.~E.}\ \bibnamefont {Economou}}, \bibinfo {author} {\bibfnamefont {E.}~\bibnamefont {Barnes}},\ and\ \bibinfo {author} {\bibfnamefont {N.~J.}\ \bibnamefont {Mayhall}},\ }\bibfield  {title} {\enquote {\bibinfo {title} {An adaptive variational algorithm for exact molecular simulations on a quantum computer},}\ }\href {https://doi.org/10.1038/s41467-019-10988-2} {\bibfield  {journal} {\bibinfo  {journal} {Nature Communications}\ }\textbf {\bibinfo {volume} {10}},\ \bibinfo {pages} {3007} (\bibinfo {year} {2019})}\BibitemShut {NoStop}%
\bibitem [{\citenamefont {Cerezo}\ \emph {et~al.}(2021)\citenamefont {Cerezo}, \citenamefont {Arrasmith}, \citenamefont {Babbush}, \citenamefont {Benjamin}, \citenamefont {Endo}, \citenamefont {Fujii}, \citenamefont {McClean}, \citenamefont {Mitarai}, \citenamefont {Yuan}, \citenamefont {Cincio},\ and\ \citenamefont {Coles}}]{Cerezo2021}%
  \BibitemOpen
  \bibfield  {author} {\bibinfo {author} {\bibfnamefont {M.}~\bibnamefont {Cerezo}}, \bibinfo {author} {\bibfnamefont {A.}~\bibnamefont {Arrasmith}}, \bibinfo {author} {\bibfnamefont {R.}~\bibnamefont {Babbush}}, \bibinfo {author} {\bibfnamefont {S.~C.}\ \bibnamefont {Benjamin}}, \bibinfo {author} {\bibfnamefont {S.}~\bibnamefont {Endo}}, \bibinfo {author} {\bibfnamefont {K.}~\bibnamefont {Fujii}}, \bibinfo {author} {\bibfnamefont {J.~R.}\ \bibnamefont {McClean}}, \bibinfo {author} {\bibfnamefont {K.}~\bibnamefont {Mitarai}}, \bibinfo {author} {\bibfnamefont {X.}~\bibnamefont {Yuan}}, \bibinfo {author} {\bibfnamefont {L.}~\bibnamefont {Cincio}},\ and\ \bibinfo {author} {\bibfnamefont {P.~J.}\ \bibnamefont {Coles}},\ }\bibfield  {title} {\enquote {\bibinfo {title} {Variational quantum algorithms},}\ }\href {https://doi.org/10.1038/s42254-021-00348-9} {\bibfield  {journal} {\bibinfo  {journal} {Nature Reviews Physics}\ }\textbf {\bibinfo {volume} {3}},\ \bibinfo {pages} {625--644} (\bibinfo {year}
  {2021})}\BibitemShut {NoStop}%
\bibitem [{\citenamefont {Anschuetz}\ and\ \citenamefont {Kiani}(2022)}]{anschuetz2022quantum}%
  \BibitemOpen
  \bibfield  {author} {\bibinfo {author} {\bibfnamefont {E.~R.}\ \bibnamefont {Anschuetz}}\ and\ \bibinfo {author} {\bibfnamefont {B.~T.}\ \bibnamefont {Kiani}},\ }\bibfield  {title} {\enquote {\bibinfo {title} {Quantum variational algorithms are swamped with traps},}\ }\href@noop {} {\bibfield  {journal} {\bibinfo  {journal} {Nature Communications}\ }\textbf {\bibinfo {volume} {13}},\ \bibinfo {pages} {7760} (\bibinfo {year} {2022})}\BibitemShut {NoStop}%
\bibitem [{\citenamefont {McClean}\ \emph {et~al.}(2018)\citenamefont {McClean}, \citenamefont {Boixo}, \citenamefont {Smelyanskiy}, \citenamefont {Babbush},\ and\ \citenamefont {Neven}}]{McClean2018}%
  \BibitemOpen
  \bibfield  {author} {\bibinfo {author} {\bibfnamefont {J.~R.}\ \bibnamefont {McClean}}, \bibinfo {author} {\bibfnamefont {S.}~\bibnamefont {Boixo}}, \bibinfo {author} {\bibfnamefont {V.~N.}\ \bibnamefont {Smelyanskiy}}, \bibinfo {author} {\bibfnamefont {R.}~\bibnamefont {Babbush}},\ and\ \bibinfo {author} {\bibfnamefont {H.}~\bibnamefont {Neven}},\ }\bibfield  {title} {\enquote {\bibinfo {title} {Barren plateaus in quantum neural network training landscapes},}\ }\href {https://doi.org/10.1038/s41467-018-07090-4} {\bibfield  {journal} {\bibinfo  {journal} {Nature Communications}\ }\textbf {\bibinfo {volume} {9}},\ \bibinfo {pages} {4812} (\bibinfo {year} {2018})}\BibitemShut {NoStop}%
\bibitem [{\citenamefont {Larocca}\ \emph {et~al.}(2025)\citenamefont {Larocca}, \citenamefont {Thanasilp}, \citenamefont {Wang}, \citenamefont {Sharma}, \citenamefont {Biamonte}, \citenamefont {Coles}, \citenamefont {Cincio}, \citenamefont {McClean}, \citenamefont {Holmes},\ and\ \citenamefont {Cerezo}}]{larocca2025barren}%
  \BibitemOpen
  \bibfield  {author} {\bibinfo {author} {\bibfnamefont {M.}~\bibnamefont {Larocca}}, \bibinfo {author} {\bibfnamefont {S.}~\bibnamefont {Thanasilp}}, \bibinfo {author} {\bibfnamefont {S.}~\bibnamefont {Wang}}, \bibinfo {author} {\bibfnamefont {K.}~\bibnamefont {Sharma}}, \bibinfo {author} {\bibfnamefont {J.}~\bibnamefont {Biamonte}}, \bibinfo {author} {\bibfnamefont {P.~J.}\ \bibnamefont {Coles}}, \bibinfo {author} {\bibfnamefont {L.}~\bibnamefont {Cincio}}, \bibinfo {author} {\bibfnamefont {J.~R.}\ \bibnamefont {McClean}}, \bibinfo {author} {\bibfnamefont {Z.}~\bibnamefont {Holmes}},\ and\ \bibinfo {author} {\bibfnamefont {M.}~\bibnamefont {Cerezo}},\ }\bibfield  {title} {\enquote {\bibinfo {title} {Barren plateaus in variational quantum computing},}\ }\href@noop {} {\bibfield  {journal} {\bibinfo  {journal} {Nature Reviews Physics}\ ,\ \bibinfo {pages} {1--16}} (\bibinfo {year} {2025})}\BibitemShut {NoStop}%
\bibitem [{\citenamefont {Grimsley}\ \emph {et~al.}(2023)\citenamefont {Grimsley}, \citenamefont {Barron}, \citenamefont {Barnes}, \citenamefont {Economou},\ and\ \citenamefont {Mayhall}}]{Grimsley2023}%
  \BibitemOpen
  \bibfield  {author} {\bibinfo {author} {\bibfnamefont {H.~R.}\ \bibnamefont {Grimsley}}, \bibinfo {author} {\bibfnamefont {G.~S.}\ \bibnamefont {Barron}}, \bibinfo {author} {\bibfnamefont {E.}~\bibnamefont {Barnes}}, \bibinfo {author} {\bibfnamefont {S.~E.}\ \bibnamefont {Economou}},\ and\ \bibinfo {author} {\bibfnamefont {N.~J.}\ \bibnamefont {Mayhall}},\ }\bibfield  {title} {\enquote {\bibinfo {title} {Adaptive, problem-tailored variational quantum eigensolver mitigates rough parameter landscapes and barren plateaus},}\ }\href@noop {} {\bibfield  {journal} {\bibinfo  {journal} {npj Quantum Information}\ }\textbf {\bibinfo {volume} {9}},\ \bibinfo {pages} {19} (\bibinfo {year} {2023})}\BibitemShut {NoStop}%
\bibitem [{\citenamefont {Matsuzawa}\ and\ \citenamefont {Kurashige}(2020)}]{zzawa2020jastrow}%
  \BibitemOpen
  \bibfield  {author} {\bibinfo {author} {\bibfnamefont {Y.}~\bibnamefont {Matsuzawa}}\ and\ \bibinfo {author} {\bibfnamefont {Y.}~\bibnamefont {Kurashige}},\ }\bibfield  {title} {\enquote {\bibinfo {title} {Jastrow-type decomposition in quantum chemistry for low-depth quantum circuits},}\ }\href@noop {} {\bibfield  {journal} {\bibinfo  {journal} {Journal of chemical theory and computation}\ }\textbf {\bibinfo {volume} {16}},\ \bibinfo {pages} {944--952} (\bibinfo {year} {2020})}\BibitemShut {NoStop}%
\bibitem [{\citenamefont {Yordanov}, \citenamefont {Arvidsson-Shukur},\ and\ \citenamefont {Barnes}(2020)}]{yordanov2020efficient}%
  \BibitemOpen
  \bibfield  {author} {\bibinfo {author} {\bibfnamefont {Y.~S.}\ \bibnamefont {Yordanov}}, \bibinfo {author} {\bibfnamefont {D.~R.}\ \bibnamefont {Arvidsson-Shukur}},\ and\ \bibinfo {author} {\bibfnamefont {C.~H.}\ \bibnamefont {Barnes}},\ }\bibfield  {title} {\enquote {\bibinfo {title} {Efficient quantum circuits for quantum computational chemistry},}\ }\href@noop {} {\bibfield  {journal} {\bibinfo  {journal} {Physical Review A}\ }\textbf {\bibinfo {volume} {102}},\ \bibinfo {pages} {062612} (\bibinfo {year} {2020})}\BibitemShut {NoStop}%
\bibitem [{\citenamefont {Rivera-Dean}\ \emph {et~al.}(2021)\citenamefont {Rivera-Dean}, \citenamefont {Huembeli}, \citenamefont {Ac{\'\i}n},\ and\ \citenamefont {Bowles}}]{rivera2021avoiding}%
  \BibitemOpen
  \bibfield  {author} {\bibinfo {author} {\bibfnamefont {J.}~\bibnamefont {Rivera-Dean}}, \bibinfo {author} {\bibfnamefont {P.}~\bibnamefont {Huembeli}}, \bibinfo {author} {\bibfnamefont {A.}~\bibnamefont {Ac{\'\i}n}},\ and\ \bibinfo {author} {\bibfnamefont {J.}~\bibnamefont {Bowles}},\ }\bibfield  {title} {\enquote {\bibinfo {title} {Avoiding local minima in variational quantum algorithms with neural networks},}\ }\href@noop {} {\bibfield  {journal} {\bibinfo  {journal} {arXiv preprint arXiv:2104.02955}\ } (\bibinfo {year} {2021})}\BibitemShut {NoStop}%
\bibitem [{\citenamefont {Mondal}\ \emph {et~al.}(2023)\citenamefont {Mondal}, \citenamefont {Halder}, \citenamefont {Halder},\ and\ \citenamefont {Maitra}}]{mondal2023development}%
  \BibitemOpen
  \bibfield  {author} {\bibinfo {author} {\bibfnamefont {D.}~\bibnamefont {Mondal}}, \bibinfo {author} {\bibfnamefont {D.}~\bibnamefont {Halder}}, \bibinfo {author} {\bibfnamefont {S.}~\bibnamefont {Halder}},\ and\ \bibinfo {author} {\bibfnamefont {R.}~\bibnamefont {Maitra}},\ }\bibfield  {title} {\enquote {\bibinfo {title} {{Development of a compact Ansatz via operator commutativity screening: Digital quantum simulation of molecular systems}},}\ }\href {https://doi.org/10.1063/5.0153182} {\bibfield  {journal} {\bibinfo  {journal} {The Journal of Chemical Physics}\ }\textbf {\bibinfo {volume} {159}},\ \bibinfo {pages} {014105} (\bibinfo {year} {2023})}\BibitemShut {NoStop}%
\bibitem [{\citenamefont {Halder}\ \emph {et~al.}(2023)\citenamefont {Halder}, \citenamefont {Patra}, \citenamefont {Mondal},\ and\ \citenamefont {Maitra}}]{sonaldeep2023}%
  \BibitemOpen
  \bibfield  {author} {\bibinfo {author} {\bibfnamefont {S.}~\bibnamefont {Halder}}, \bibinfo {author} {\bibfnamefont {C.}~\bibnamefont {Patra}}, \bibinfo {author} {\bibfnamefont {D.}~\bibnamefont {Mondal}},\ and\ \bibinfo {author} {\bibfnamefont {R.}~\bibnamefont {Maitra}},\ }\bibfield  {title} {\enquote {\bibinfo {title} {{Machine learning aided dimensionality reduction toward a resource efficient projective quantum eigensolver: Formal development and pilot applications}},}\ }\href@noop {} {\bibfield  {journal} {\bibinfo  {journal} {The Journal of Chemical Physics}\ }\textbf {\bibinfo {volume} {158}} (\bibinfo {year} {2023})},\ \bibinfo {note} {244101}\BibitemShut {NoStop}%
\bibitem [{\citenamefont {Halder}, \citenamefont {Mondal},\ and\ \citenamefont {Maitra}(2024)}]{halder2024noise}%
  \BibitemOpen
  \bibfield  {author} {\bibinfo {author} {\bibfnamefont {D.}~\bibnamefont {Halder}}, \bibinfo {author} {\bibfnamefont {D.}~\bibnamefont {Mondal}},\ and\ \bibinfo {author} {\bibfnamefont {R.}~\bibnamefont {Maitra}},\ }\bibfield  {title} {\enquote {\bibinfo {title} {Noise-independent route toward the genesis of a compact ansatz for molecular energetics: A dynamic approach},}\ }\href@noop {} {\bibfield  {journal} {\bibinfo  {journal} {The Journal of Chemical Physics}\ }\textbf {\bibinfo {volume} {160}} (\bibinfo {year} {2024})}\BibitemShut {NoStop}%
\bibitem [{\citenamefont {Patra}, \citenamefont {Halder},\ and\ \citenamefont {Maitra}(2024)}]{patra2024projective}%
  \BibitemOpen
  \bibfield  {author} {\bibinfo {author} {\bibfnamefont {C.}~\bibnamefont {Patra}}, \bibinfo {author} {\bibfnamefont {S.}~\bibnamefont {Halder}},\ and\ \bibinfo {author} {\bibfnamefont {R.}~\bibnamefont {Maitra}},\ }\bibfield  {title} {\enquote {\bibinfo {title} {Projective quantum eigensolver via adiabatically decoupled subsystem evolution: A resource efficient approach to molecular energetics in noisy quantum computers},}\ }\href@noop {} {\bibfield  {journal} {\bibinfo  {journal} {The Journal of Chemical Physics}\ }\textbf {\bibinfo {volume} {160}} (\bibinfo {year} {2024})}\BibitemShut {NoStop}%
\bibitem [{\citenamefont {Patra}\ \emph {et~al.}(2024)\citenamefont {Patra}, \citenamefont {Mukherjee}, \citenamefont {Halder}, \citenamefont {Mondal},\ and\ \citenamefont {Maitra}}]{patra2024toward}%
  \BibitemOpen
  \bibfield  {author} {\bibinfo {author} {\bibfnamefont {C.}~\bibnamefont {Patra}}, \bibinfo {author} {\bibfnamefont {D.}~\bibnamefont {Mukherjee}}, \bibinfo {author} {\bibfnamefont {S.}~\bibnamefont {Halder}}, \bibinfo {author} {\bibfnamefont {D.}~\bibnamefont {Mondal}},\ and\ \bibinfo {author} {\bibfnamefont {R.}~\bibnamefont {Maitra}},\ }\bibfield  {title} {\enquote {\bibinfo {title} {Toward a resource-optimized dynamic quantum algorithm via non-iterative auxiliary subspace corrections},}\ }\href@noop {} {\bibfield  {journal} {\bibinfo  {journal} {The Journal of Chemical Physics}\ }\textbf {\bibinfo {volume} {161}} (\bibinfo {year} {2024})}\BibitemShut {NoStop}%
\bibitem [{\citenamefont {Halder}\ \emph {et~al.}(2024)\citenamefont {Halder}, \citenamefont {Dey}, \citenamefont {Shrikhande},\ and\ \citenamefont {Maitra}}]{10.1039/d3sc05807g}%
  \BibitemOpen
  \bibfield  {author} {\bibinfo {author} {\bibfnamefont {S.}~\bibnamefont {Halder}}, \bibinfo {author} {\bibfnamefont {A.}~\bibnamefont {Dey}}, \bibinfo {author} {\bibfnamefont {C.}~\bibnamefont {Shrikhande}},\ and\ \bibinfo {author} {\bibfnamefont {R.}~\bibnamefont {Maitra}},\ }\bibfield  {title} {\enquote {\bibinfo {title} {Machine learning assisted construction of a shallow depth dynamic ansatz for noisy quantum hardware},}\ }\href {https://doi.org/10.1039/d3sc05807g} {\bibfield  {journal} {\bibinfo  {journal} {Chemical Science}\ }\textbf {\bibinfo {volume} {15}},\ \bibinfo {pages} {3279--3289} (\bibinfo {year} {2024})},\ \Eprint {https://arxiv.org/abs/https://pubs.rsc.org/sc/article-pdf/15/9/3279/9402549/d3sc05807g.pdf} {https://pubs.rsc.org/sc/article-pdf/15/9/3279/9402549/d3sc05807g.pdf} \BibitemShut {NoStop}%
\bibitem [{\citenamefont {Patra}\ and\ \citenamefont {Maitra}(2025)}]{patra2025energy}%
  \BibitemOpen
  \bibfield  {author} {\bibinfo {author} {\bibfnamefont {C.}~\bibnamefont {Patra}}\ and\ \bibinfo {author} {\bibfnamefont {R.}~\bibnamefont {Maitra}},\ }\bibfield  {title} {\enquote {\bibinfo {title} {Energy landscape plummeting in variational quantum eigensolver: Subspace optimization, non-iterative corrections, and generator-informed initialization for improved quantum efficiency},}\ }\href@noop {} {\bibfield  {journal} {\bibinfo  {journal} {The Journal of Chemical Physics}\ }\textbf {\bibinfo {volume} {163}},\ \bibinfo {pages} {024112} (\bibinfo {year} {2025})}\BibitemShut {NoStop}%
\bibitem [{\citenamefont {Ding}\ \emph {et~al.}(2026)\citenamefont {Ding}, \citenamefont {Zhan}, \citenamefont {Preskill},\ and\ \citenamefont {Lin}}]{ding2026simple}%
  \BibitemOpen
  \bibfield  {author} {\bibinfo {author} {\bibfnamefont {Z.}~\bibnamefont {Ding}}, \bibinfo {author} {\bibfnamefont {Y.}~\bibnamefont {Zhan}}, \bibinfo {author} {\bibfnamefont {J.}~\bibnamefont {Preskill}},\ and\ \bibinfo {author} {\bibfnamefont {L.}~\bibnamefont {Lin}},\ }\bibfield  {title} {\enquote {\bibinfo {title} {Simple and efficient end-to-end quantum thermal and ground state preparation},}\ }\href@noop {} {\bibfield  {journal} {\bibinfo  {journal} {Nature Physics}\ ,\ \bibinfo {pages} {1--5}} (\bibinfo {year} {2026})}\BibitemShut {NoStop}%
\bibitem [{\citenamefont {Wang}, \citenamefont {Avdic},\ and\ \citenamefont {Mazziotti}(2025)}]{wang2025shadow}%
  \BibitemOpen
  \bibfield  {author} {\bibinfo {author} {\bibfnamefont {Y.}~\bibnamefont {Wang}}, \bibinfo {author} {\bibfnamefont {I.}~\bibnamefont {Avdic}},\ and\ \bibinfo {author} {\bibfnamefont {D.~A.}\ \bibnamefont {Mazziotti}},\ }\bibfield  {title} {\enquote {\bibinfo {title} {Shadow ansatz for the many-fermion wave function in scalable molecular simulations on quantum computing devices},}\ }\href@noop {} {\bibfield  {journal} {\bibinfo  {journal} {Physical Review A}\ }\textbf {\bibinfo {volume} {112}},\ \bibinfo {pages} {022432} (\bibinfo {year} {2025})}\BibitemShut {NoStop}%
\bibitem [{\citenamefont {Patel}\ \emph {et~al.}(2026)\citenamefont {Patel}, \citenamefont {Jayakumar}, \citenamefont {Huang}, \citenamefont {Zeng},\ and\ \citenamefont {Izmaylov}}]{patel2026quantum}%
  \BibitemOpen
  \bibfield  {author} {\bibinfo {author} {\bibfnamefont {S.}~\bibnamefont {Patel}}, \bibinfo {author} {\bibfnamefont {P.}~\bibnamefont {Jayakumar}}, \bibinfo {author} {\bibfnamefont {R.}~\bibnamefont {Huang}}, \bibinfo {author} {\bibfnamefont {T.}~\bibnamefont {Zeng}},\ and\ \bibinfo {author} {\bibfnamefont {A.~F.}\ \bibnamefont {Izmaylov}},\ }\bibfield  {title} {\enquote {\bibinfo {title} {Quantum seniority-based subspace expansion: Linear combinations of short-circuit unitary transformations for the electronic structure problem},}\ }\href@noop {} {\bibfield  {journal} {\bibinfo  {journal} {Journal of Chemical Theory and Computation}\ }\textbf {\bibinfo {volume} {22}},\ \bibinfo {pages} {3937--3949} (\bibinfo {year} {2026})}\BibitemShut {NoStop}%
\bibitem [{\citenamefont {Mondal}, \citenamefont {Patra},\ and\ \citenamefont {Maitra}(2026)}]{mondal2026advancing}%
  \BibitemOpen
  \bibfield  {author} {\bibinfo {author} {\bibfnamefont {D.}~\bibnamefont {Mondal}}, \bibinfo {author} {\bibfnamefont {A.~K.}\ \bibnamefont {Patra}},\ and\ \bibinfo {author} {\bibfnamefont {R.}~\bibnamefont {Maitra}},\ }\bibfield  {title} {\enquote {\bibinfo {title} {Advancing practical quantum embedding simulations via operator commutativity based state preparation for complex chemical systems},}\ }\href@noop {} {\bibfield  {journal} {\bibinfo  {journal} {Chemical Science}\ } (\bibinfo {year} {2026})}\BibitemShut {NoStop}%
\bibitem [{\citenamefont {Yoshioka}\ \emph {et~al.}(2025)\citenamefont {Yoshioka}, \citenamefont {Amico}, \citenamefont {Kirby}, \citenamefont {Jurcevic}, \citenamefont {Dutt}, \citenamefont {Fuller}, \citenamefont {Garion}, \citenamefont {Haas}, \citenamefont {Hamamura}, \citenamefont {Ivrii}, \citenamefont {Majumdar}, \citenamefont {Minev}, \citenamefont {Motta}, \citenamefont {Pokharel}, \citenamefont {Rivero}, \citenamefont {Sharma}, \citenamefont {Wood}, \citenamefont {Javadi-Abhari},\ and\ \citenamefont {Mezzacapo}}]{Yoshioka2025}%
  \BibitemOpen
  \bibfield  {author} {\bibinfo {author} {\bibfnamefont {N.}~\bibnamefont {Yoshioka}}, \bibinfo {author} {\bibfnamefont {M.}~\bibnamefont {Amico}}, \bibinfo {author} {\bibfnamefont {W.}~\bibnamefont {Kirby}}, \bibinfo {author} {\bibfnamefont {P.}~\bibnamefont {Jurcevic}}, \bibinfo {author} {\bibfnamefont {A.}~\bibnamefont {Dutt}}, \bibinfo {author} {\bibfnamefont {B.}~\bibnamefont {Fuller}}, \bibinfo {author} {\bibfnamefont {S.}~\bibnamefont {Garion}}, \bibinfo {author} {\bibfnamefont {H.}~\bibnamefont {Haas}}, \bibinfo {author} {\bibfnamefont {I.}~\bibnamefont {Hamamura}}, \bibinfo {author} {\bibfnamefont {A.}~\bibnamefont {Ivrii}}, \bibinfo {author} {\bibfnamefont {R.}~\bibnamefont {Majumdar}}, \bibinfo {author} {\bibfnamefont {Z.}~\bibnamefont {Minev}}, \bibinfo {author} {\bibfnamefont {M.}~\bibnamefont {Motta}}, \bibinfo {author} {\bibfnamefont {B.}~\bibnamefont {Pokharel}}, \bibinfo {author} {\bibfnamefont {P.}~\bibnamefont {Rivero}}, \bibinfo {author} {\bibfnamefont {K.}~\bibnamefont {Sharma}}, \bibinfo
  {author} {\bibfnamefont {C.~J.}\ \bibnamefont {Wood}}, \bibinfo {author} {\bibfnamefont {A.}~\bibnamefont {Javadi-Abhari}},\ and\ \bibinfo {author} {\bibfnamefont {A.}~\bibnamefont {Mezzacapo}},\ }\bibfield  {title} {\enquote {\bibinfo {title} {Krylov diagonalization of large many-body hamiltonians on a quantum processor},}\ }\href {https://doi.org/10.1038/s41467-025-59716-z} {\bibfield  {journal} {\bibinfo  {journal} {Nature Communications}\ }\textbf {\bibinfo {volume} {16}},\ \bibinfo {pages} {5014} (\bibinfo {year} {2025})}\BibitemShut {NoStop}%
\bibitem [{\citenamefont {Kanno}\ \emph {et~al.}(2026)\citenamefont {Kanno}, \citenamefont {Kohda}, \citenamefont {Imai}, \citenamefont {Koh}, \citenamefont {Mitarai}, \citenamefont {Mizukami},\ and\ \citenamefont {Nakagawa}}]{kanno2026quantum}%
  \BibitemOpen
  \bibfield  {author} {\bibinfo {author} {\bibfnamefont {K.}~\bibnamefont {Kanno}}, \bibinfo {author} {\bibfnamefont {M.}~\bibnamefont {Kohda}}, \bibinfo {author} {\bibfnamefont {R.}~\bibnamefont {Imai}}, \bibinfo {author} {\bibfnamefont {S.}~\bibnamefont {Koh}}, \bibinfo {author} {\bibfnamefont {K.}~\bibnamefont {Mitarai}}, \bibinfo {author} {\bibfnamefont {W.}~\bibnamefont {Mizukami}},\ and\ \bibinfo {author} {\bibfnamefont {Y.~O.}\ \bibnamefont {Nakagawa}},\ }\bibfield  {title} {\enquote {\bibinfo {title} {Quantum-selected configuration interaction: Classical diagonalization of hamiltonians in subspaces selected by quantum computers},}\ }\href@noop {} {\bibfield  {journal} {\bibinfo  {journal} {Physical Review Research}\ }\textbf {\bibinfo {volume} {8}},\ \bibinfo {pages} {023268} (\bibinfo {year} {2026})}\BibitemShut {NoStop}%
\bibitem [{\citenamefont {Robledo-Moreno}\ \emph {et~al.}(2025)\citenamefont {Robledo-Moreno}, \citenamefont {Motta}, \citenamefont {Haas}, \citenamefont {Javadi-Abhari}, \citenamefont {Jurcevic}, \citenamefont {Kirby}, \citenamefont {Martiel}, \citenamefont {Sharma}, \citenamefont {Sharma}, \citenamefont {Shirakawa}, \citenamefont {Sitdikov}, \citenamefont {Sun}, \citenamefont {Sung}, \citenamefont {Takita}, \citenamefont {Tran}, \citenamefont {Yunoki},\ and\ \citenamefont {Mezzacapo}}]{doi:10.1126/sciadv.adu9991}%
  \BibitemOpen
  \bibfield  {author} {\bibinfo {author} {\bibfnamefont {J.}~\bibnamefont {Robledo-Moreno}}, \bibinfo {author} {\bibfnamefont {M.}~\bibnamefont {Motta}}, \bibinfo {author} {\bibfnamefont {H.}~\bibnamefont {Haas}}, \bibinfo {author} {\bibfnamefont {A.}~\bibnamefont {Javadi-Abhari}}, \bibinfo {author} {\bibfnamefont {P.}~\bibnamefont {Jurcevic}}, \bibinfo {author} {\bibfnamefont {W.}~\bibnamefont {Kirby}}, \bibinfo {author} {\bibfnamefont {S.}~\bibnamefont {Martiel}}, \bibinfo {author} {\bibfnamefont {K.}~\bibnamefont {Sharma}}, \bibinfo {author} {\bibfnamefont {S.}~\bibnamefont {Sharma}}, \bibinfo {author} {\bibfnamefont {T.}~\bibnamefont {Shirakawa}}, \bibinfo {author} {\bibfnamefont {I.}~\bibnamefont {Sitdikov}}, \bibinfo {author} {\bibfnamefont {R.-Y.}\ \bibnamefont {Sun}}, \bibinfo {author} {\bibfnamefont {K.~J.}\ \bibnamefont {Sung}}, \bibinfo {author} {\bibfnamefont {M.}~\bibnamefont {Takita}}, \bibinfo {author} {\bibfnamefont {M.~C.}\ \bibnamefont {Tran}}, \bibinfo {author} {\bibfnamefont
  {S.}~\bibnamefont {Yunoki}},\ and\ \bibinfo {author} {\bibfnamefont {A.}~\bibnamefont {Mezzacapo}},\ }\bibfield  {title} {\enquote {\bibinfo {title} {Chemistry beyond the scale of exact diagonalization on a quantum-centric supercomputer},}\ }\href {https://doi.org/10.1126/sciadv.adu9991} {\bibfield  {journal} {\bibinfo  {journal} {Science Advances}\ }\textbf {\bibinfo {volume} {11}},\ \bibinfo {pages} {eadu9991} (\bibinfo {year} {2025})},\ \Eprint {https://arxiv.org/abs/https://www.science.org/doi/pdf/10.1126/sciadv.adu9991} {https://www.science.org/doi/pdf/10.1126/sciadv.adu9991} \BibitemShut {NoStop}%
\bibitem [{\citenamefont {Shajan}\ \emph {et~al.}(2025)\citenamefont {Shajan}, \citenamefont {Kaliakin}, \citenamefont {Mitra}, \citenamefont {Robledo~Moreno}, \citenamefont {Li}, \citenamefont {Motta}, \citenamefont {Johnson}, \citenamefont {Saki}, \citenamefont {Das}, \citenamefont {Sitdikov} \emph {et~al.}}]{shajan2025toward}%
  \BibitemOpen
  \bibfield  {author} {\bibinfo {author} {\bibfnamefont {A.}~\bibnamefont {Shajan}}, \bibinfo {author} {\bibfnamefont {D.}~\bibnamefont {Kaliakin}}, \bibinfo {author} {\bibfnamefont {A.}~\bibnamefont {Mitra}}, \bibinfo {author} {\bibfnamefont {J.}~\bibnamefont {Robledo~Moreno}}, \bibinfo {author} {\bibfnamefont {Z.}~\bibnamefont {Li}}, \bibinfo {author} {\bibfnamefont {M.}~\bibnamefont {Motta}}, \bibinfo {author} {\bibfnamefont {C.}~\bibnamefont {Johnson}}, \bibinfo {author} {\bibfnamefont {A.~A.}\ \bibnamefont {Saki}}, \bibinfo {author} {\bibfnamefont {S.}~\bibnamefont {Das}}, \bibinfo {author} {\bibfnamefont {I.}~\bibnamefont {Sitdikov}}, \emph {et~al.},\ }\bibfield  {title} {\enquote {\bibinfo {title} {Toward quantum-centric simulations of extended molecules: Sample-based quantum diagonalization enhanced with density matrix embedding theory},}\ }\href@noop {} {\bibfield  {journal} {\bibinfo  {journal} {Journal of Chemical Theory and Computation}\ }\textbf {\bibinfo {volume} {21}},\ \bibinfo {pages}
  {6801--6810} (\bibinfo {year} {2025})}\BibitemShut {NoStop}%
\bibitem [{\citenamefont {Danilov}\ \emph {et~al.}(2025)\citenamefont {Danilov}, \citenamefont {Robledo-Moreno}, \citenamefont {Sung}, \citenamefont {Motta},\ and\ \citenamefont {Shee}}]{danilov2025enhancing}%
  \BibitemOpen
  \bibfield  {author} {\bibinfo {author} {\bibfnamefont {D.}~\bibnamefont {Danilov}}, \bibinfo {author} {\bibfnamefont {J.}~\bibnamefont {Robledo-Moreno}}, \bibinfo {author} {\bibfnamefont {K.~J.}\ \bibnamefont {Sung}}, \bibinfo {author} {\bibfnamefont {M.}~\bibnamefont {Motta}},\ and\ \bibinfo {author} {\bibfnamefont {J.}~\bibnamefont {Shee}},\ }\bibfield  {title} {\enquote {\bibinfo {title} {Enhancing the accuracy and efficiency of sample-based quantum diagonalization with phaseless auxiliary-field quantum monte carlo},}\ }\href@noop {} {\bibfield  {journal} {\bibinfo  {journal} {Journal of Chemical Theory and Computation}\ }\textbf {\bibinfo {volume} {21}},\ \bibinfo {pages} {11585--11594} (\bibinfo {year} {2025})}\BibitemShut {NoStop}%
\bibitem [{\citenamefont {Mikkelsen}\ and\ \citenamefont {Nakagawa}(2025)}]{mikkelsen2025quantum}%
  \BibitemOpen
  \bibfield  {author} {\bibinfo {author} {\bibfnamefont {M.}~\bibnamefont {Mikkelsen}}\ and\ \bibinfo {author} {\bibfnamefont {Y.~O.}\ \bibnamefont {Nakagawa}},\ }\bibfield  {title} {\enquote {\bibinfo {title} {Quantum-selected configuration interaction with time-evolved state},}\ }\href@noop {} {\bibfield  {journal} {\bibinfo  {journal} {Physical Review Research}\ }\textbf {\bibinfo {volume} {7}},\ \bibinfo {pages} {043043} (\bibinfo {year} {2025})}\BibitemShut {NoStop}%
\bibitem [{\citenamefont {Merz~Jr}\ \emph {et~al.}(2026)\citenamefont {Merz~Jr}, \citenamefont {Shajan}, \citenamefont {Kaliakin}, \citenamefont {Liang}, \citenamefont {Otsuka}, \citenamefont {Shirakawa}, \citenamefont {Broers}, \citenamefont {Xu}, \citenamefont {Tsuji}, \citenamefont {Sato} \emph {et~al.}}]{merz2026crossing}%
  \BibitemOpen
  \bibfield  {author} {\bibinfo {author} {\bibfnamefont {K.~M.}\ \bibnamefont {Merz~Jr}}, \bibinfo {author} {\bibfnamefont {A.}~\bibnamefont {Shajan}}, \bibinfo {author} {\bibfnamefont {D.}~\bibnamefont {Kaliakin}}, \bibinfo {author} {\bibfnamefont {F.}~\bibnamefont {Liang}}, \bibinfo {author} {\bibfnamefont {Y.}~\bibnamefont {Otsuka}}, \bibinfo {author} {\bibfnamefont {T.}~\bibnamefont {Shirakawa}}, \bibinfo {author} {\bibfnamefont {L.}~\bibnamefont {Broers}}, \bibinfo {author} {\bibfnamefont {H.}~\bibnamefont {Xu}}, \bibinfo {author} {\bibfnamefont {M.}~\bibnamefont {Tsuji}}, \bibinfo {author} {\bibfnamefont {M.}~\bibnamefont {Sato}}, \emph {et~al.},\ }\bibfield  {title} {\enquote {\bibinfo {title} {Crossing the 12,000-atom barrier with heterogeneous quantum-classical supercomputing: quantum chemistry of protein-ligand complexes},}\ }\href@noop {} {\bibfield  {journal} {\bibinfo  {journal} {arXiv preprint arXiv:2605.01138}\ } (\bibinfo {year} {2026})}\BibitemShut {NoStop}%
\bibitem [{\citenamefont {Shirakawa}\ \emph {et~al.}(2026)\citenamefont {Shirakawa}, \citenamefont {Robledo-Moreno}, \citenamefont {Itoko}, \citenamefont {Tripathi}, \citenamefont {Ueda}, \citenamefont {Kawashima}, \citenamefont {Broers}, \citenamefont {Kirby}, \citenamefont {Pathak}, \citenamefont {Paik} \emph {et~al.}}]{shirakawa2026closed}%
  \BibitemOpen
  \bibfield  {author} {\bibinfo {author} {\bibfnamefont {T.}~\bibnamefont {Shirakawa}}, \bibinfo {author} {\bibfnamefont {J.}~\bibnamefont {Robledo-Moreno}}, \bibinfo {author} {\bibfnamefont {T.}~\bibnamefont {Itoko}}, \bibinfo {author} {\bibfnamefont {V.}~\bibnamefont {Tripathi}}, \bibinfo {author} {\bibfnamefont {K.}~\bibnamefont {Ueda}}, \bibinfo {author} {\bibfnamefont {Y.}~\bibnamefont {Kawashima}}, \bibinfo {author} {\bibfnamefont {L.}~\bibnamefont {Broers}}, \bibinfo {author} {\bibfnamefont {W.}~\bibnamefont {Kirby}}, \bibinfo {author} {\bibfnamefont {H.}~\bibnamefont {Pathak}}, \bibinfo {author} {\bibfnamefont {H.}~\bibnamefont {Paik}}, \emph {et~al.},\ }\bibfield  {title} {\enquote {\bibinfo {title} {Closed-loop calculations of electronic structure on a quantum processor and a classical supercomputer at full scale},}\ }\href@noop {} {\bibfield  {journal} {\bibinfo  {journal} {Future Generation Computer Systems}\ ,\ \bibinfo {pages} {108731}} (\bibinfo {year} {2026})}\BibitemShut {NoStop}%
\bibitem [{\citenamefont {Shajan}\ \emph {et~al.}(2026)\citenamefont {Shajan}, \citenamefont {Kaliakin}, \citenamefont {Liang}, \citenamefont {Pellegrini}, \citenamefont {Doga}, \citenamefont {Bhowmik}, \citenamefont {Das}, \citenamefont {Mezzacapo}, \citenamefont {Motta},\ and\ \citenamefont {Merz~Jr}}]{shajan2026molecular}%
  \BibitemOpen
  \bibfield  {author} {\bibinfo {author} {\bibfnamefont {A.}~\bibnamefont {Shajan}}, \bibinfo {author} {\bibfnamefont {D.}~\bibnamefont {Kaliakin}}, \bibinfo {author} {\bibfnamefont {F.}~\bibnamefont {Liang}}, \bibinfo {author} {\bibfnamefont {T.}~\bibnamefont {Pellegrini}}, \bibinfo {author} {\bibfnamefont {H.}~\bibnamefont {Doga}}, \bibinfo {author} {\bibfnamefont {S.}~\bibnamefont {Bhowmik}}, \bibinfo {author} {\bibfnamefont {S.}~\bibnamefont {Das}}, \bibinfo {author} {\bibfnamefont {A.}~\bibnamefont {Mezzacapo}}, \bibinfo {author} {\bibfnamefont {M.}~\bibnamefont {Motta}},\ and\ \bibinfo {author} {\bibfnamefont {K.~M.}\ \bibnamefont {Merz~Jr}},\ }\bibfield  {title} {\enquote {\bibinfo {title} {Molecular quantum computations on a protein},}\ }\href@noop {} {\bibfield  {journal} {\bibinfo  {journal} {Journal of Chemical Theory and Computation}\ }\textbf {\bibinfo {volume} {22}},\ \bibinfo {pages} {6041} (\bibinfo {year} {2026})}\BibitemShut {NoStop}%
\bibitem [{\citenamefont {Wang}\ \emph {et~al.}(2026)\citenamefont {Wang}, \citenamefont {Sung}, \citenamefont {D’Cunha}, \citenamefont {Hermes}, \citenamefont {Gujarati}, \citenamefont {Kawashima}, \citenamefont {Ohnishi}, \citenamefont {Jones}, \citenamefont {Motta},\ and\ \citenamefont {Gagliardi}}]{wang2026localized}%
  \BibitemOpen
  \bibfield  {author} {\bibinfo {author} {\bibfnamefont {Q.}~\bibnamefont {Wang}}, \bibinfo {author} {\bibfnamefont {K.~J.}\ \bibnamefont {Sung}}, \bibinfo {author} {\bibfnamefont {R.}~\bibnamefont {D’Cunha}}, \bibinfo {author} {\bibfnamefont {M.~R.}\ \bibnamefont {Hermes}}, \bibinfo {author} {\bibfnamefont {T.}~\bibnamefont {Gujarati}}, \bibinfo {author} {\bibfnamefont {Y.}~\bibnamefont {Kawashima}}, \bibinfo {author} {\bibfnamefont {Y.-y.}\ \bibnamefont {Ohnishi}}, \bibinfo {author} {\bibfnamefont {G.~O.}\ \bibnamefont {Jones}}, \bibinfo {author} {\bibfnamefont {M.}~\bibnamefont {Motta}},\ and\ \bibinfo {author} {\bibfnamefont {L.}~\bibnamefont {Gagliardi}},\ }\bibfield  {title} {\enquote {\bibinfo {title} {Localized sample-based quantum diagonalization for strongly correlated chemistry},}\ }\href@noop {} {\bibfield  {journal} {\bibinfo  {journal} {Proceedings of the National Academy of Sciences}\ }\textbf {\bibinfo {volume} {123}},\ \bibinfo {pages} {e2603914123} (\bibinfo {year} {2026})}\BibitemShut
  {NoStop}%
\bibitem [{\citenamefont {Yamamoto}\ \emph {et~al.}(2026)\citenamefont {Yamamoto}, \citenamefont {Masui}, \citenamefont {Nakajima}, \citenamefont {Tsuji}, \citenamefont {Sato}, \citenamefont {Schow}, \citenamefont {Heidemann}, \citenamefont {Burke}, \citenamefont {Seitz}, \citenamefont {Backhouse} \emph {et~al.}}]{yamamoto2026quantum}%
  \BibitemOpen
  \bibfield  {author} {\bibinfo {author} {\bibfnamefont {K.}~\bibnamefont {Yamamoto}}, \bibinfo {author} {\bibfnamefont {R.}~\bibnamefont {Masui}}, \bibinfo {author} {\bibfnamefont {T.}~\bibnamefont {Nakajima}}, \bibinfo {author} {\bibfnamefont {M.}~\bibnamefont {Tsuji}}, \bibinfo {author} {\bibfnamefont {M.}~\bibnamefont {Sato}}, \bibinfo {author} {\bibfnamefont {P.}~\bibnamefont {Schow}}, \bibinfo {author} {\bibfnamefont {L.}~\bibnamefont {Heidemann}}, \bibinfo {author} {\bibfnamefont {M.}~\bibnamefont {Burke}}, \bibinfo {author} {\bibfnamefont {P.}~\bibnamefont {Seitz}}, \bibinfo {author} {\bibfnamefont {O.~J.}\ \bibnamefont {Backhouse}}, \emph {et~al.},\ }\bibfield  {title} {\enquote {\bibinfo {title} {Quantum-hpc hybrid computation of biomolecular excited-state energies},}\ }\href@noop {} {\bibfield  {journal} {\bibinfo  {journal} {arXiv preprint arXiv:2601.15677}\ } (\bibinfo {year} {2026})}\BibitemShut {NoStop}%
\bibitem [{\citenamefont {Kamoshita}\ and\ \citenamefont {Mitarai}(2026)}]{kamoshita2026qsci}%
  \BibitemOpen
  \bibfield  {author} {\bibinfo {author} {\bibfnamefont {M.}~\bibnamefont {Kamoshita}}\ and\ \bibinfo {author} {\bibfnamefont {K.}~\bibnamefont {Mitarai}},\ }\bibfield  {title} {\enquote {\bibinfo {title} {Qsci-cmp: Quantum-selected configuration interaction with chemically motivated preselection},}\ }\href@noop {} {\bibfield  {journal} {\bibinfo  {journal} {arXiv preprint arXiv:2608.05766}\ } (\bibinfo {year} {2026})}\BibitemShut {NoStop}%
\bibitem [{\citenamefont {Yu}\ \emph {et~al.}(2025)\citenamefont {Yu}, \citenamefont {Moreno}, \citenamefont {Iosue}, \citenamefont {Bertels}, \citenamefont {Claudino}, \citenamefont {Fuller}, \citenamefont {Groszkowski}, \citenamefont {Humble}, \citenamefont {Jurcevic}, \citenamefont {Kirby}, \citenamefont {Maier}, \citenamefont {Motta}, \citenamefont {Pokharel}, \citenamefont {Seif}, \citenamefont {Shehata}, \citenamefont {Sung}, \citenamefont {Tran}, \citenamefont {Tripathi}, \citenamefont {Mezzacapo},\ and\ \citenamefont {Sharma}}]{yu2025quantumcentricalgorithmsamplebasedkrylov}%
  \BibitemOpen
  \bibfield  {author} {\bibinfo {author} {\bibfnamefont {J.}~\bibnamefont {Yu}}, \bibinfo {author} {\bibfnamefont {J.~R.}\ \bibnamefont {Moreno}}, \bibinfo {author} {\bibfnamefont {J.~T.}\ \bibnamefont {Iosue}}, \bibinfo {author} {\bibfnamefont {L.}~\bibnamefont {Bertels}}, \bibinfo {author} {\bibfnamefont {D.}~\bibnamefont {Claudino}}, \bibinfo {author} {\bibfnamefont {B.}~\bibnamefont {Fuller}}, \bibinfo {author} {\bibfnamefont {P.}~\bibnamefont {Groszkowski}}, \bibinfo {author} {\bibfnamefont {T.~S.}\ \bibnamefont {Humble}}, \bibinfo {author} {\bibfnamefont {P.}~\bibnamefont {Jurcevic}}, \bibinfo {author} {\bibfnamefont {W.}~\bibnamefont {Kirby}}, \bibinfo {author} {\bibfnamefont {T.~A.}\ \bibnamefont {Maier}}, \bibinfo {author} {\bibfnamefont {M.}~\bibnamefont {Motta}}, \bibinfo {author} {\bibfnamefont {B.}~\bibnamefont {Pokharel}}, \bibinfo {author} {\bibfnamefont {A.}~\bibnamefont {Seif}}, \bibinfo {author} {\bibfnamefont {A.}~\bibnamefont {Shehata}}, \bibinfo {author} {\bibfnamefont {K.~J.}\
  \bibnamefont {Sung}}, \bibinfo {author} {\bibfnamefont {M.~C.}\ \bibnamefont {Tran}}, \bibinfo {author} {\bibfnamefont {V.}~\bibnamefont {Tripathi}}, \bibinfo {author} {\bibfnamefont {A.}~\bibnamefont {Mezzacapo}},\ and\ \bibinfo {author} {\bibfnamefont {K.}~\bibnamefont {Sharma}},\ }\href {https://arxiv.org/abs/2501.09702} {\enquote {\bibinfo {title} {Quantum-centric algorithm for sample-based krylov diagonalization},}\ } (\bibinfo {year} {2025}),\ \Eprint {https://arxiv.org/abs/2501.09702} {arXiv:2501.09702 [quant-ph]} \BibitemShut {NoStop}%
\bibitem [{\citenamefont {Piccinelli}\ \emph {et~al.}(2026)\citenamefont {Piccinelli}, \citenamefont {Baiardi}, \citenamefont {Barison}, \citenamefont {Rossmannek}, \citenamefont {Vazquez}, \citenamefont {Tacchino}, \citenamefont {Mensa}, \citenamefont {Altamura}, \citenamefont {Alavi}, \citenamefont {Motta}, \citenamefont {Robledo-Moreno}, \citenamefont {Kirby}, \citenamefont {Sharma}, \citenamefont {Mezzacapo},\ and\ \citenamefont {Tavernelli}}]{piccinelli2026quantumchemistryprovableconvergence}%
  \BibitemOpen
  \bibfield  {author} {\bibinfo {author} {\bibfnamefont {S.}~\bibnamefont {Piccinelli}}, \bibinfo {author} {\bibfnamefont {A.}~\bibnamefont {Baiardi}}, \bibinfo {author} {\bibfnamefont {S.}~\bibnamefont {Barison}}, \bibinfo {author} {\bibfnamefont {M.}~\bibnamefont {Rossmannek}}, \bibinfo {author} {\bibfnamefont {A.~C.}\ \bibnamefont {Vazquez}}, \bibinfo {author} {\bibfnamefont {F.}~\bibnamefont {Tacchino}}, \bibinfo {author} {\bibfnamefont {S.}~\bibnamefont {Mensa}}, \bibinfo {author} {\bibfnamefont {E.}~\bibnamefont {Altamura}}, \bibinfo {author} {\bibfnamefont {A.}~\bibnamefont {Alavi}}, \bibinfo {author} {\bibfnamefont {M.}~\bibnamefont {Motta}}, \bibinfo {author} {\bibfnamefont {J.}~\bibnamefont {Robledo-Moreno}}, \bibinfo {author} {\bibfnamefont {W.}~\bibnamefont {Kirby}}, \bibinfo {author} {\bibfnamefont {K.}~\bibnamefont {Sharma}}, \bibinfo {author} {\bibfnamefont {A.}~\bibnamefont {Mezzacapo}},\ and\ \bibinfo {author} {\bibfnamefont {I.}~\bibnamefont {Tavernelli}},\ }\href
  {https://arxiv.org/abs/2508.02578} {\enquote {\bibinfo {title} {Quantum chemistry with provable convergence via randomized sample-based krylov quantum diagonalization},}\ } (\bibinfo {year} {2026}),\ \Eprint {https://arxiv.org/abs/2508.02578} {arXiv:2508.02578 [quant-ph]} \BibitemShut {NoStop}%
\bibitem [{\citenamefont {Reinholdt}\ \emph {et~al.}(2025)\citenamefont {Reinholdt}, \citenamefont {Ziems}, \citenamefont {Kjellgren}, \citenamefont {Coriani}, \citenamefont {Sauer},\ and\ \citenamefont {Kongsted}}]{doi:10.1021/acs.jctc.5c00375}%
  \BibitemOpen
  \bibfield  {author} {\bibinfo {author} {\bibfnamefont {P.}~\bibnamefont {Reinholdt}}, \bibinfo {author} {\bibfnamefont {K.~M.}\ \bibnamefont {Ziems}}, \bibinfo {author} {\bibfnamefont {E.~R.}\ \bibnamefont {Kjellgren}}, \bibinfo {author} {\bibfnamefont {S.}~\bibnamefont {Coriani}}, \bibinfo {author} {\bibfnamefont {S.~P.~A.}\ \bibnamefont {Sauer}},\ and\ \bibinfo {author} {\bibfnamefont {J.}~\bibnamefont {Kongsted}},\ }\bibfield  {title} {\enquote {\bibinfo {title} {Critical limitations in quantum-selected configuration interaction methods},}\ }\href {https://doi.org/10.1021/acs.jctc.5c00375} {\bibfield  {journal} {\bibinfo  {journal} {Journal of Chemical Theory and Computation}\ }\textbf {\bibinfo {volume} {21}},\ \bibinfo {pages} {6811--6822} (\bibinfo {year} {2025})},\ \bibinfo {note} {pMID: 40586729},\ \Eprint {https://arxiv.org/abs/https://doi.org/10.1021/acs.jctc.5c00375} {https://doi.org/10.1021/acs.jctc.5c00375} \BibitemShut {NoStop}%
\bibitem [{\citenamefont {Holmes}, \citenamefont {Tubman},\ and\ \citenamefont {Umrigar}(2016{\natexlab{a}})}]{10.1021/acs.jctc.6b00407}%
  \BibitemOpen
  \bibfield  {author} {\bibinfo {author} {\bibfnamefont {A.~A.}\ \bibnamefont {Holmes}}, \bibinfo {author} {\bibfnamefont {N.~M.}\ \bibnamefont {Tubman}},\ and\ \bibinfo {author} {\bibfnamefont {C.~J.}\ \bibnamefont {Umrigar}},\ }\bibfield  {title} {\enquote {\bibinfo {title} {Heat-bath configuration interaction: An efficient selected configuration interaction algorithm inspired by heat-bath sampling},}\ }\href {https://doi.org/10.1021/acs.jctc.6b00407} {\bibfield  {journal} {\bibinfo  {journal} {Journal of Chemical Theory and Computation}\ }\textbf {\bibinfo {volume} {12}},\ \bibinfo {pages} {3674--3680} (\bibinfo {year} {2016}{\natexlab{a}})},\ \Eprint {https://arxiv.org/abs/https://pubs.acs.org/jctcce/article-pdf/12/8/3674/13458861/ct6b00407.pdf} {https://pubs.acs.org/jctcce/article-pdf/12/8/3674/13458861/ct6b00407.pdf} \BibitemShut {NoStop}%
\bibitem [{\citenamefont {Sharma}\ \emph {et~al.}(2017)\citenamefont {Sharma}, \citenamefont {Holmes}, \citenamefont {Jeanmairet}, \citenamefont {Alavi},\ and\ \citenamefont {Umrigar}}]{10.1021/acs.jctc.6b01028}%
  \BibitemOpen
  \bibfield  {author} {\bibinfo {author} {\bibfnamefont {S.}~\bibnamefont {Sharma}}, \bibinfo {author} {\bibfnamefont {A.~A.}\ \bibnamefont {Holmes}}, \bibinfo {author} {\bibfnamefont {G.}~\bibnamefont {Jeanmairet}}, \bibinfo {author} {\bibfnamefont {A.}~\bibnamefont {Alavi}},\ and\ \bibinfo {author} {\bibfnamefont {C.~J.}\ \bibnamefont {Umrigar}},\ }\bibfield  {title} {\enquote {\bibinfo {title} {Semistochastic heat-bath configuration interaction method: Selected configuration interaction with semistochastic perturbation theory},}\ }\href {https://doi.org/10.1021/acs.jctc.6b01028} {\bibfield  {journal} {\bibinfo  {journal} {Journal of Chemical Theory and Computation}\ }\textbf {\bibinfo {volume} {13}},\ \bibinfo {pages} {1595--1604} (\bibinfo {year} {2017})},\ \Eprint {https://arxiv.org/abs/https://pubs.acs.org/jctcce/article-pdf/13/4/1595/6792391/ct6b01028.pdf} {https://pubs.acs.org/jctcce/article-pdf/13/4/1595/6792391/ct6b01028.pdf} \BibitemShut {NoStop}%
\bibitem [{\citenamefont {Carleo}\ and\ \citenamefont {Troyer}(2017{\natexlab{a}})}]{doi:10.1126/science.aag2302}%
  \BibitemOpen
  \bibfield  {author} {\bibinfo {author} {\bibfnamefont {G.}~\bibnamefont {Carleo}}\ and\ \bibinfo {author} {\bibfnamefont {M.}~\bibnamefont {Troyer}},\ }\bibfield  {title} {\enquote {\bibinfo {title} {Solving the quantum many-body problem with artificial neural networks},}\ }\href {https://doi.org/10.1126/science.aag2302} {\bibfield  {journal} {\bibinfo  {journal} {Science}\ }\textbf {\bibinfo {volume} {355}},\ \bibinfo {pages} {602--606} (\bibinfo {year} {2017}{\natexlab{a}})},\ \Eprint {https://arxiv.org/abs/https://www.science.org/doi/pdf/10.1126/science.aag2302} {https://www.science.org/doi/pdf/10.1126/science.aag2302} \BibitemShut {NoStop}%
\bibitem [{\citenamefont {Halder}, \citenamefont {Anand},\ and\ \citenamefont {Maitra}(2025)}]{10.1021/acs.jpca.5c02346}%
  \BibitemOpen
  \bibfield  {author} {\bibinfo {author} {\bibfnamefont {S.}~\bibnamefont {Halder}}, \bibinfo {author} {\bibfnamefont {K.}~\bibnamefont {Anand}},\ and\ \bibinfo {author} {\bibfnamefont {R.}~\bibnamefont {Maitra}},\ }\bibfield  {title} {\enquote {\bibinfo {title} {Construction of chemistry-inspired dynamic ansatz utilizing generative machine learning},}\ }\href {https://doi.org/10.1021/acs.jpca.5c02346} {\bibfield  {journal} {\bibinfo  {journal} {The Journal of Physical Chemistry A}\ }\textbf {\bibinfo {volume} {129}},\ \bibinfo {pages} {5889--5900} (\bibinfo {year} {2025})},\ \Eprint {https://arxiv.org/abs/https://pubs.acs.org/jpcafh/article-pdf/129/26/5889/40721968/jp5c02346.pdf} {https://pubs.acs.org/jpcafh/article-pdf/129/26/5889/40721968/jp5c02346.pdf} \BibitemShut {NoStop}%
\bibitem [{\citenamefont {Herzog}\ \emph {et~al.}(2023)\citenamefont {Herzog}, \citenamefont {Casier}, \citenamefont {Lebègue},\ and\ \citenamefont {Rocca}}]{doi:10.1021/acs.jctc.2c01216}%
  \BibitemOpen
  \bibfield  {author} {\bibinfo {author} {\bibfnamefont {B.}~\bibnamefont {Herzog}}, \bibinfo {author} {\bibfnamefont {B.}~\bibnamefont {Casier}}, \bibinfo {author} {\bibfnamefont {S.}~\bibnamefont {Lebègue}},\ and\ \bibinfo {author} {\bibfnamefont {D.}~\bibnamefont {Rocca}},\ }\bibfield  {title} {\enquote {\bibinfo {title} {Solving the schrödinger equation in the configuration space with generative machine learning},}\ }\href {https://doi.org/10.1021/acs.jctc.2c01216} {\bibfield  {journal} {\bibinfo  {journal} {Journal of Chemical Theory and Computation}\ }\textbf {\bibinfo {volume} {19}},\ \bibinfo {pages} {2484--2490} (\bibinfo {year} {2023})},\ \bibinfo {note} {pMID: 37043718},\ \Eprint {https://arxiv.org/abs/https://doi.org/10.1021/acs.jctc.2c01216} {https://doi.org/10.1021/acs.jctc.2c01216} \BibitemShut {NoStop}%
\bibitem [{\citenamefont {Patra}\ \emph {et~al.}(2026{\natexlab{a}})\citenamefont {Patra}, \citenamefont {Mondal}, \citenamefont {Halder}, \citenamefont {Halder}, \citenamefont {Laskar}, \citenamefont {Goel},\ and\ \citenamefont {Maitra}}]{patra2026physicsinformedgenerativemachinelearning}%
  \BibitemOpen
  \bibfield  {author} {\bibinfo {author} {\bibfnamefont {C.}~\bibnamefont {Patra}}, \bibinfo {author} {\bibfnamefont {D.}~\bibnamefont {Mondal}}, \bibinfo {author} {\bibfnamefont {S.}~\bibnamefont {Halder}}, \bibinfo {author} {\bibfnamefont {D.}~\bibnamefont {Halder}}, \bibinfo {author} {\bibfnamefont {M.~R.}\ \bibnamefont {Laskar}}, \bibinfo {author} {\bibfnamefont {R.}~\bibnamefont {Goel}},\ and\ \bibinfo {author} {\bibfnamefont {R.}~\bibnamefont {Maitra}},\ }\bibfield  {title} {\enquote {\bibinfo {title} {Accelerated quantum-centric supercomputing through perturbation-theoretic measures and generative machine learning},}\ }\href {https://doi.org/10.1088/2058-9565/ae917f} {\bibfield  {journal} {\bibinfo  {journal} {Quantum Science and Technology}\ }\textbf {\bibinfo {volume} {11}},\ \bibinfo {pages} {035075} (\bibinfo {year} {2026}{\natexlab{a}})}\BibitemShut {NoStop}%
\bibitem [{\citenamefont {Patra}\ \emph {et~al.}(2026{\natexlab{b}})\citenamefont {Patra}, \citenamefont {V.}, \citenamefont {Bhat}, \citenamefont {P.}, \citenamefont {Maitra},\ and\ \citenamefont {G}}]{patra2026machinelearnedcompactsubspacegeneration}%
  \BibitemOpen
  \bibfield  {author} {\bibinfo {author} {\bibfnamefont {A.~K.}\ \bibnamefont {Patra}}, \bibinfo {author} {\bibfnamefont {A.~K.~S.}\ \bibnamefont {V.}}, \bibinfo {author} {\bibfnamefont {R.}~\bibnamefont {Bhat}}, \bibinfo {author} {\bibfnamefont {S.~S.}\ \bibnamefont {P.}}, \bibinfo {author} {\bibfnamefont {R.}~\bibnamefont {Maitra}},\ and\ \bibinfo {author} {\bibfnamefont {J.}~\bibnamefont {G}},\ }\href {https://arxiv.org/abs/2607.20585} {\enquote {\bibinfo {title} {Machine-learned compact subspace generation for quantum selected configuration interaction within density matrix embedding framework},}\ } (\bibinfo {year} {2026}{\natexlab{b}}),\ \Eprint {https://arxiv.org/abs/2607.20585} {arXiv:2607.20585 [quant-ph]} \BibitemShut {NoStop}%
\bibitem [{\citenamefont {Vargas}(2026)}]{vargas2026machinelearningsamplebasedquantum}%
  \BibitemOpen
  \bibfield  {author} {\bibinfo {author} {\bibfnamefont {N.~B.}\ \bibnamefont {Vargas}},\ }\href {https://arxiv.org/abs/2608.05314} {\enquote {\bibinfo {title} {Machine learning for sample-based quantum diagonalization: generative configuration recovery and the classical-simulability frontier},}\ } (\bibinfo {year} {2026}),\ \Eprint {https://arxiv.org/abs/2608.05314} {arXiv:2608.05314 [quant-ph]} \BibitemShut {NoStop}%
\bibitem [{\citenamefont {Coe}(2019)}]{10.1021/acs.jctc.9b00828}%
  \BibitemOpen
  \bibfield  {author} {\bibinfo {author} {\bibfnamefont {J.~P.}\ \bibnamefont {Coe}},\ }\bibfield  {title} {\enquote {\bibinfo {title} {Machine learning configuration interaction for ab initio potential energy curves},}\ }\href {https://doi.org/10.1021/acs.jctc.9b00828} {\bibfield  {journal} {\bibinfo  {journal} {Journal of Chemical Theory and Computation}\ }\textbf {\bibinfo {volume} {15}},\ \bibinfo {pages} {6179--6189} (\bibinfo {year} {2019})},\ \Eprint {https://arxiv.org/abs/https://pubs.acs.org/jctcce/article-pdf/15/11/6179/6176093/ct9b00828.pdf} {https://pubs.acs.org/jctcce/article-pdf/15/11/6179/6176093/ct9b00828.pdf} \BibitemShut {NoStop}%
\bibitem [{\citenamefont {Kingma}\ and\ \citenamefont {Welling}(2022)}]{kingma2022autoencodingvariationalbayes}%
  \BibitemOpen
  \bibfield  {author} {\bibinfo {author} {\bibfnamefont {D.~P.}\ \bibnamefont {Kingma}}\ and\ \bibinfo {author} {\bibfnamefont {M.}~\bibnamefont {Welling}},\ }\href {https://arxiv.org/abs/1312.6114} {\enquote {\bibinfo {title} {Auto-encoding variational bayes},}\ } (\bibinfo {year} {2022}),\ \Eprint {https://arxiv.org/abs/1312.6114} {arXiv:1312.6114 [stat.ML]} \BibitemShut {NoStop}%
\bibitem [{\citenamefont {Rezende}, \citenamefont {Mohamed},\ and\ \citenamefont {Wierstra}(2014)}]{rezende2014stochasticbackpropagationapproximateinference}%
  \BibitemOpen
  \bibfield  {author} {\bibinfo {author} {\bibfnamefont {D.~J.}\ \bibnamefont {Rezende}}, \bibinfo {author} {\bibfnamefont {S.}~\bibnamefont {Mohamed}},\ and\ \bibinfo {author} {\bibfnamefont {D.}~\bibnamefont {Wierstra}},\ }\href {https://arxiv.org/abs/1401.4082} {\enquote {\bibinfo {title} {Stochastic backpropagation and approximate inference in deep generative models},}\ } (\bibinfo {year} {2014}),\ \Eprint {https://arxiv.org/abs/1401.4082} {arXiv:1401.4082 [stat.ML]} \BibitemShut {NoStop}%
\bibitem [{\citenamefont {Higgins}\ \emph {et~al.}(2017)\citenamefont {Higgins}, \citenamefont {Matthey}, \citenamefont {Pal}, \citenamefont {Burgess}, \citenamefont {Glorot}, \citenamefont {Botvinick}, \citenamefont {Mohamed},\ and\ \citenamefont {Lerchner}}]{higgins2017betavae}%
  \BibitemOpen
  \bibfield  {author} {\bibinfo {author} {\bibfnamefont {I.}~\bibnamefont {Higgins}}, \bibinfo {author} {\bibfnamefont {L.}~\bibnamefont {Matthey}}, \bibinfo {author} {\bibfnamefont {A.}~\bibnamefont {Pal}}, \bibinfo {author} {\bibfnamefont {C.}~\bibnamefont {Burgess}}, \bibinfo {author} {\bibfnamefont {X.}~\bibnamefont {Glorot}}, \bibinfo {author} {\bibfnamefont {M.}~\bibnamefont {Botvinick}}, \bibinfo {author} {\bibfnamefont {S.}~\bibnamefont {Mohamed}},\ and\ \bibinfo {author} {\bibfnamefont {A.}~\bibnamefont {Lerchner}},\ }\bibfield  {title} {\enquote {\bibinfo {title} {beta-{VAE}: Learning basic visual concepts with a constrained variational framework},}\ }in\ \href {https://openreview.net/forum?id=Sy2fzU9gl} {\emph {\bibinfo {booktitle} {International Conference on Learning Representations}}}\ (\bibinfo {year} {2017})\BibitemShut {NoStop}%
\bibitem [{\citenamefont {Bowman}\ \emph {et~al.}(2016)\citenamefont {Bowman}, \citenamefont {Vilnis}, \citenamefont {Vinyals}, \citenamefont {Dai}, \citenamefont {Jozefowicz},\ and\ \citenamefont {Bengio}}]{bowman2016generatingsentencescontinuousspace}%
  \BibitemOpen
  \bibfield  {author} {\bibinfo {author} {\bibfnamefont {S.~R.}\ \bibnamefont {Bowman}}, \bibinfo {author} {\bibfnamefont {L.}~\bibnamefont {Vilnis}}, \bibinfo {author} {\bibfnamefont {O.}~\bibnamefont {Vinyals}}, \bibinfo {author} {\bibfnamefont {A.~M.}\ \bibnamefont {Dai}}, \bibinfo {author} {\bibfnamefont {R.}~\bibnamefont {Jozefowicz}},\ and\ \bibinfo {author} {\bibfnamefont {S.}~\bibnamefont {Bengio}},\ }\href {https://arxiv.org/abs/1511.06349} {\enquote {\bibinfo {title} {Generating sentences from a continuous space},}\ } (\bibinfo {year} {2016}),\ \Eprint {https://arxiv.org/abs/1511.06349} {arXiv:1511.06349 [cs.LG]} \BibitemShut {NoStop}%
\bibitem [{\citenamefont {Olivares-Amaya}\ \emph {et~al.}(2015)\citenamefont {Olivares-Amaya}, \citenamefont {Hu}, \citenamefont {Nakatani}, \citenamefont {Sharma}, \citenamefont {Yang},\ and\ \citenamefont {Chan}}]{10.1063/1.4905329}%
  \BibitemOpen
  \bibfield  {author} {\bibinfo {author} {\bibfnamefont {R.}~\bibnamefont {Olivares-Amaya}}, \bibinfo {author} {\bibfnamefont {W.}~\bibnamefont {Hu}}, \bibinfo {author} {\bibfnamefont {N.}~\bibnamefont {Nakatani}}, \bibinfo {author} {\bibfnamefont {S.}~\bibnamefont {Sharma}}, \bibinfo {author} {\bibfnamefont {J.}~\bibnamefont {Yang}},\ and\ \bibinfo {author} {\bibfnamefont {G.~K.-L.}\ \bibnamefont {Chan}},\ }\bibfield  {title} {\enquote {\bibinfo {title} {The ab-initio density matrix renormalization group in practice},}\ }\href {https://doi.org/10.1063/1.4905329} {\bibfield  {journal} {\bibinfo  {journal} {The Journal of Chemical Physics}\ }\textbf {\bibinfo {volume} {142}},\ \bibinfo {pages} {034102} (\bibinfo {year} {2015})}\BibitemShut {NoStop}%
\bibitem [{\citenamefont {Abraham~\emph{et. al}}(2021)}]{Qiskit}%
  \BibitemOpen
  \bibfield  {author} {\bibinfo {author} {\bibfnamefont {H.}~\bibnamefont {Abraham~\emph{et. al}}},\ }\href {https://doi.org/10.5281/zenodo.2573505} {\enquote {\bibinfo {title} {Qiskit: An open-source framework for quantum computing},}\ } (\bibinfo {year} {2021})\BibitemShut {NoStop}%
\bibitem [{\citenamefont {Sun}\ \emph {et~al.}(2020)\citenamefont {Sun}, \citenamefont {Zhang}, \citenamefont {Banerjee}, \citenamefont {Bao}, \citenamefont {Barbry}, \citenamefont {Blunt}, \citenamefont {Bogdanov}, \citenamefont {Booth}, \citenamefont {Chen}, \citenamefont {Cui}, \citenamefont {Eriksen}, \citenamefont {Gao}, \citenamefont {Guo}, \citenamefont {Hermann}, \citenamefont {Hermes}, \citenamefont {Koh}, \citenamefont {Koval}, \citenamefont {Lehtola}, \citenamefont {Li}, \citenamefont {Liu}, \citenamefont {Mardirossian}, \citenamefont {McClain}, \citenamefont {Motta}, \citenamefont {Mussard}, \citenamefont {Pham}, \citenamefont {Pulkin}, \citenamefont {Purwanto}, \citenamefont {Robinson}, \citenamefont {Ronca}, \citenamefont {Sayfutyarova}, \citenamefont {Scheurer}, \citenamefont {Schurkus}, \citenamefont {Smith}, \citenamefont {Sun}, \citenamefont {Sun}, \citenamefont {Upadhyay}, \citenamefont {Wagner}, \citenamefont {Wang}, \citenamefont {White}, \citenamefont {Whitfield}, \citenamefont
  {Williamson}, \citenamefont {Wouters}, \citenamefont {Yang}, \citenamefont {Yu}, \citenamefont {Zhu}, \citenamefont {Berkelbach}, \citenamefont {Sharma}, \citenamefont {Sokolov},\ and\ \citenamefont {Chan}}]{10.1063/5.0006074}%
  \BibitemOpen
  \bibfield  {author} {\bibinfo {author} {\bibfnamefont {Q.}~\bibnamefont {Sun}}, \bibinfo {author} {\bibfnamefont {X.}~\bibnamefont {Zhang}}, \bibinfo {author} {\bibfnamefont {S.}~\bibnamefont {Banerjee}}, \bibinfo {author} {\bibfnamefont {P.}~\bibnamefont {Bao}}, \bibinfo {author} {\bibfnamefont {M.}~\bibnamefont {Barbry}}, \bibinfo {author} {\bibfnamefont {N.~S.}\ \bibnamefont {Blunt}}, \bibinfo {author} {\bibfnamefont {N.~A.}\ \bibnamefont {Bogdanov}}, \bibinfo {author} {\bibfnamefont {G.~H.}\ \bibnamefont {Booth}}, \bibinfo {author} {\bibfnamefont {J.}~\bibnamefont {Chen}}, \bibinfo {author} {\bibfnamefont {Z.-H.}\ \bibnamefont {Cui}}, \bibinfo {author} {\bibfnamefont {J.~J.}\ \bibnamefont {Eriksen}}, \bibinfo {author} {\bibfnamefont {Y.}~\bibnamefont {Gao}}, \bibinfo {author} {\bibfnamefont {S.}~\bibnamefont {Guo}}, \bibinfo {author} {\bibfnamefont {J.}~\bibnamefont {Hermann}}, \bibinfo {author} {\bibfnamefont {M.~R.}\ \bibnamefont {Hermes}}, \bibinfo {author} {\bibfnamefont {K.}~\bibnamefont {Koh}},
  \bibinfo {author} {\bibfnamefont {P.}~\bibnamefont {Koval}}, \bibinfo {author} {\bibfnamefont {S.}~\bibnamefont {Lehtola}}, \bibinfo {author} {\bibfnamefont {Z.}~\bibnamefont {Li}}, \bibinfo {author} {\bibfnamefont {J.}~\bibnamefont {Liu}}, \bibinfo {author} {\bibfnamefont {N.}~\bibnamefont {Mardirossian}}, \bibinfo {author} {\bibfnamefont {J.~D.}\ \bibnamefont {McClain}}, \bibinfo {author} {\bibfnamefont {M.}~\bibnamefont {Motta}}, \bibinfo {author} {\bibfnamefont {B.}~\bibnamefont {Mussard}}, \bibinfo {author} {\bibfnamefont {H.~Q.}\ \bibnamefont {Pham}}, \bibinfo {author} {\bibfnamefont {A.}~\bibnamefont {Pulkin}}, \bibinfo {author} {\bibfnamefont {W.}~\bibnamefont {Purwanto}}, \bibinfo {author} {\bibfnamefont {P.~J.}\ \bibnamefont {Robinson}}, \bibinfo {author} {\bibfnamefont {E.}~\bibnamefont {Ronca}}, \bibinfo {author} {\bibfnamefont {E.~R.}\ \bibnamefont {Sayfutyarova}}, \bibinfo {author} {\bibfnamefont {M.}~\bibnamefont {Scheurer}}, \bibinfo {author} {\bibfnamefont {H.~F.}\ \bibnamefont {Schurkus}},
  \bibinfo {author} {\bibfnamefont {J.~E.~T.}\ \bibnamefont {Smith}}, \bibinfo {author} {\bibfnamefont {C.}~\bibnamefont {Sun}}, \bibinfo {author} {\bibfnamefont {S.-N.}\ \bibnamefont {Sun}}, \bibinfo {author} {\bibfnamefont {S.}~\bibnamefont {Upadhyay}}, \bibinfo {author} {\bibfnamefont {L.~K.}\ \bibnamefont {Wagner}}, \bibinfo {author} {\bibfnamefont {X.}~\bibnamefont {Wang}}, \bibinfo {author} {\bibfnamefont {A.}~\bibnamefont {White}}, \bibinfo {author} {\bibfnamefont {J.~D.}\ \bibnamefont {Whitfield}}, \bibinfo {author} {\bibfnamefont {M.~J.}\ \bibnamefont {Williamson}}, \bibinfo {author} {\bibfnamefont {S.}~\bibnamefont {Wouters}}, \bibinfo {author} {\bibfnamefont {J.}~\bibnamefont {Yang}}, \bibinfo {author} {\bibfnamefont {J.~M.}\ \bibnamefont {Yu}}, \bibinfo {author} {\bibfnamefont {T.}~\bibnamefont {Zhu}}, \bibinfo {author} {\bibfnamefont {T.~C.}\ \bibnamefont {Berkelbach}}, \bibinfo {author} {\bibfnamefont {S.}~\bibnamefont {Sharma}}, \bibinfo {author} {\bibfnamefont {A.~Y.}\ \bibnamefont
  {Sokolov}},\ and\ \bibinfo {author} {\bibfnamefont {G.~K.-L.}\ \bibnamefont {Chan}},\ }\bibfield  {title} {\enquote {\bibinfo {title} {Recent developments in the pyscf program package},}\ }\href {https://doi.org/10.1063/5.0006074} {\bibfield  {journal} {\bibinfo  {journal} {The Journal of Chemical Physics}\ }\textbf {\bibinfo {volume} {153}},\ \bibinfo {pages} {024109} (\bibinfo {year} {2020})}\BibitemShut {NoStop}%
\bibitem [{\citenamefont {Holmes}, \citenamefont {Tubman},\ and\ \citenamefont {Umrigar}(2016{\natexlab{b}})}]{doi:10.1021/acs.jctc.6b00407}%
  \BibitemOpen
  \bibfield  {author} {\bibinfo {author} {\bibfnamefont {A.~A.}\ \bibnamefont {Holmes}}, \bibinfo {author} {\bibfnamefont {N.~M.}\ \bibnamefont {Tubman}},\ and\ \bibinfo {author} {\bibfnamefont {C.~J.}\ \bibnamefont {Umrigar}},\ }\bibfield  {title} {\enquote {\bibinfo {title} {Heat-bath configuration interaction: An efficient selected configuration interaction algorithm inspired by heat-bath sampling},}\ }\href {https://doi.org/10.1021/acs.jctc.6b00407} {\bibfield  {journal} {\bibinfo  {journal} {Journal of Chemical Theory and Computation}\ }\textbf {\bibinfo {volume} {12}},\ \bibinfo {pages} {3674--3680} (\bibinfo {year} {2016}{\natexlab{b}})},\ \bibinfo {note} {pMID: 27428771},\ \Eprint {https://arxiv.org/abs/https://doi.org/10.1021/acs.jctc.6b00407} {https://doi.org/10.1021/acs.jctc.6b00407} \BibitemShut {NoStop}%
\bibitem [{\citenamefont {Pellow-Jarman}\ \emph {et~al.}(2025)\citenamefont {Pellow-Jarman}, \citenamefont {McFarthing}, \citenamefont {Kang}, \citenamefont {Yoo}, \citenamefont {Elala}, \citenamefont {Pellow-Jarman}, \citenamefont {Nakliang}, \citenamefont {Kim},\ and\ \citenamefont {Rhee}}]{pellowjarman2025hivqehandoveriterativevariational}%
  \BibitemOpen
  \bibfield  {author} {\bibinfo {author} {\bibfnamefont {A.}~\bibnamefont {Pellow-Jarman}}, \bibinfo {author} {\bibfnamefont {S.}~\bibnamefont {McFarthing}}, \bibinfo {author} {\bibfnamefont {D.~H.}\ \bibnamefont {Kang}}, \bibinfo {author} {\bibfnamefont {P.}~\bibnamefont {Yoo}}, \bibinfo {author} {\bibfnamefont {E.~E.}\ \bibnamefont {Elala}}, \bibinfo {author} {\bibfnamefont {R.}~\bibnamefont {Pellow-Jarman}}, \bibinfo {author} {\bibfnamefont {P.~M.}\ \bibnamefont {Nakliang}}, \bibinfo {author} {\bibfnamefont {J.}~\bibnamefont {Kim}},\ and\ \bibinfo {author} {\bibfnamefont {J.-K.~K.}\ \bibnamefont {Rhee}},\ }\href {https://arxiv.org/abs/2503.06292} {\enquote {\bibinfo {title} {Hivqe: Handover iterative variational quantum eigensolver for efficient quantum chemistry calculations},}\ } (\bibinfo {year} {2025}),\ \Eprint {https://arxiv.org/abs/2503.06292} {arXiv:2503.06292 [quant-ph]} \BibitemShut {NoStop}%
\bibitem [{\citenamefont {Carleo}\ and\ \citenamefont {Troyer}(2017{\natexlab{b}})}]{carleo2017solving}%
  \BibitemOpen
  \bibfield  {author} {\bibinfo {author} {\bibfnamefont {G.}~\bibnamefont {Carleo}}\ and\ \bibinfo {author} {\bibfnamefont {M.}~\bibnamefont {Troyer}},\ }\bibfield  {title} {\enquote {\bibinfo {title} {Solving the quantum many-body problem with artificial neural networks},}\ }\href@noop {} {\bibfield  {journal} {\bibinfo  {journal} {Science}\ }\textbf {\bibinfo {volume} {355}},\ \bibinfo {pages} {602--606} (\bibinfo {year} {2017}{\natexlab{b}})}\BibitemShut {NoStop}%
\bibitem [{\citenamefont {Choo}, \citenamefont {Mezzacapo},\ and\ \citenamefont {Carleo}(2020)}]{choo2020fermionic}%
  \BibitemOpen
  \bibfield  {author} {\bibinfo {author} {\bibfnamefont {K.}~\bibnamefont {Choo}}, \bibinfo {author} {\bibfnamefont {A.}~\bibnamefont {Mezzacapo}},\ and\ \bibinfo {author} {\bibfnamefont {G.}~\bibnamefont {Carleo}},\ }\bibfield  {title} {\enquote {\bibinfo {title} {Fermionic neural-network states for ab-initio electronic structure},}\ }\href@noop {} {\bibfield  {journal} {\bibinfo  {journal} {Nature communications}\ }\textbf {\bibinfo {volume} {11}},\ \bibinfo {pages} {2368} (\bibinfo {year} {2020})}\BibitemShut {NoStop}%
\end{thebibliography}
\end{document}